\documentclass[aip,amsmath,amssymb,reprint]{revtex4-2}

\usepackage{graphicx}
\usepackage{dcolumn}
\usepackage{bm}

\usepackage[utf8]{inputenc}
\usepackage[T1]{fontenc}
\usepackage{mathptmx}
\usepackage[version=3]{mhchem} 
\usepackage{braket}
\usepackage{amsmath,amssymb}
\usepackage{physics}
\usepackage{amsmath}
\usepackage{bbold}
\usepackage{braket}
\usepackage{xcolor}
\usepackage{placeins}
\usepackage{ulem}
\usepackage{natbib}

\begin{document}


\title[]{A Semiclassical Framework for Mixed Quantum Classical Dynamics}
\author{Shreyas Malpathak}
\affiliation{Department of Chemistry and Chemical Biology, Baker Laboratory, Cornell University Ithaca,14853 NY, USA}
\author{Matthew S. Church}
\affiliation{Department of Chemistry, Brown University, Providence, RI 02906, USA}
\author{Nandini Ananth}
\affiliation{Department of Chemistry and Chemical Biology, Baker Laboratory, Cornell University Ithaca,14853 NY, USA}
\email{ananth@cornell.edu}


\date{\today}

\begin{abstract}
    Semiclassical approximations for quantum dynamic 
    simulations in complex chemical systems range from 
    rigorously accurate methods that are computationally expensive
    to methods that exhibit near-classical scaling with 
    system size but are limited in their ability to describe
    quantum effects. In practical studies of high-dimensional
    reactions, neither extreme is the best choice: frequently
    a high-level quantum mechanical description is only required 
    for a handful of modes, while the majority of environment
    modes that do not play a key role in the reactive event of 
    interest are well served with a lower level of theory. In 
    this feature we introduce Modified Filinov filtration as 
    a powerful tool for mixed quantum-classical simulations in 
    a uniform semiclassical framework.
\end{abstract}

\maketitle

\section{Introduction}
The development of approximate quantum dynamic methods 
for the simulation of complex reactions 
remains an outstanding challenge despite several decades 
of active research. 
In the search for a rigorous method capable of predictive simulations,
approximations based on Semiclassical (SC) theory have emerged as a 
practical route to the calculation of real-time quantum correlation
functions.~\cite{Miller2001a,Tannor2000,
Thoss2004,Kay2005,Stock2005,Heller2006,Kapral2006,Pollak2007,Miller2009,McRobbie2009b,
Liu2015,Lee2016,Lu2019,Conte2020c, Bonfanti2020,Vanicek2021,Loring2022}
Classifying SC methods by the extent to which they can 
capture quantum mechanical behaviors including nonadiabatic effects, 
deep tunneling, 
and nuclear coherence, we find a clear trend: 
the more rigorous, `quantum-limit' approximations are computationally expensive
and limited to dynamic calculations in model systems and small molecules,
while the less accurate `classical-limit' methods lend themselves
to complex system simulations.
In order to mitigate this inverse relationship between computational
cost and accuracy, several methods have also been developed that 
further approximate quantum-limit SC approximations, and these have been 
used to successfully calculate reaction rates~\cite{Venkataraman2007,Sun1998,Wang1998,Shi2003c,
BeingJ.Ka2005,Navrotskaya2006,Being2006,Being2006a,Vazquez2010,Wang2000a,Tannor2000,Grossmann2000,Burant2002}, 
compute linear and non-linear spectra~\cite{Sepulveda1996,Liu2011a,Poulsen2005b,Poulsen2006b,Beutier2015,
Kaledin2003c,Kaledin2003d,Kaledin2004,Kuhn1999,Issack2007b,Wong2011c,
Liberto2016b,Ceotto2009c,Ceotto2009a,Ceotto2010c,Conte2013c,Gabas2017,
Ceotto2011,Aieta2020,Aieta2020a,Micciarelli2019,Ceotto2017c,Liberto2018,
Liberto2018a,Gabas2018,Gabas2019,Bertaina2019c,Gandolfi2020c,Ovchinnikov1996e,
Ovchinnikov1998g,Grossmann2016,Buchholz2012,Buchholz2016,Buchholz2017,Buchholz2018}, 
and simulate nonadiabatic dynamics.~\cite{Sun1997,Sun1998b,Batista1998,Rabani1999,Wang1999,
    Thoss2000,Shi2004,Shi2005,Bonella2005,Ananth2007,Shi2008,McRobbie2009b,Miller2009,Miller2010,
Venkataraman2011,Huo2011,Huo2012,Cotton2013,Tao2013b,Sun2015,Lee2016,Sun2016,Sun2016b,
Sun2016c,Teh2017,Kananenka2017,Sun2018,Provazza2018,Provazza2019,
Mulvihill2019,Gao2020,Gao2020b,Dodin2022,Kumar2021} 
Despite these advances, rigorous quantum-limit 
SC studies of high-dimensional chemical systems 
remains largely out of reach.

Within SC methods, Semiclassical Initial Value Representation (SC-IVR)
has, arguably, shown the most sustained promise
in the simulation of chemical systems.~\cite{Miller2001a, Miller2009} 
In the path integral representation of quantum mechanics, 
the real-time quantum propagator is obtained by summing over 
the phase contributions of all possible paths connecting an initial
and final configuration in time $t$.~\cite{Feynman1965}
The original Van-Vleck (VV) SC approximation to the propagator 
is derived using a stationary phase approximation that truncates
this sum over all paths to include only paths of stationary action (classical paths)
with a prefactor that accounts for contributions from near-classical 
paths.~\cite{VanVleck1928} However, finding all classical paths 
between a fixed initial and final configuration is still challenging 
(the boundary value problem). The IVR framework addresses this
by replacing the integrals over fixed final configurations with 
integrals over initial momenta, reframing the problem as one of 
sampling initial phase space conditions from which deterministic
classical trajectories can be generated.~\cite{Miller1970,Heller1991,Heller1991a,Miller1991} 
The VV-IVR propagator employs a position-state basis, while using 
an over-complete coherent state basis results in the popular Herman-Kluk
(HK)-IVR approximation to the propagator.~\cite{Herman1984}

SC-IVR based approximations to real-time quantum correlation functions
involve two propagators \textemdash\, one forward in time and one 
backward in time. Employing the HK-IVR approximation for both propagators, 
leads to the Double HK (DHK)-IVR approximation to the correlation function,
and this method, along with its near-relation the DVV-IVR approximation,~\cite{Venkataraman2007}
represent quantum-limit SC approximations capable
of capturing almost almost all quantum effects in 
low-dimensional systems.~\cite{Miller2001a}
As system dimensionality is increased, both these quantum-limit approximations 
become computationally intractable due to the infamous SC sign problem. 
Succinctly, the sign problem refers to the difficulty
numerically converging an integral over a multidimensional,
rapidly oscillating function.
There have been many efforts over the years to mitigate
this sign problem in SC theory and here we describe a few 
strategies employed in the calculation of SC-IVR based 
approximations to correlation functions.

Computing a quantum-limit correlation 
function involves generating classical forward and backward 
trajectories from properly sampled initial conditions,
evaluating a complex SC prefactor associated with each 
trajectory, and then evaluating an oscillatory integrand with
a phase that is largely attributed to the difference in action 
between the forward and backward trajectories. Linearizing
in this forward-backward action difference mitigates the 
sign problem by significantly reducing the oscillatory 
structure of the integrand and leading 
to the well established, classical-limit 
Linearized Semiclassical (LSC)-IVR approximation.~\cite{Sun1997c,Sun1998,Wang1998,Shi2003,Liu2015}  
However, while LSC-IVR and the related Husimi-IVR,~\cite{YiZhao2002,Wright2004} 
can account for some quantum effects like zero-point energy and 
shallow tunneling, these methods cannot be used to describe 
processes where deep tunneling or interference effects play
a key role.~\cite{Miller2001a, Gelabert2001,Liu2015} 
Despite this limitation, LSC-IVR in particular finds 
extensive application in the calculation of dynamic observables
for high-dimensional systems, and in the condensed 
phase~\cite{Shi2003c,BeingJ.Ka2005,Navrotskaya2006,Being2006,Being2006a,Vazquez2010,Liu2006,Liu2007,Liu2008,Liu2011a,Liu2011,Monteferrante2013,Poulsen2003,Poulsen2004,Poulsen2004b,Poulsen2005b,Poulsen2006b,Poulsen2007,Beutier2015}
where they provide quantum dynamic information
at a computational cost comparable to classical 
molecular dynamics (MD) simulations.

Strategies to mitigate the sign problem while 
retaining the ability to capture some interference 
effects have focused on improved phase space sampling
for initial conditions.
For instance, it has been shown that correlated sampling
the forward and backward trajectory 
initial conditions or employing a time-dependent sampling function 
can lead to more rapid convergence of the oscillatory 
integrand.~\cite{Pan2013,Tao2011,Tao2012,Tao2013,Tao2014c}
Similarly, time-averaging (TA)-SC-IVR,~\cite{Elran1999,Elran1999a,Kaledin2003c,Kaledin2003d,Kaledin2004,Issack2007} 
implemented by drawing initial conditions from
phase space configurations along a classical trajectory 
has been shown to significantly improve convergence.~\cite{Kaledin2003c}
Other efforts to tame the phase of the oscillatory integrand 
have resulted in methods like the Forward-Backward IVR~\cite{Makri1998a,
Thompson1999a,Thompson1999b,Wang2000a,Gelabert2001,Nakayama2003,Wright2003,
Makri2004,Lawrence2004,Makri2011,Kuhn1999,Tao2009} 
that control the extent to which forward and backward trajectories
differ rather than linearizing the action to make them coincide.
This gives rise to methods that can still capture some coherence 
effects unlike the classical-limit SC-IVR approximations.~\cite{Gelabert2001} 

The severity of the sign problem increases with the number 
of system degrees of freedom motivating the development 
of hybrid/fragment SC methods, where the number of modes that contribute to 
the overall phase is limited by partitioning the system into subsystems that are 
then described using different levels of SC (or related) theory. 
Proposed multi-physics methods include the use of LSC-IVR for the more
classical subsystem\cite{Sun1997c}, combining VV-IVR for the quantum subsystem with a prefactor-free 
VV-IVR for the rest\cite{Ovchinnikov1996e,Ovchinnikov1998g}, 
and the Semiclassical hybrid dynamics method~\cite{Grossmann2006,Grossmann2016,Goletz2009,
Goletz2010,Buchholz2012,Buchholz2016,Buchholz2017,Buchholz2018} 
that employs HK-IVR for the quantum subsystem and its near-relation, Thawed Gaussian Wavepacket 
dynamics,~\cite{Heller1975, Littlejohn1986, Deshpande2006}
for the rest of the system.
In general, these multi-physics approaches 
introduce additional approximations in describing 
inter-subsystem interactions and work best when this coupling
is very weak.

We note that a second computational bottleneck in 
quantum-limit SC-IVR simulations arises from the complex prefactor
that is evaluated by taking the square root of a complex 
determinant constructed from elements of the Monodromy 
matrices.~\cite{Herman1984, Venkataraman2007}
The log-derivative prefactor approach side-steps 
evaluating the complex square root in the HK-IVR 
prefactor and simplifies the equations of 
motion considerably.~\cite{Gelabert2000d} 
Other methods to reduce the cost of 
computing the SC prefactor include 
the Johnson multi-channel 
approximation,~\cite{Issack2005,Issack2007,Issack2007a,Issack2007b,Wong2011c}
the adiabatic approximation,\cite{Guallar1999,Guallar2000}
the so-called poor person's approximation,
\cite{Tatchen2011} 
and others introduced specifically to deal 
with chaotic systems.~\cite{Liberto2016b} 
A different approach that circumvents the calculation of pre-factors 
is the pre-factor-free SC-IVR series approach~\cite{Zhang2004} 
that is derived from more general SC-IVR series formalism developed
systematically to correct for differences between the 
HK-IVR propagator and the exact quantum propagator~\cite{Pollak2003,Zhang2003,Zhang2003a}. 
On-the-fly simulations of high dimensional systems have been undertaken 
in the SC framework, some 
leveraging both prefactor approximations 
and a modified implementation of the TA-SC-IVR approach.~\cite{Ceotto2009c,
    Ceotto2009a,Ceotto2010c,Wong2011c,Conte2013c,Gabas2017,Ceotto2011,Aieta2020,Aieta2020a,
Micciarelli2019,Ceotto2017c,Liberto2018,Liberto2018a,Gabas2018,Gabas2019,
Bertaina2019c,Gandolfi2020c} 

In this feature, we describe a new SC framework that allows for different 
degrees of freedom to be treated at different levels of SC theory without 
introducing any {\it adhoc} approximations to capture interactions between them. 
This is achieved through a mode-specific phase filtration scheme, Modified Filinov
Filtration (MFF), that when applied to DHK-IVR effectively controls the extent to which 
each degrees of freedom contributes to the phase of the integrand. 
By varying the filter strength, it is possible to limit contributions
to the overall phase to a handful of important degrees of freedom,
while the rest are treated in the classical-limit, 
or on a continuum between these two limits. 
We introduce the derivation of this Mixed Quantum-Classical (MQC)-IVR method 
and discuss the applications to-date, as well as some of the limitations. 
We demonstrate that, for linear operator, changing the strength of the phase filter 
systematically changes the MQC-IVR expression for a real-time correlation function from 
the quantum-limit DHK-IVR to the classical-limit Husimi IVR.~\cite{Antipov2015,Church2017}
We show numerical results obtained by using this method to characterize
the dynamics in multidimensional systems with strong and weak inter-mode coupling,~\cite{Antipov2015, Church2017, Church2019a} 
to model multi-channel scattering in nonadiabatic model systems,~\cite{Church2018} 
and to calculate condensed phase reaction rates.~\cite{Church2019a} 
We note the limitations of the MQC-IVR as it currently stands for 
the evaluation of non-linear operator correlation functions 
and the computational cost associated with the MQC prefactor. 
We then introduce a new prefactor-free MQC method obtained by 
applying MFF to the exact, real-time, path integral 
expression for correlation functions. 
The resulting Filinov-Filtered Path Integral (FFPI) expression for the correlation function 
moves systematically from an exact path integral expression to a classical-limit
LSC-IVR correlation function as the phase of the Filinov filter is increased. 
We conclude by discussing the challenges in
sampling path space in this new framework, potential implementation strategies,
and more generally, the future of MQC-SC methods.

\section{Mixed Quantum Classical Initial Value Representation}

As described earlier, 
the oscillatory phase of the DHK-IVR integrand is primarily 
due to the action difference between pairs of forward-backward trajectories.
This motivates the derivation of the MQC-IVR approximation
using MFF to control the extent to which individual degrees 
of freedom contribute to the action
difference, and therefore the overall phase of the DHK-IVR
integrand.

\subsection{DHK-IVR approximation for real-time correlation functions}
Experimentally measured dynamic observables 
like time-dependent expectation values, reaction rates, and spectra
frequently correspond to real-time quantum correlation functions,
\begin{align}
    C_{AB}(t) &= \text{Tr}\left[\hat{A}e^{i\hat{H}t/\hbar}\hat{B}e^{-i\hat{H}t/\hbar}\right],
   \label{eq:cf1}
\end{align}
where Tr signifies the trace, $\hat{H}$ is the system Hamiltonian,
and $\hat{A}$ and $\hat{B}$ are the operators of interest.
For condensed phase systems, 
the density operator, $\hat\rho$, is typically grouped with the observable, 
$\hat{A}_\rho = \hat{\rho}\hat{A}$.

\begin{figure}
    \includegraphics[width=0.45\textwidth]
    {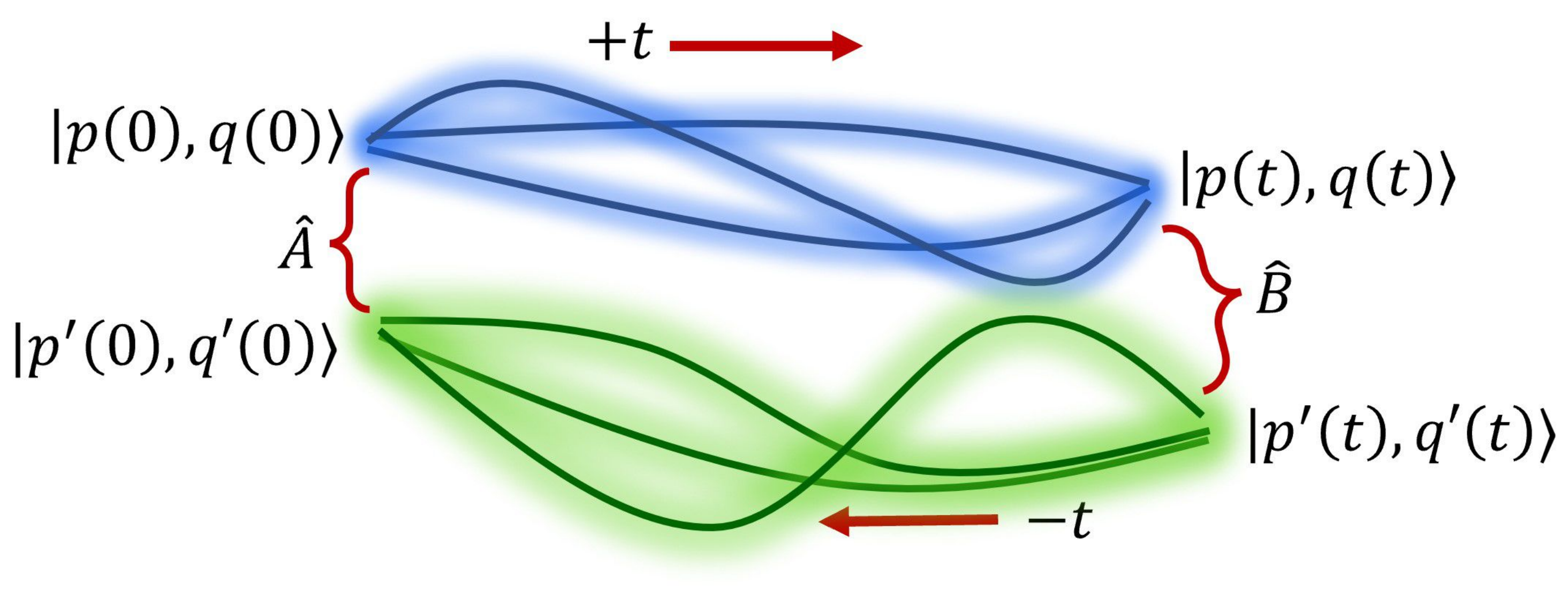}
    \caption{A schematic showing the structure of a DHK-IVR 
        correlation function $C_{AB}(t)$. The initial coherent 
        state of the system at time zero, $\ket{p_0,q_0}$, 
        is time-evolved forward to state $\ket{p_t, q_t}$. 
        Operator $\hat{B}$ acts on $\ket{p_t, q_t}$ to 
        generate a new state, $\ket{p_t^\prime, q_t^\prime}$, 
        which is then evolved backward in time to a final 
        state, $\ket{p_0^\prime,q_0^\prime}$.
     Matrix elements of operators $\hat A$ and $\hat B$ are evaluated
     at time $t=0$ and $t$ respectively.
      The classical paths that contribute to the DHK-IVR 
  integrand are shown with solid lines and the surrounding
  region highlighted to indicate that the HK prefactor captures
  contributions from near-classical paths as well.}
    \label{fig:scheme1}
\end{figure}

The DHK-IVR approximation to the real-time quantum correlation function is,
\begin{align}
   C^{\text{DHK}}_{AB}(t) &= \frac{1}{\left(2\pi\hbar\right)^{2F}} \int d\textbf{p}_{0}\int d\textbf{q}_{0}\int d\textbf{p}_{t}^{'}\int d\textbf{q}_{t}^{'}C_{t}\left(\textbf{p}_{0},\textbf{q}_{0}\right) \notag \\
   & \times C_{-t}(\textbf{p}_{t}^{'},\textbf{q}_{t}^{'})\mel{\textbf{p}_{0}\textbf{q}_{0}}{\hat{A}}{\textbf{p}_{0}^{'}\textbf{q}_{0}^{'}}
   \mel{\textbf{p}_{t}^{'}\textbf{q}_{t}^{'}}{\hat{B}}{\textbf{p}_{t}\textbf{q}_{t}} \notag  \\
   & \times e^{i\left[S_{t}(\textbf{p}_{0},\textbf{q}_{0})+
   S_{-t}(\textbf{p}_{t}^{'},\textbf{q}_{t}^{'})\right]/\hbar},
   \label{eq:cf-dhk}
\end{align}
where $F$ is the dimensionality of the system, the forward 
path is described by initial phase space variables,
($\textbf{p}_0, \textbf{q}_0$), that are time evolved classically 
to the final variables, ($\textbf{p}_t, \textbf{q}_t$), and backward 
path defined by inverse time evolution from ($\textbf{p}_t^\prime, \textbf{q}_t^\prime$)
to ($\textbf{p}_0^\prime, \textbf{q}_0^\prime$) as sketched 
in Fig~\ref{fig:scheme1}.
In Eq.~\ref{eq:cf-dhk}, the phase of the integrand 
is determined by the forward-backward action difference,
$S_{t}(\textbf{p}_{0},\textbf{q}_{0})+S_{-t}(\textbf{p}_{t}^{'},\textbf{q}_{t}^{'})$, and 
the Herman-Kluk prefactor for the forward trajectory 
defined as

\begin{align}
    C_{t}^{2}\left(\textbf{p}_{0},\textbf{q}_{0}\right) &= \text{det}\left|\frac{1}{2}\left[\gamma_{t}^{\frac{1}{2}}\textbf{M}_{qq}^{f}\gamma_{0}^{-\frac{1}{2}}+\gamma_{t}^{-\frac{1}{2}}\textbf{M}_{pp}^{f}\gamma_{0}^{\frac{1}{2}} \right. \right. \notag \\
    & \left. \left. -i\hbar\gamma_{t}^{\frac{1}{2}}\textbf{M}_{qp}^{f}\gamma_{0}^{\frac{1}{2}}+\frac{i}{\hbar}\gamma_{t}^{-\frac{1}{2}}\textbf{M}_{pq}^{f}\gamma_{0}^{-\frac{1}{2}}\right]\right|,
    \label{eq:hk_pre}
\end{align}
where $\textbf{M}_{\alpha\beta}^{f}=\frac{\partial\alpha_t}{\partial\beta_0}$ 
are the Monodromy matrices along the forward path, and $C_{-t}$ is similarly defined
in terms of backward path Monodromy matrices.~\cite{Herman1984} 

\subsection{Modified Filinov Filtration}
Consider the integral over an $F$-dimensional oscillatory function,
\begin{equation}
    I=\int d\textbf{r} ~g(\textbf{r}) e^{i\phi(\textbf{r})},
    \label{eq:osc-int}
\end{equation}
where $g(\textbf{r})$ is, in general, complex-valued 
and $\phi(\textbf{r})$ is a real-valued phase.
The Modified Filinov filtration (MFF)\cite{Filinov1986a,Makri1987c,Makri1988b} 
approximation to the integral in Eq.~\ref{eq:osc-int} 
is 
\begin{equation}
     I(\textbf{c})=\int d\textbf{r} ~g(\textbf{r}) e^{i\phi(\textbf{r})}
     F(\textbf{r};\textbf{c}), 
     \label{eq:fil-int}
\end{equation}
where $F(\textbf{r};\textbf{c})$ is the smoothing factor,
\begin{equation}
 F(\textbf{r};\textbf{c})=\text{det}\left|\mathbb{1}+
 i\textbf{c}\frac{\partial^2\phi}{\partial\textbf{r}^2}\right|^{\frac{1}{2}}
 e^{-\frac{1}{2}\frac{\partial \phi}{\partial\textbf{r}}^T.
 \textbf{c}.\frac{\partial \phi}{\partial\textbf{r}}},
 \label{eq:fil-sf}
\end{equation} 
and elements of the diagonal, $F \times F$, Filinov parameter matrix, $\textbf{c}$,
determine the strength of the phase filter.
When all the Filinov parameters are zero, 
the integral $I$ in Eq.~\ref{eq:fil-int} is 
identical to the original integrand.
In the limit of infinitely large Filinov parameters,
it can be shown that $I({\textbf{c}})$ in Eq.~\ref{eq:fil-int} 
corresponds to a stationary phase 
approximation to the integral in Eq.~\ref{eq:osc-int}.~\cite{Makri1987c}
For finite, non-zero values of the Filinov parameters, the 
MFF integrand is less oscillatory than the original,
and by choosing the elements of the Filinov parameter matrix 
to be distinct, it becomes possible to tune 
the extent to which an individual degree of freedom 
contribute to the overall phase.

\subsection{Deriving MQC-IVR}
By filtering the phase of the DHK-IVR integrand using 
MFF, we obtain the MQC-IVR correlation function,
\begin{widetext}
\begin{align}
    C^{\text{MQC}}_{AB}(t;{\bf c}) &= \frac{1}{\left(2\pi\hbar\right)^{2F}} \int d\textbf{p}_{0}\int d\textbf{q}_{0}\int d\bm{\Delta}_{p_t}\int d\bm{\Delta}_{q_t}D\left(\textbf{p}_{0},\textbf{q}_{0},\bm{\Delta}_{p_t},\bm{\Delta}_{q_t},\textbf{c}_{p},\textbf{c}_{q}\right)\mel{\textbf{p}_{0}\textbf{q}_{0}}{\hat{A}}{\textbf{p}_{0}^{'}\textbf{q}_{0}^{'}}\notag \\ 
    & \times \mel{\textbf{p}_{t}^{'}\textbf{q}_{t}^{'}}{\hat{B}}{\textbf{p}_{t}\textbf{q}_{t}} e^{i\left[S_{t}(\textbf{p}_{0},\textbf{q}_{0})+S_{-t}(\textbf{p}_{t}^{'},\textbf{q}_{t}^{'})\right]/\hbar}\,e^{-\frac{1}{2}\bm{\Delta}_{q_t}^{T}\cdot\textbf{c}_{q}\cdot\bm{\Delta}_{q_t}}\,e^{-\frac{1}{2}\bm{\Delta}_{p_t}^{T}\cdot\textbf{c}_{p}\cdot\bm{\Delta}_{p_t}},
   \label{eq:cf-mqc-fb}    
\end{align}
\end{widetext}
where the complex prefactor, $D$, is defined in Appendix A. As previously discussed, the classical-limit corresponds to forward and backward trajectories that coincide, 
$\bm{\Delta}_{p_{t}}~=~\bm{\Delta_{q_t}}~=~0$,
with net zero action. Trajectory pairs with finite displacements,
$\bm{\Delta}_{p_{t}} = \textbf{p}_{t}^{'}-\textbf{p}_{t}$ and 
$\bm{\Delta}_{q_{t}} = \textbf{q}_{t}^{'}-\textbf{q}_{t}$,
contribute a significant non-zero phase $S_t+S_{-t}$. 
In Eq.~\ref{eq:cf-mqc-fb}, we see that the $F\times F$ diagonal Filinov
matrices, $\textbf{c}_{p}$ and $\textbf{c}_{q}$, parameterize
the width of Gaussians in $\bm{\Delta}_{p_t}$ and $\bm{\Delta}_{q_t}$, 
respectively, and therefore, determine the types 
of forward-backward paths that contribute to the overall 
phase of the MQC-IVR integrand. 
In the limit $\{\textbf{c}_{p}$,$\textbf{c}_{q}\}$ $\to$ 0, 
Eq.~\eqref{eq:cf-mqc-fb} reduces to the DHK-IVR expression 
Eq.~\eqref{eq:cf-dhk} for the correlation function and 
in the $\{\textbf{c}_{p}$,$\textbf{c}_{q}\} \to \infty$ limit, 
$\{\bm{\Delta}_{p_t},\bm{\Delta}_{q_t}\} \to 0$, 
and for {\it linear operators} we obtain the classical limit 
Husimi-IVR expression,~\cite{Antipov2015}
\begin{align}
    C^{\text{Hus}}_{AB}(t) &= \frac{1}{\left(2\pi\hbar\right)^{2F}} 
    \int d\textbf{p}_{0}\int d\textbf{q}_{0}
    \mel{\textbf{p}_{0}\textbf{q}_{0}}{\hat{A}}{\textbf{p}_{0}^{'}\textbf{q}_{0}^{'}} \notag \\
    & \times \mel{\textbf{p}_{t}^{'}\textbf{q}_{t}^{'}}{\hat{B}}{\textbf{p}_{t}\textbf{q}_{t}}.
    \label{eq:cf-husimi}
\end{align}
We note that a previous work applying MFF to DHK-IVR resulted in 
the Generalized Forward-Backward (GFB)-IVR.~\cite{Thoss2001a} 
Formally, the difference between the GFB-IVR and MQC-IVR correlation
functions is simply a matter of the choice of phase to be filtered: 
MQC-IVR filters only the phase due to the action difference, whereas
GFB-IVR also filters phase contributions from the matrix elements 
of operator $\hat B$. In the zero filter limit, both GFB-IVR and MQC-IVR 
correspond to DHK-IVR, however, when the Filinov parameter is 
set to large values, the GFB-IVR correlation function becomes identical
to the FB-IVR correlation function, a quantum-limit method in our classification
with only marginally improved numerical convergence 
properties.~\cite{Thoss2001a}

The implementation of the MQC-IVR correlation function is similar 
to other quantum-limit SC methods like the DHK-IVR. Trajectory
initial conditions are obtained by Monte Carlo sampling initial
phase space positions $(\textbf{p}_0, \textbf{q}_0)$ 
for the forward trajectories and the 
difference variables, $(\bm{\Delta}_{p_t}, \bm{\Delta}_{q_t})$, 
at time $t$ to generate initial conditions for the backward
trajectories. Note that this `forward-backward' implementation described here 
is less efficient than a more recently introduced
double-forward (DF) implementation where the phase space variables at time zero, 
$\left( \bm{p}_0, \bm{q}_0, \bm{p}_0^\prime, \bm{q}_0^\prime \right)$ are Monte Carlo sampled, 
allowing for two independent forward trajectories to be generated.~\cite{Church2017}
All MQC-IVR results presented in this article
were generated using the SC-Corr code package, an open-source program developed 
in-house.~\cite{corrcode}

\subsection{Numerical Study of Phase Filtering in MQC-IVR}
The phase filtration achieved in the MQC-IVR framework is best 
understood and demonstrated with a model system.
We choose to work with the real-time position correlation function
for a 1D anharmonic oscillator initially in a non-stationary state,
where the expected quantum-limit amplitude recurrences are
systematically damped as the filter strength is increased 
and the MQC-IVR correlation function approaches the classical-limit.~\cite{Antipov2015}
Detailed analysis establishes that the Filinov filter 
effectively acts to reduce noise arising from numerically 
integrating over regions where the integrand has near-zero 
amplitude but highly oscillatory phase.~\cite{Church2017}

The specific model details are as follows: a particle of mass 
$m_x=1$ a.u. is subject to an anharmonic potential,
\begin{align}
    V(x)=\frac{1}{2}m_x \omega_x x^2 - 0.1 x^3 + 0.1x^4
    \label{eq:1dao}
\end{align}
where $\omega_x = \sqrt{2}$ a.u. 
For the position correlation function, we define 
operator $\hat{A}\equiv\ket{p_i q_i}\bra{p_i q_i}$
and $\hat B \equiv \hat x$ where 
the initial coherent state wavefunction is
\begin{align}
    \braket{x}{p_iq_i} = \left( \frac{\gamma_x}{\pi} \right)^{\frac{1}{4}}
    e^{-\frac{\gamma_x}{2}\left( x - q_i \right)^2 + ip_i \left( x-q_i \right)},
\end{align}
with $\gamma_x=\sqrt{2}$, $p_i = 0$ and $q_i=1$, all in atomic units.

\begin{figure}
    \centering
    \includegraphics[width=0.45\textwidth]{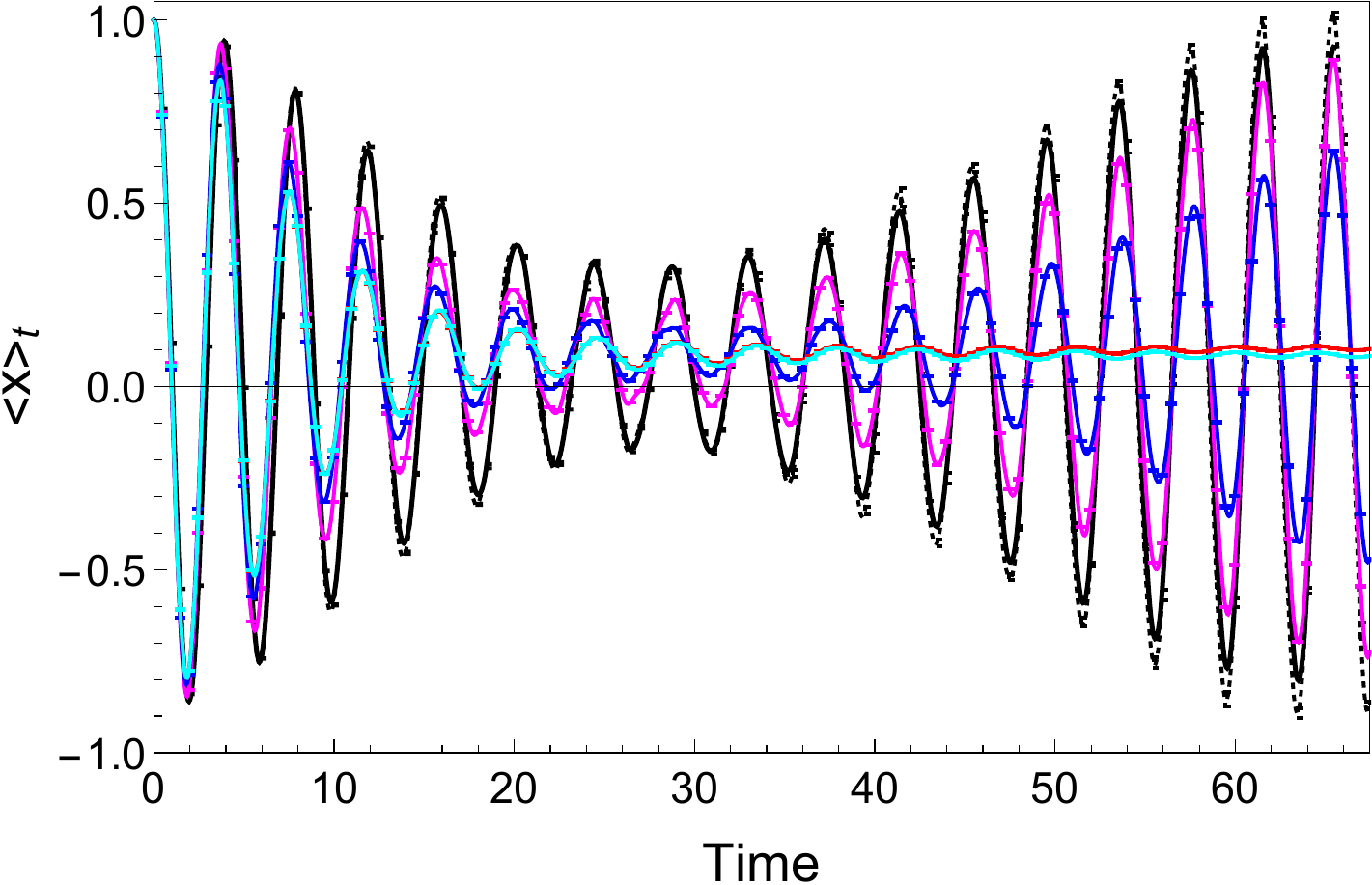}
    \caption{The exact position correlation function for a 1D anharmonic oscillator,
        obtained by diagonalizing the Hamiltonian on a 
        Discrete Variable Representation (DVR) grid~\cite{Colbert1992a} 
        is shown as a solid black line. 
        DHK-IVR results are plotted with dashed black lines and the 
        classical-limit Husimi-IVR is shown with a red solid line. 
        MQC-IVR correlation function results are shown for different 
        values of the Filinov parameter, $c_p = 0.7$ (pink), 
        $c_p = 3.0$ (blue), $c_p = 500$ (cyan). The plots demonstrate
        that the amplitude of oscillations at long times moves gradually
        from the quantum recurrences in DHK-IVR to the fully damped 
    classical limit Husimi-IVR. Figure adapted from Ref.~\citenum{Church2017}.}
        \label{fig:mqc-1d}
\end{figure}

The MQC-IVR position correlation functions obtained for 
different values of the Filinov parameter are shown 
in Fig.~\ref{fig:mqc-1d}. We note that there is only
one Filinov parameter, $c_p$, in this calculation 
since operator $\hat B$ is a position operator, 
the change in momentum, $\Delta p_t$, is finite 
while $\Delta q_t=0$. 
As shown in Fig.~\ref{fig:mqc-1d},
for small values of this Filinov parameter, MQC-IVR captures
long-time quantum recurrence with accuracy comparable to DHK-IVR.
As the strength of the Filinov phase filter is increased, the coherent
structure is damped and the resulting correlation function 
coincides with the classical-limit result, specifically the 
Husimi-IVR.~\cite{Church2017}
Table ~\ref{tbl:fb-mqc-1d} demonstrates one of the advantages of the MQC-IVR approach: 
for finite but near zero values of the Filinov parameter, the number of trajectories 
required to obtain a converged result shows improvement over the corresponding 
DHK-IVR simulation, and this number further decreases as the filter strength
is increased.

\begin{table}
    \caption{The number of trajectories $N_{traj}$ needed to achieve 
    numerical convergence for calculating the position correlation 
function with different formulations. Data from Ref.~\citenum{Church2017}}
  \label{tbl:fb-mqc-1d}
  \begin{ruledtabular}
  \begin{tabular}{lll}
    IVR formulation     & $c_p$     & $N_{traj}$  \\
    \hline
    DHK                 & 0         & $3.0 \times 10^6$   \\
                        & 0.7       & $2.4 \times 10^4$   \\
    MQC              & 3.0       & $9.6 \times 10^3$   \\
                        & 500.0     & $6.0 \times 10^2$   \\
    Husimi              & $\infty$  & $2.4 \times 10^2$   \\ 
  \end{tabular}
  \end{ruledtabular}
\end{table}

\begin{figure}
    \centering
    \includegraphics[width=0.45\textwidth]{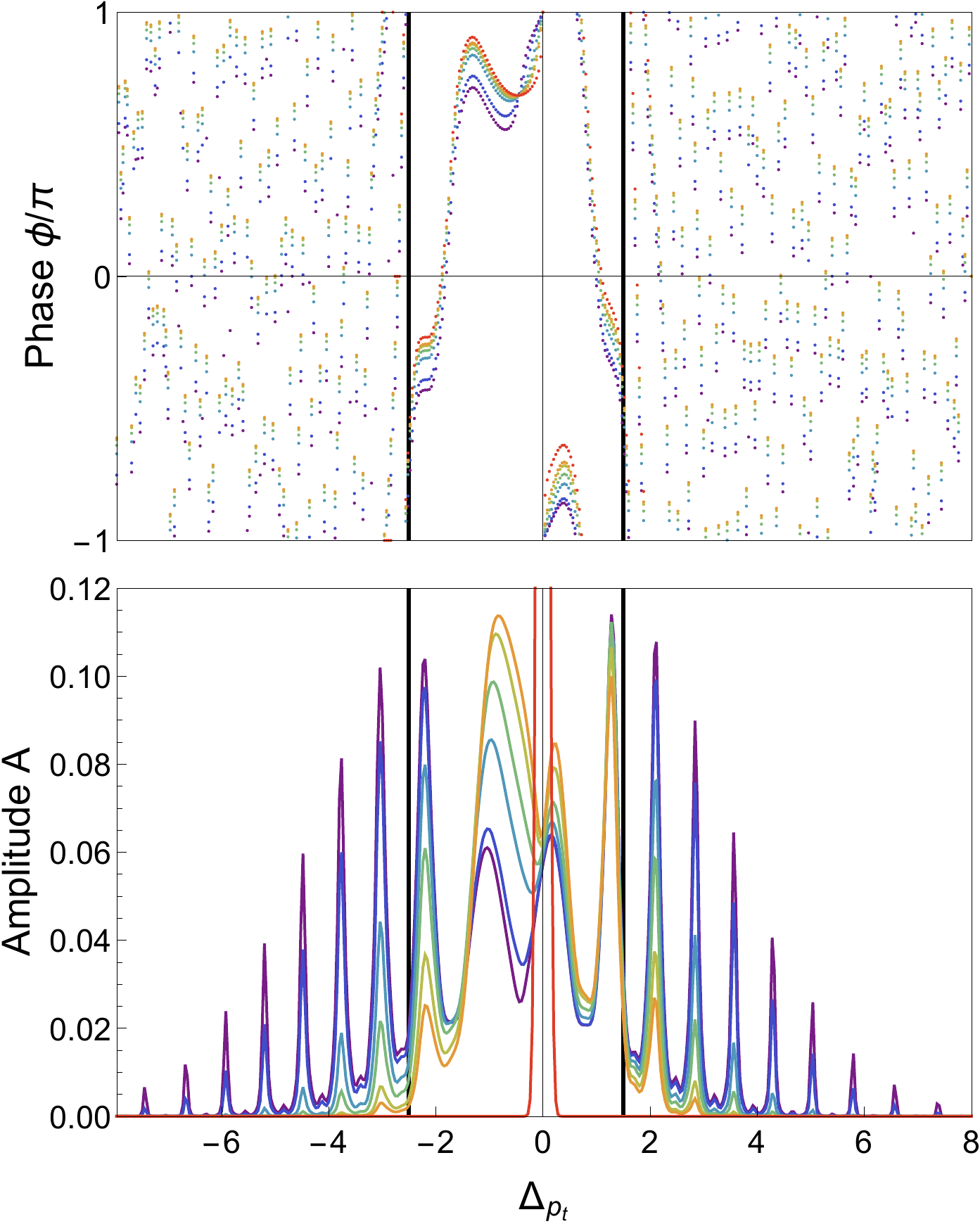}
    \caption{The phase (top) and amplitude (bottom) contribution to the 
        MQC integrand for the 1D anharmonic oscillator position correlation function
        generated from a single initial phase space point for the forward trajectory
        with all possible backward trajectories as a function of $\Delta p_t$.
        In both plots, the colors correspond to different values of the Filinov parameter, $c_p = 0.05$ (purple), $c_p = 0.1$ (blue), $c_p = 0.3$ (blue-green), $c_p = 0.5$ (green), $c_p = 0.7$ (yellow), $c_p = 1.0$ (orange), and $c_p = 200$ (red). The vertical black lines in both plots enclose the regions of slowly varying phase.
        We find that by increasing the strength of the filter, it is possible to confine
        the amplitude to have non-zero values only in the region of relatively stationary
    phase. Figure adapted from Ref.~\citenum{Church2017}.}
    \label{fig:amp-phase1}
\end{figure}

\begin{figure}
    \centering
    \includegraphics[width=0.45\textwidth]{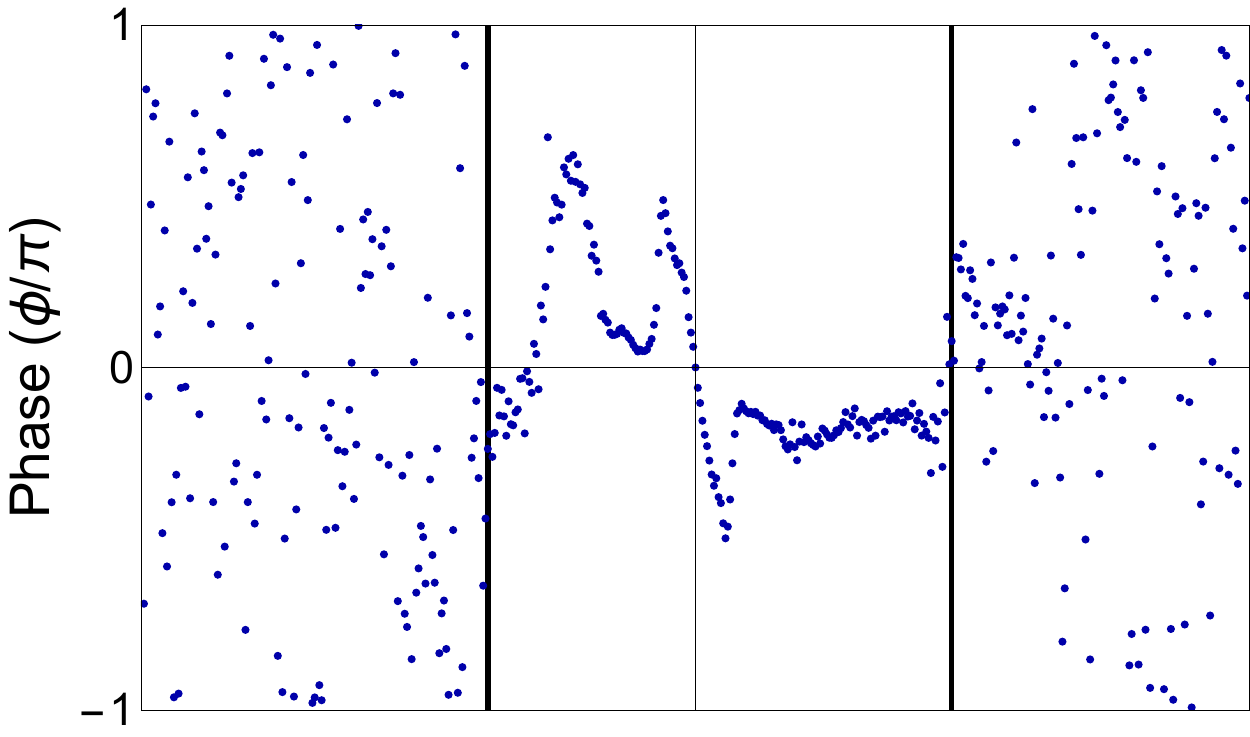}
    \includegraphics[width=0.45\textwidth]{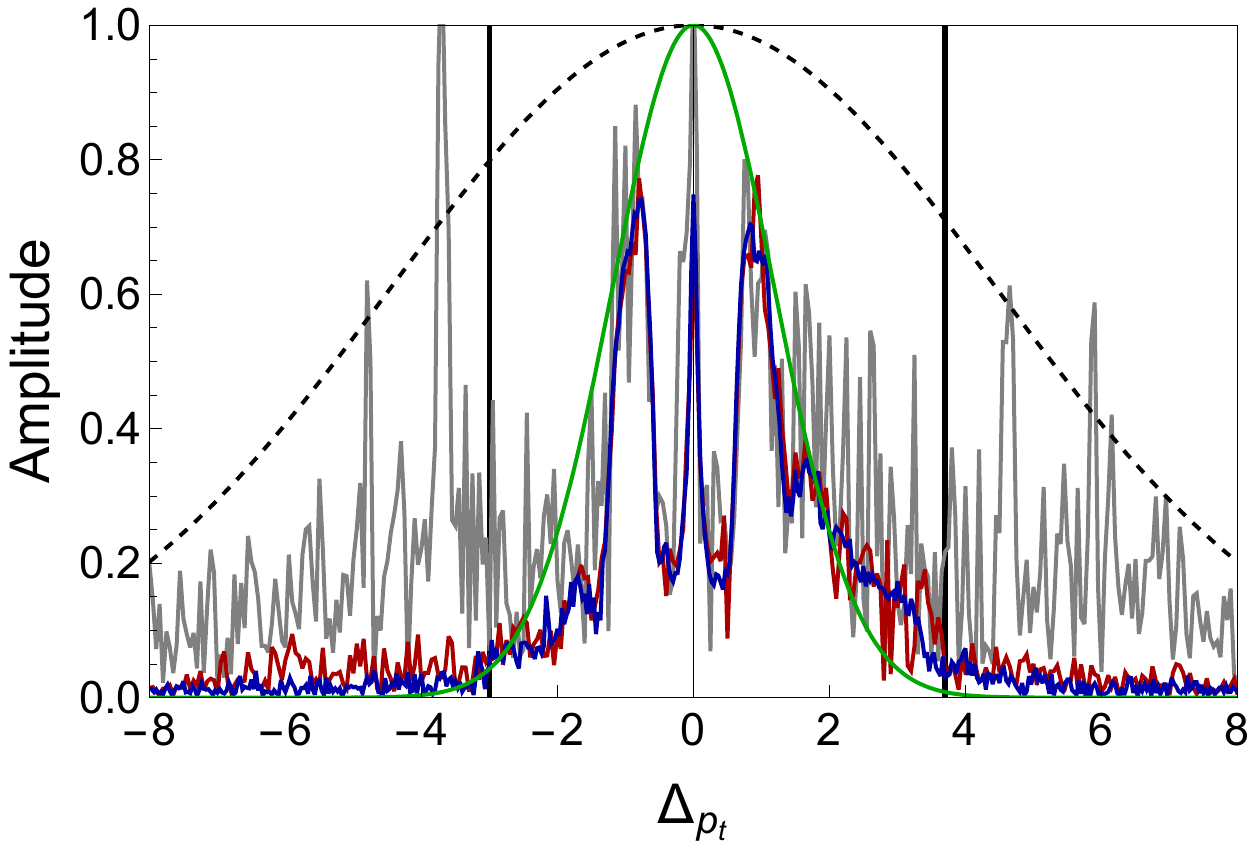}
    \caption{The phase (top) and amplitude (bottom) of the MQC-IVR integrand 
        for the position correlation function are plotted against the 
        momentum displacement between forward-backward trajectories 
        in the quantum limit, ensemble averaged over $1.2 \times 10^3$ trajectories (gray), 
   $6.0 \times 10^4$ trajectories (red), 
    and $2.4 \times 10^5$ trajectories (blue). Overlaid
    on the amplitude plot, we also show the Filinov filter Gaussians
    using a black dashed line for the weak filter $(c_p = 0.05)$ and a 
    green dashed-line for the `optimal' filter strength $(c_p = 0.7)$.
Figure adapted from Ref.~\citenum{Church2017}.}
\label{fig:amp-phase2}
\end{figure}

Careful numerical analysis of the MQC-IVR integrand for the position correlation
function shown in Fig.~\ref{fig:mqc-1d} serves to establish the efficacy of the 
MFF scheme. In Fig.~\ref{fig:amp-phase1} (top), we show the contribution of 
each pair of forward-backward trajectories to the phase and amplitude of the MQC integrand,
as a function of the momentum displacement, $\Delta p_t$.
We find that there exists a range of $\Delta p_t$ 
values that define a region of relatively slowly varying phase, and outside 
this range, corresponding to larger values of $|\Delta p_t|$, the phase oscillates
rapidly. Plotting the amplitude of the integrand for each of these pairs 
in Fig.~\ref{fig:amp-phase1} (bottom), we find that, 
in the quantum limit, the small but non-zero amplitudes in regions
of highly oscillatory phase result in a very noisy integrand. 
As the Filinov parameter value is increased, we find that the amplitude is
increasingly constrained to be non-zero only in the vicinity of the $\Delta p_t\rightarrow 0$,
resulting in a much less oscillatory integrand. 
~\cite{Church2017} 
Having established how changing the filter strength modifies components 
of the integrand, we examine the phase and amplitude averaged over an ensemble
of trajectory pairs. Figure~\ref{fig:amp-phase2} (bottom) shows the amplitude of 
a small-filter MQC-IVR integrand ensemble averaged over an increasing number of 
trajectories; as we approach the number required to numerically converge the 
DHK-IVR integral ($\sim 10^6$), the amplitude in regions of non-stationary 
phase becomes negligible, but does not completely vanish giving rise to 
residual noise in the integrand. Overlaying the Filinov filter parameterized
Gaussian function in $\Delta p_t$ over the MQC-IVR integrand 
in Fig.~\ref{fig:amp-phase2}, we find that increasing $c_p$ from the small-filter
limit to a larger value ($c_p=0.7$ in this case) is sufficient to zero
amplitude in regions of non-stationary phase, reducing the noise in the integrand,
and allowing for more rapid numerical convergence without significant loss in 
accuracy.

\subsection{Nonadiabatic MQC-IVR}
Nonadiabatic dynamic processes, where nuclear motion is coupled to and drives transitions
between electronic states, have been studied using a variety of SC-IVR 
methods.~\cite{Sun1997,Sun1998b,Batista1998,Rabani1999,Wang1999,Thoss2000,
Shi2004,Shi2005,Bonella2005,Shi2008,McRobbie2009b,Miller2009,Miller2010,
Venkataraman2011,Huo2011,Huo2012,Cotton2013,Tao2013b,Sun2015,Lee2016,Sun2016,Sun2016b,
Sun2016c,Teh2017,Kananenka2017,Sun2018,Provazza2018,Provazza2019,Mulvihill2019,
Gao2020,Gao2020b,Dodin2022,Kumar2021}
This is facilitated by the Meyer-Miller-Stock-Thoss (MMST) mapping\cite{Meyer1979,Stock1997} 
where discrete electronic state variables are mapped to continuous Cartesian electronic phase
space variables that can undergo approximate time-evolution under a classical analog
Hamiltonian.
Using the MMST mapping, a nonadiabatic MQC-IVR expression can be derived to enable 
independent control over the extent of quantization of the nuclear and electronic 
degrees of freedom.~\cite{Church2018} In addition, a novel symplectic 
integration scheme has been proposed to enable accurate time evolution of the 
nuclear and electronic degrees of freedom as well as the Monodromy matrices 
that appear in the MQC prefactor.~\cite{Kelly2012,Church2018} 

\subsection{Analytic Mixed Limit MQC-IVR}
Although choosing a small but non-zero value of the Filinov parameter
can offer a significant mitigation of the sign problem in low-dimensional
systems, for complex system studies, it is expedient to 
simply choose a handful of degrees of freedom to describe 
in the quantum limit while treating the rest of the system in the classical limit.
This motivates the derivation of an analytic, mixed-limit 
AMQC-IVR correlation function,~\cite{Church2019a}
\begin{align}
   C_{AB}^\text{AMQC}(t) &= \frac{1}{\left(2\pi\hbar\right)^{F+F_Q}} 
   \int d\textbf{p}_{0}\int d\textbf{q}_{0}
   \int d\textbf{p}_{Q}^{'}\int d\textbf{q}_{Q}^{'} \notag \\
   & \times C_{t}\left(\textbf{p}_{0},\textbf{q}_{0}\right)
   C_{t}^{*}(\textbf{p}_{0}^{'},\textbf{q}_{0}^{'})
   \mel{\textbf{p}_{0}\textbf{q}_{0}}{\hat{A}}{\textbf{p}_{0}^{'}\textbf{q}_{0}^{'}} \notag \\
   & \times \mel{\textbf{p}_{t}^{'}\textbf{q}_{t}^{'}}{\hat{B}}{\textbf{p}_{t}\textbf{q}_{t}} e^{i\left[S_{t}(\textbf{p}_{0},\textbf{q}_{0})-
       S_{t}(\textbf{p}_{0}^{'},\textbf{q}_{0}^{'})\right]/\hbar} \notag \\
   & \times \Lambda_{t}(\textbf{p}_{0},\textbf{q}_{0},\textbf{p}_{Q}^{'},\textbf{q}_{Q}^{'}),
   \label{eq:amqc}
\end{align}
where $F$ is the total system degrees of freedom (dofs),
$(\textbf{p}_{0},\textbf{q}_{0})$ are $F$-dimensional 
position and momentum vectors, $F_Q$ is the number of quantized dofs for 
which the Filinov parameters are set to exactly zero,
and $(\textbf{p}_{Q}^{'},\textbf{q}_{Q}^{'})$ are the phase
space variables of the subset of quantized dofs.
The AMQC-IVR prefactor in Eq.~\ref{eq:amqc} is the product of the $F_Q$ dimensional 
HK-IVR prefactors corresponding to the forward ($C_t$) and backward ($C_t^*$)
propagators, and $\Lambda_{t}$ represents an $F$ dimensional determinant
corresponding to coupling between the quantum and classical dofs,
defined in detail in Ref.~\citenum{Church2019a}.
In the limit where the quantum subsystem is only weakly coupled to the 
rest of the system, a further separable prefactor approximation 
can be made by setting $\Lambda_t=1$.~\cite{Church2019a}

\subsection{MQC-IVR Results and Discussion}

\begin{figure}
    \centering
    \includegraphics[width=0.3\textwidth]{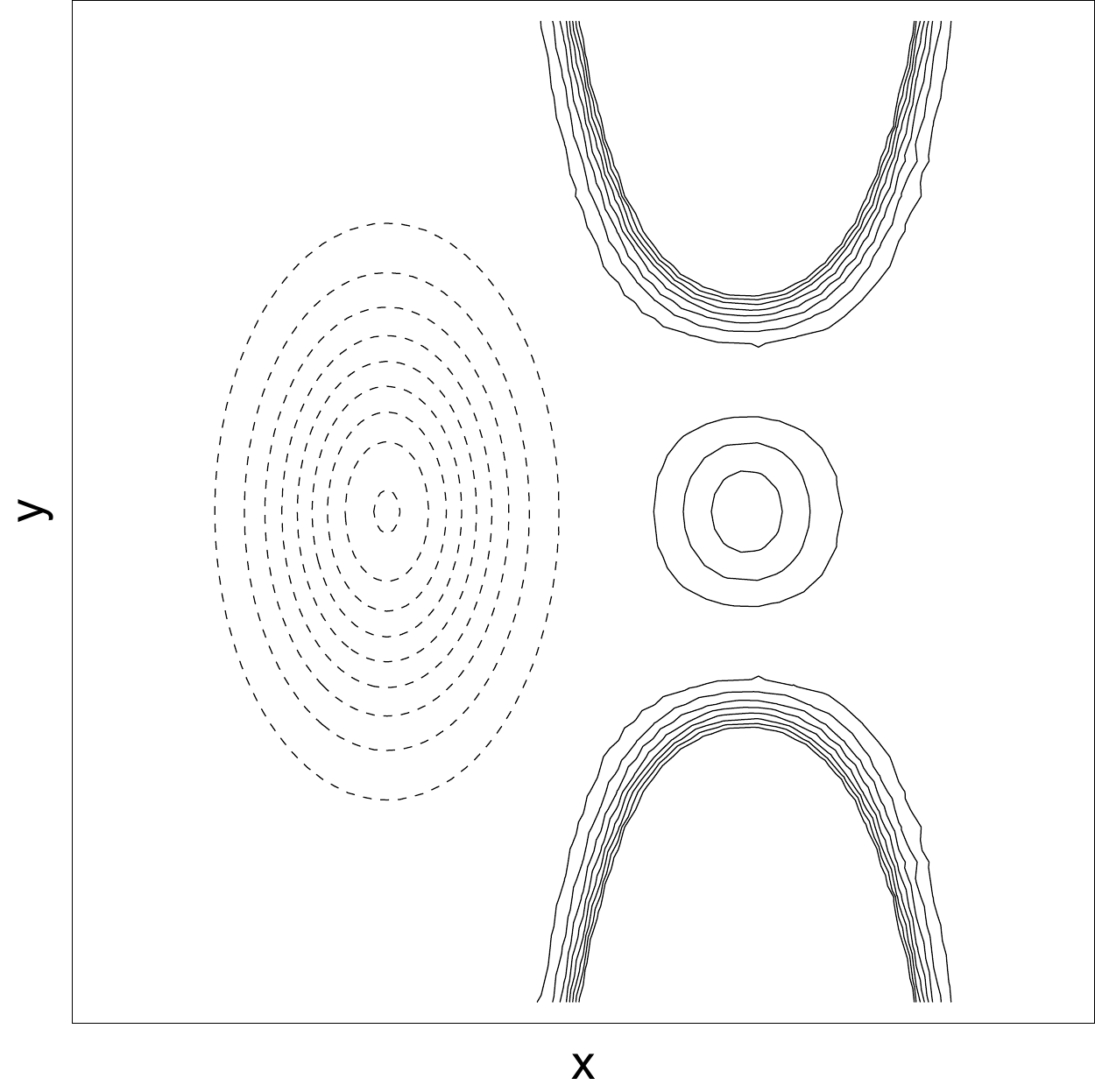}
    \includegraphics[width=0.45\textwidth]{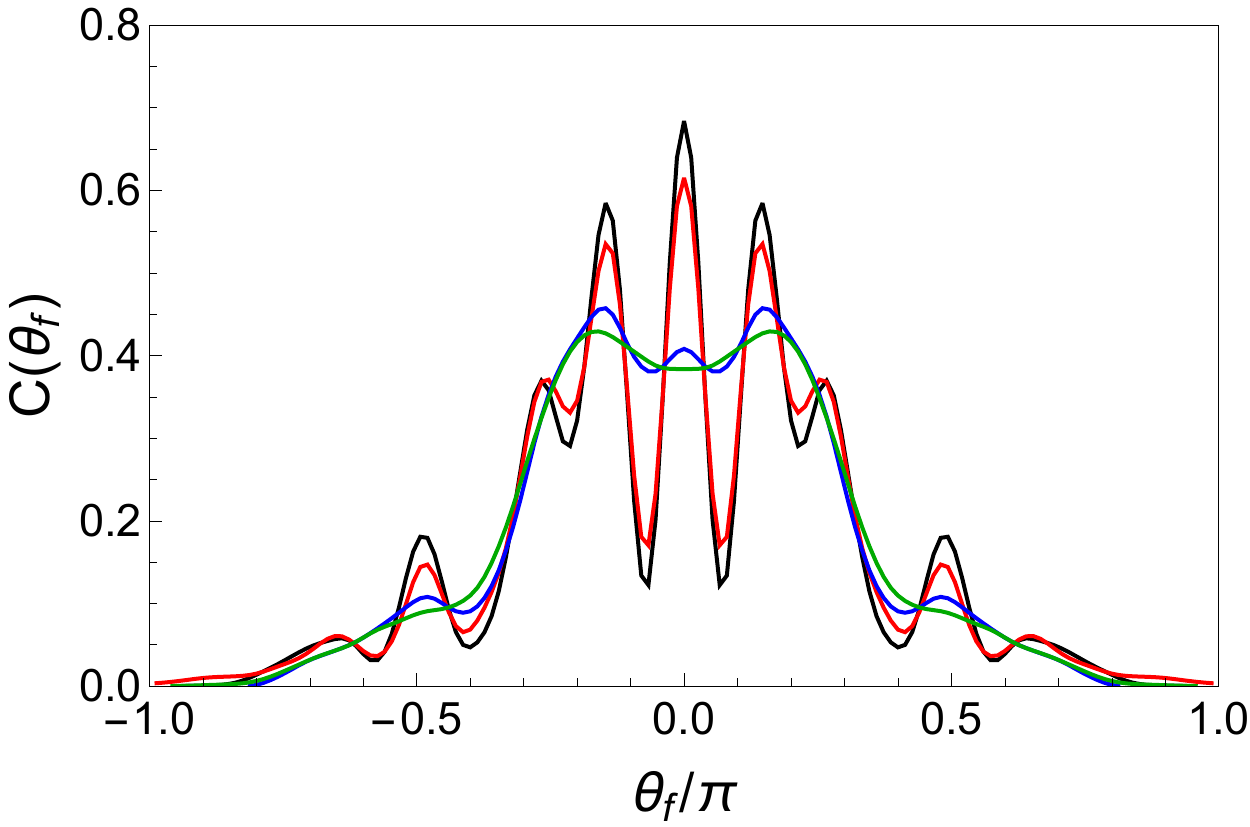}
    \caption{The potential (top) for the double-slit Hamiltonian is shown as a 
        contour plot with solid black lines and the initial wavepacket density
        is plotted with black dashed lines.  The angular distribution (bottom) calculated 
        using MQC-IVR with $c_x = c_y = 1\times 10^{-6}$ (black), 
        $c_x = c_y = 1\times 10^{-5}$ (red), $c_x = c_y = 1
        \times 10^{-4}$ (blue), $c_x = c_y = 5\times 10^{-4}$ (green). 
        We see the interference pattern observed in the quantum limit
        changes to the classical scattering result as the value of the Filinov 
        parameter is increased.}
        \label{fig:mqc-2slit}
\end{figure}

The quantum interference pattern obtained when a particle is incident
upon a screen with two slits, the double slit experiment, highlights
the non-additive nature of probability in quantum mechanics and serves
as our first test-case for MQC-IVR.
Specifically, we use MQC-IVR to obtain the probability distribution for 
particle subject to a quantum double-slit experiment mimicked 
by a 2D Hamiltonian,~\cite{Gelabert2001}
\begin{align}
    H &= \frac{p_x^2}{2m}+\frac{p_y^2}{2m} +\left(\bar{V} -
    \frac{1}{2}m\omega^2 y^2 + \frac{m^2\omega^4 y^4}{16 \bar{V}} \right) e^{-(x/\alpha)^2},
\end{align}
with  $m = 1$ a.u., $\alpha = 50$ a.u., 
$\omega = 600\ \text{cm}^{-1}$ and $\bar{V} = 8000 \,\text{cm}^{-1}$. 
The initial wavefunction is a 2D coherent state,
\begin{align}
  \braket{x,y}{\psi_i} &= \left(\frac{\gamma_x\gamma_y}{\pi^2}\right)^{\frac{1}{4}}
  e^{-\frac{\gamma_x}{2}(x-q_x)^2+ip_x(x-q_x)} \notag \\
  & \times e^{-\frac{\gamma_y}{2}(y-q_y)^2+ip_y(y-q_y)},  
  \label{2d-cs}
\end{align}
with parameters $\gamma_{x} = 1/2\alpha^2$ and $\gamma_{y}=m\omega^2/8V_{0}$, 
initial positions centered at $q_x = -220$ a.u. and $q_y = 0$ a.u., 
and initial momenta centered at $p_y = 0$ a.u., 
and $p_x^2/2m = 2048 \,\text{cm}^{-1}$, where the positive 
sign indicates motion in the direction of the double slit. 
The double slit potential is plotted in Fig.~\ref{fig:mqc-2slit}
along with the initial wavepacket with average energy 
significantly less than the barrier height to ensure that 
the particle must pass through the slits.

The MQC-IVR angular distributions 
obtained as the long time limit of a real-time correlation 
function with $\hat A\equiv \ket{\psi_i} \bra{ \psi_i}$ and 
$\hat B\equiv \delta(\theta_f-\hat{\theta})$ are shown 
in Fig.~\ref{fig:mqc-2slit}.
For small values of the Filinov parameter, the MQC-IVR 
integrand includes pair of forward-backward paths where
the particle can pass through different slits giving 
rise to the quantum interference pattern.
In the classical strong filter limit, the MQC-IVR
correlation function is dominated by contributions from forward
and backward paths that are close together, requiring the particle
pass through the same slit, collapsing the interference pattern 
to a classical binodal distribution.

\begin{figure}
    \centering
    \includegraphics[width=0.45\textwidth]{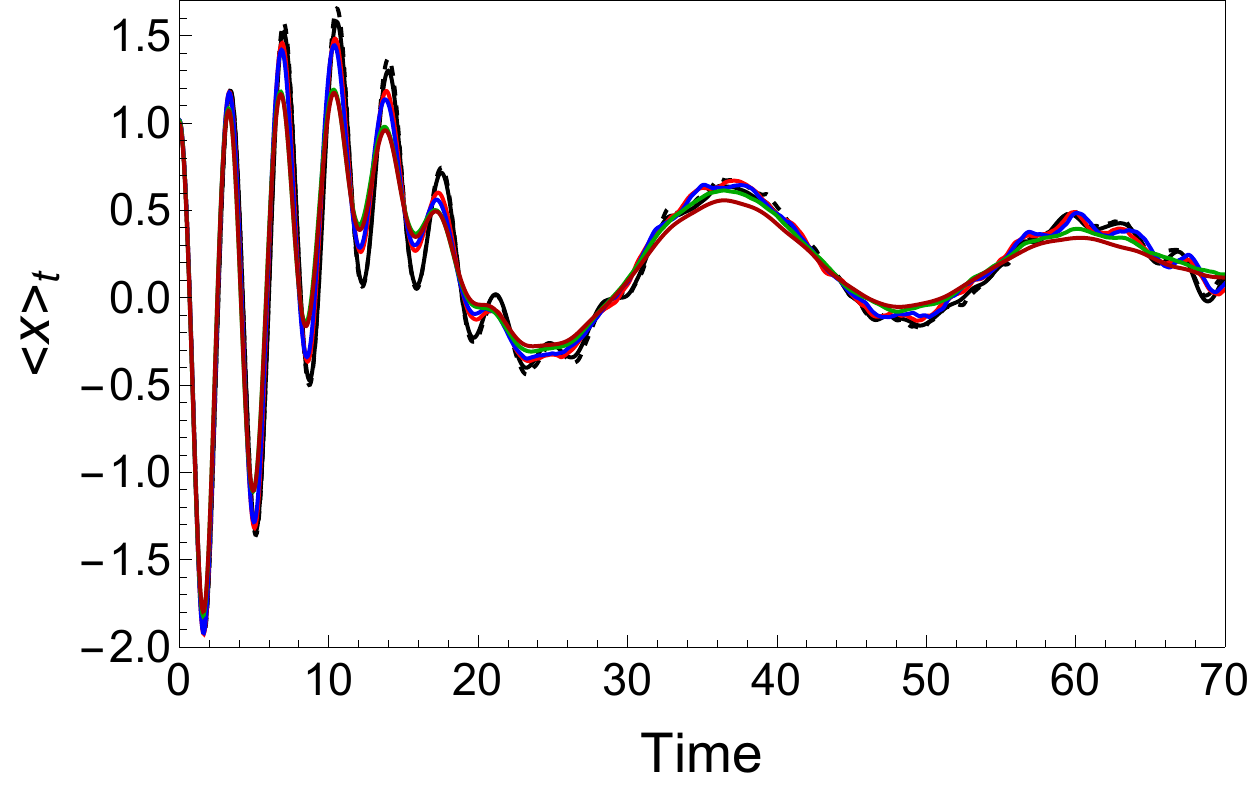}
    \caption{The $x$-mode position correlation function for a 2D model system obtained
    from DVR (solid black), DHK-IVR (dashed black), MQC-IVR with $c_x = 0.2, c_y = 0.2$ (red), $c_x = 0.2, c_y = 50$ (blue), $c_x = 50, c_y = 50$ (green), and Husimi-IVR (brown).}
    \label{fig:mqc-2d}
\end{figure}

A key feature of MQC-IVR is the ability to treat certain degrees of
freedom in the quantum limit and others in the classical limit.
We numerically demonstrate this for a 
2D potential constructed by coupling the 1D anharmonic oscillator
previously defined in Eq.~\ref{eq:1dao} to a heavy harmonic oscillator mode,
\begin{align}
    V(x,y) = \frac{1}{2}m_{x}\omega_{x}^2 x^2 - 0.1 x^3 + 0.1 x^4 +\frac{1}{2}m_{y}\omega_{y}^2 y^2 + kxy
\end{align}
where $m_x = 1$, $m_y = 25$, $\omega_x = \sqrt{2}$, $\omega_y = 1/3$ and $k=2.0$ in atomic units.
In Fig.~\ref{fig:mqc-2d}, we show the MQC-IVR position correlation function
for the anharmonic mode, where $\hat A \equiv \ket{\psi_i}\bra{\psi_i}$ and $\hat{B}\equiv \hat x$
compared against exact, DHK-IVR, and Husimi-IVR results.
As expected, MQC-IVR results with both $c_x$ and $c_y$ set to low values show good 
agreement with DHK-IVR,  and the results generated with large value Filinov
parameters for both degrees of freedom agree well with the Husimi-IVR 
correlation function. Interestingly, we find that a mixed limit calculation 
where only the anharmonic mode is treated in the quantum limit yields results
that are nearly indistinguishable from DHK-IVR, and with 
far fewer trajectories as shown in Table~\ref{tbl:fb-mqc-2d}. 

We note that operator $\hat B$ appears to play 
an important role in determining which degree of freedom should be treated 
in the quantum limit. In Fig~\ref{fig:mqc-2d}, $\hat B \equiv \hat x$, so
at time $t$ the operator induces a momentum jump in the anharmonic ($x$) dimension,
suggesting that treating only the anharmonic mode in the quantum limit is 
sufficient to achieve good agreement with exact quantum results. Similarly,
it has been shown that for this same model system when $\hat B \equiv \hat y$,
it is possible to achieve good agreement with the exact quantum correlation
function by quantizing only the heavier harmonic ($y$) degree of freedom.~\cite{Antipov2015}
This leads us to conclude that there is not an inherent need to treat specific 
modes as more or less quantum, rather the observable determines the necessary
level of theory. 

\begin{table}
    \caption{The number of trajectories $N_\text{traj}$ required establish numerical 
    convergence for the $x$-mode position correlation function of a 2D model system with 
     different levels of SC theory. }
  \label{tbl:fb-mqc-2d}
  \begin{ruledtabular}
  \begin{tabular}{lccc}
    IVR formulation     & $c_x$     & $c_y$ &  $N_\text{traj}$  \\
    \hline
    DHK                 & 0         & 0      & $6\times 10^6$ \\
                        & 0.2       & 0.2    & $4\times 10^6$ \\
    DF-MQC              & 0.2       & 50     & $1\times 10^6$ \\
                        & 50        & 50     & $2\times 10^5$   \\
    Husimi              & $\infty$  &$\infty$& $8\times 10^4$   \\ 
  \end{tabular}
  \end{ruledtabular}
\end{table}

\begin{figure}
    \centering
    \includegraphics[width=0.45\textwidth]{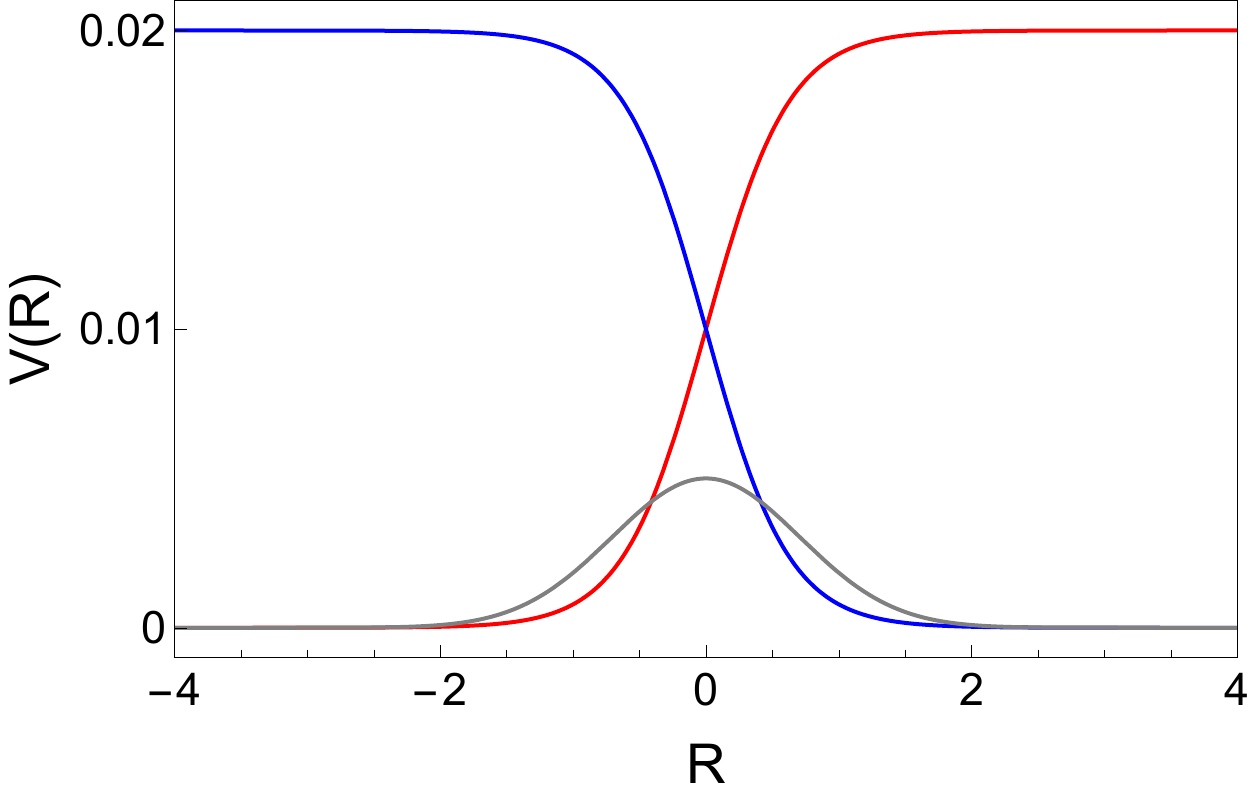}
    \caption{Elements of the scattering potential, showing the diabatic potential energy matrix elements corresponding to electronic state 1 in red, 
electronic state 2 in blue, and the coupling between states is shown in grey. }
\label{fig:na-models}
\end{figure}

The importance of the observable is also highlighted in a study of nonadiabatic
dynamics, where the electronic states are strongly coupled to nuclear degrees 
of freedom. Specifically, we calculate the final nuclear momentum distribution 
when a particle is transmitted through a scattering potential 
pictured in Fig.~\ref{fig:na-models} where two diabatic electronic 
states are coupled to a single nuclear degree of freedom.~\cite{Church2018}
The MQC-IVR correlation function is obtained by histogramming the final
nuclear momentum with operator $\hat B = \delta (\hat P - P_f)$.
This model is particularly interesting because previous studies have established 
that the classical limit LSC-IVR simulations reproduce the various electronic
state population correlation functions reasonably accurately,~\cite{Gao2020}
however, quantum limit methods
like DHK-IVR are necessary to correctly capture the final nuclear momentum
distribution.~\cite{Ananth2007}

\begin{figure}
    \centering
    \includegraphics[width=0.45\textwidth]{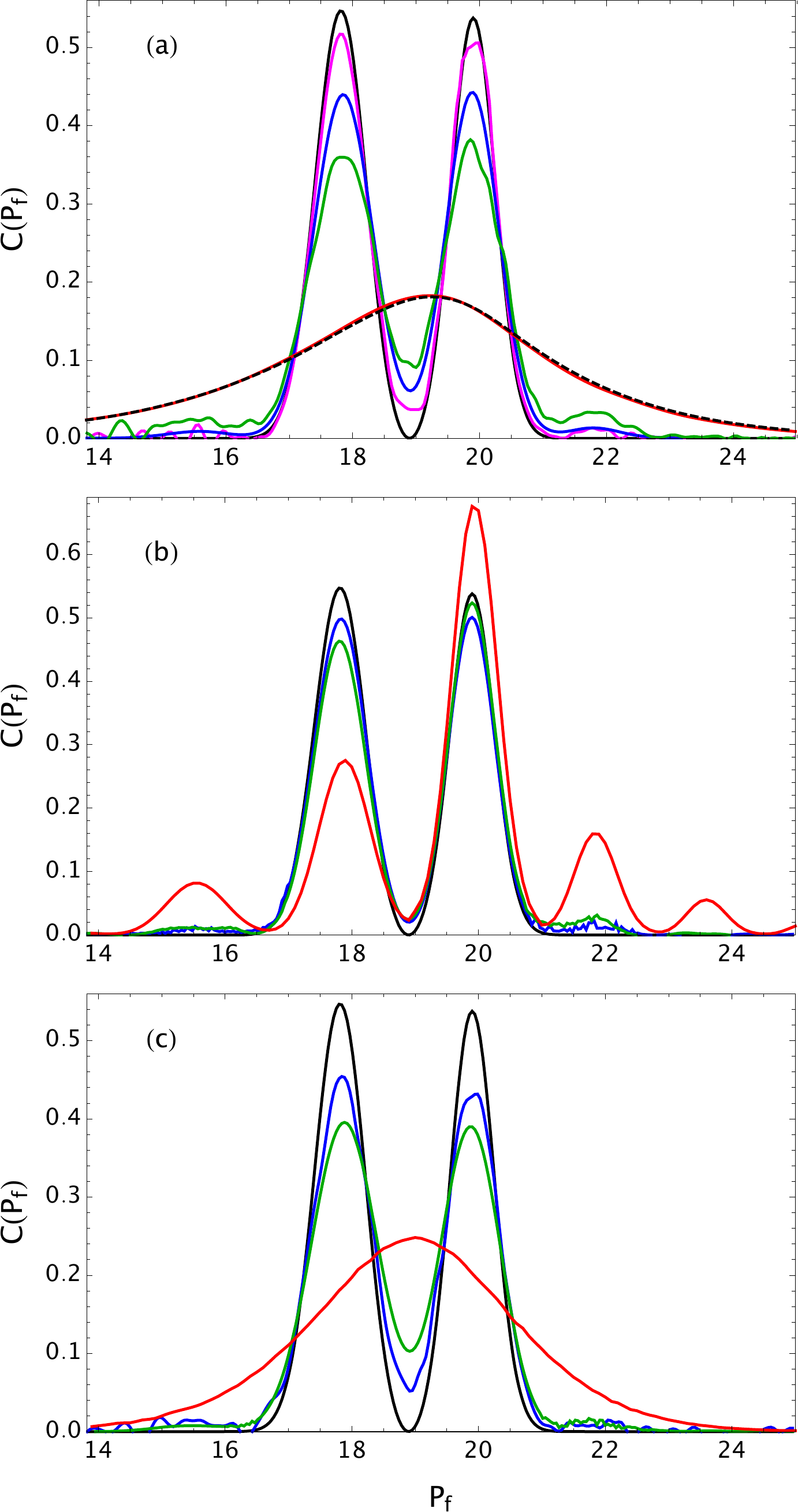}
    \caption{Final nuclear momentum distribution for a particle incident on the scattering potential 
pictured in Fig.~\ref{fig:na-models} with initial energy $0.1$ a.u. 
The exact quantum result (black, solid) is shown in each panel along 
with (a) the Husimi-IVR (black, dashed) and MQC-IVR where the Filinov filter strength for the nuclear
and electronic degrees of freedom is set to be equal and to take values ranging from $c = 0.01$ (pink), 
$c = 0.05$ (blue), $c = 0.1$ (green), to $c = 10.0$ (red). 
(b) Mixed MQC-IVR simulation results where the nuclear degree of freedom is treated 
in the quantum limit, $c_\text{nuc} = 0.01$, and the level of theory to describe electronic
states is varied from the quantum limit to the classical limit, with $c_{el} = 0.05$ (blue), 
$c_{el} = 0.1$ (green), and $c_{el} = 10.0$ (red) 
(c) Mixed MQC-IVR simulation results where the electronic states are treated in the quantum limit, $c_{el} = 0.01$, and 
the level of theory for the nuclear degree of freedom is varied from the quantum limit to the classical limit, 
with $c_{nuc} = 0.05$ (blue), $c_{nuc} = 0.1$ (green), and $c_{nuc} = 10.0$ (red).
Figure adapted from Ref.~\citenum{Church2018}.}
    \label{fig:na-1}
\end{figure}

In Fig.~\ref{fig:na-1}, we show the final nuclear momentum distribution from
MQC-IVR simulations with different values of the Filinov parameter.~\cite{Church2019a} 
When the Filinov 
parameters for all degrees of freedom are tuned from the quantum to the classical limit, 
we find the expected momentum distribution corresponding to two scattering channels 
collapses to a single, broad, 'mean-field' like distribution in Fig.~\ref{fig:na-1}a.
Following our previous observation that the extent of quantization is tied to the 
degree of freedom directly perturbed by operator $\hat B$, Fig.~\ref{fig:na-1}b presents
the distributions obtained when the nuclear degree of freedom is treated in the quantum limit,
while the Filinov filter is increased systematically for the electronic degrees of freedom.
In this mixed limit, we find that although treating the electronic modes in the classical limit
introduces additional spurious peaks in the MQC distribution, it does not cause the distribution
to collapse completely to a mean-field result. Conversely, when the electronic degrees of 
freedom are treated in the quantum limit and the nuclear Filinov parameter increased to 
the classical-limit, we see the distribution collapse to the mean-field result. This serves
to demonstrate that the choice of operator $\hat B$ does indeed play an important role
in the choice of mode(s) to treat in the quantum limit, however, it is best to treat 
all strongly coupled degrees of freedom at the same level of theory.

We finish our exploration of the role of operators, coupling, and mixed quantization
with AMQC-IVR simulations of high-dimensional system-bath models where the system modes
are treated in the analytic quantum limit and the remaining bath modes treated in
the classical limit.~\cite{Church2019a}
We calculate the AMQC-IVR anharmonic-mode position correlation function 
for a system-bath model where an anharmonic oscillator is bilinearly 
coupled to a bath with an ohmic spectral density and 
an exponential cut-off,~\cite{Church2019a}
\begin{align}
    V(x,\bm{y}) &= \frac{1}{2}m\omega^2x^2 - 0.1 x^3 + 0.1 x^4 \notag \\
    & + \sum_{j=1}^{N}\left[ \frac{1}{2}m_{j}\omega_{j}^2 
    \left(y_j^2 - \frac{c_jx}{m_j\omega_j^2}\right)^2\right],
    \label{eq:sys-bath-pot}
\end{align}
where the number of discretized bath modes $N=12$.

\begin{figure}
    \centering
    \includegraphics[width=0.45\textwidth]{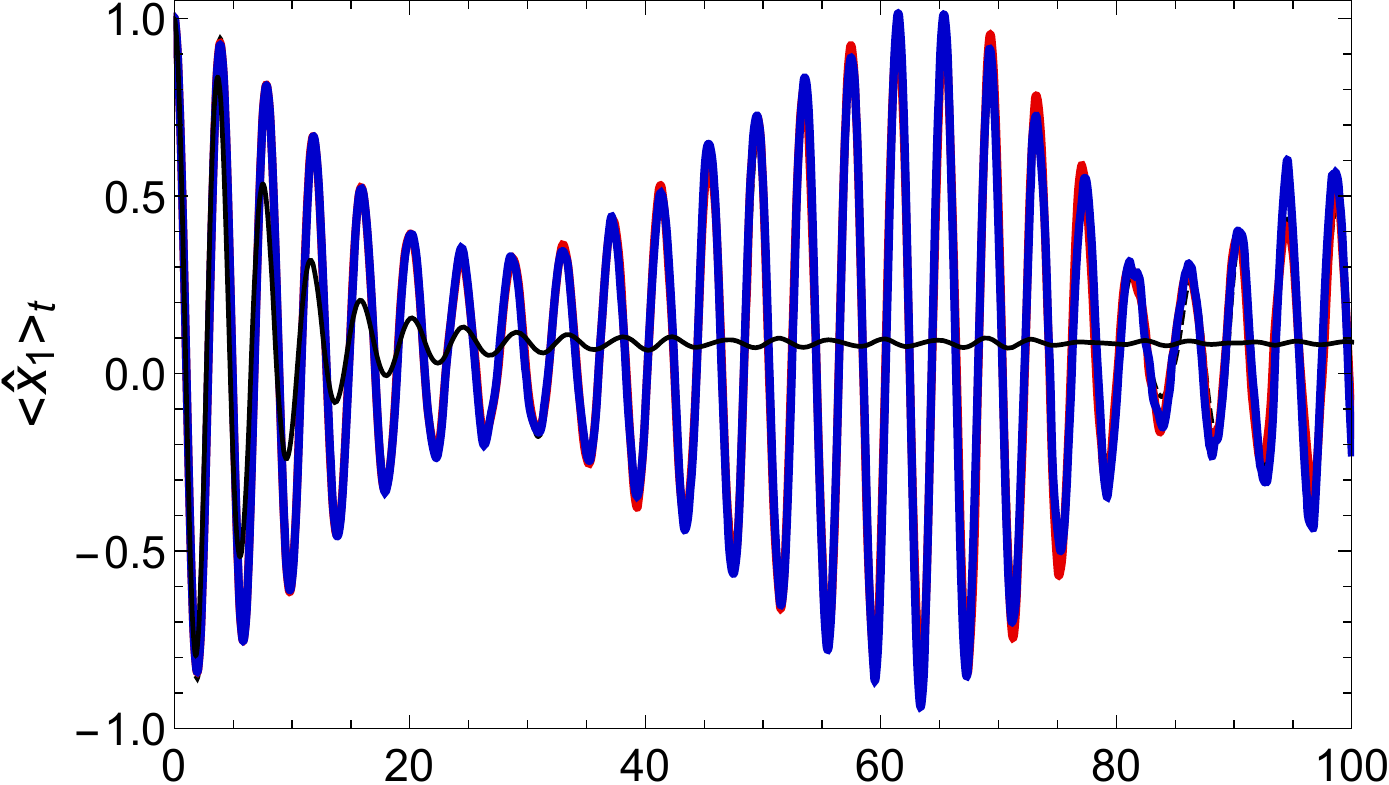}
    \includegraphics[width=0.45\textwidth]{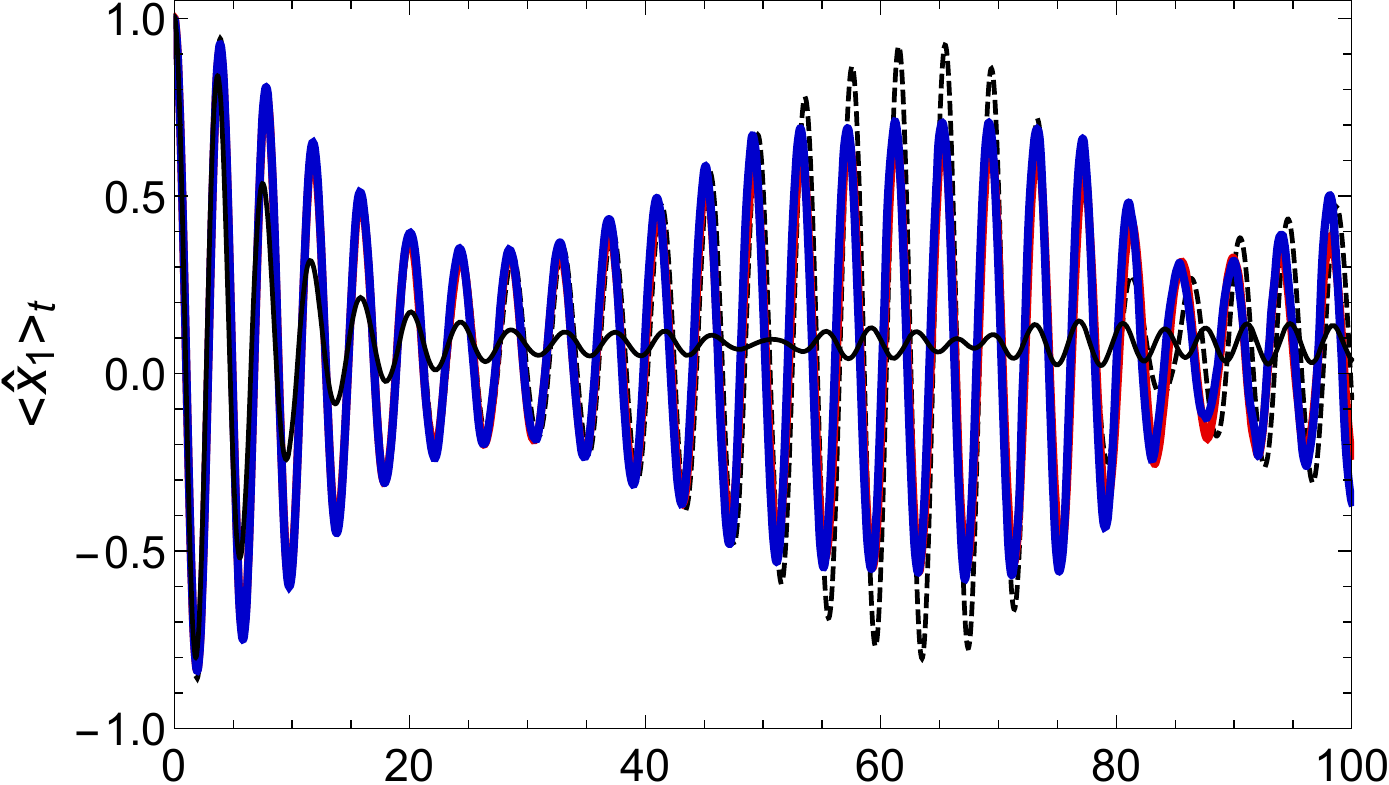}
    \includegraphics[width=0.45\textwidth]{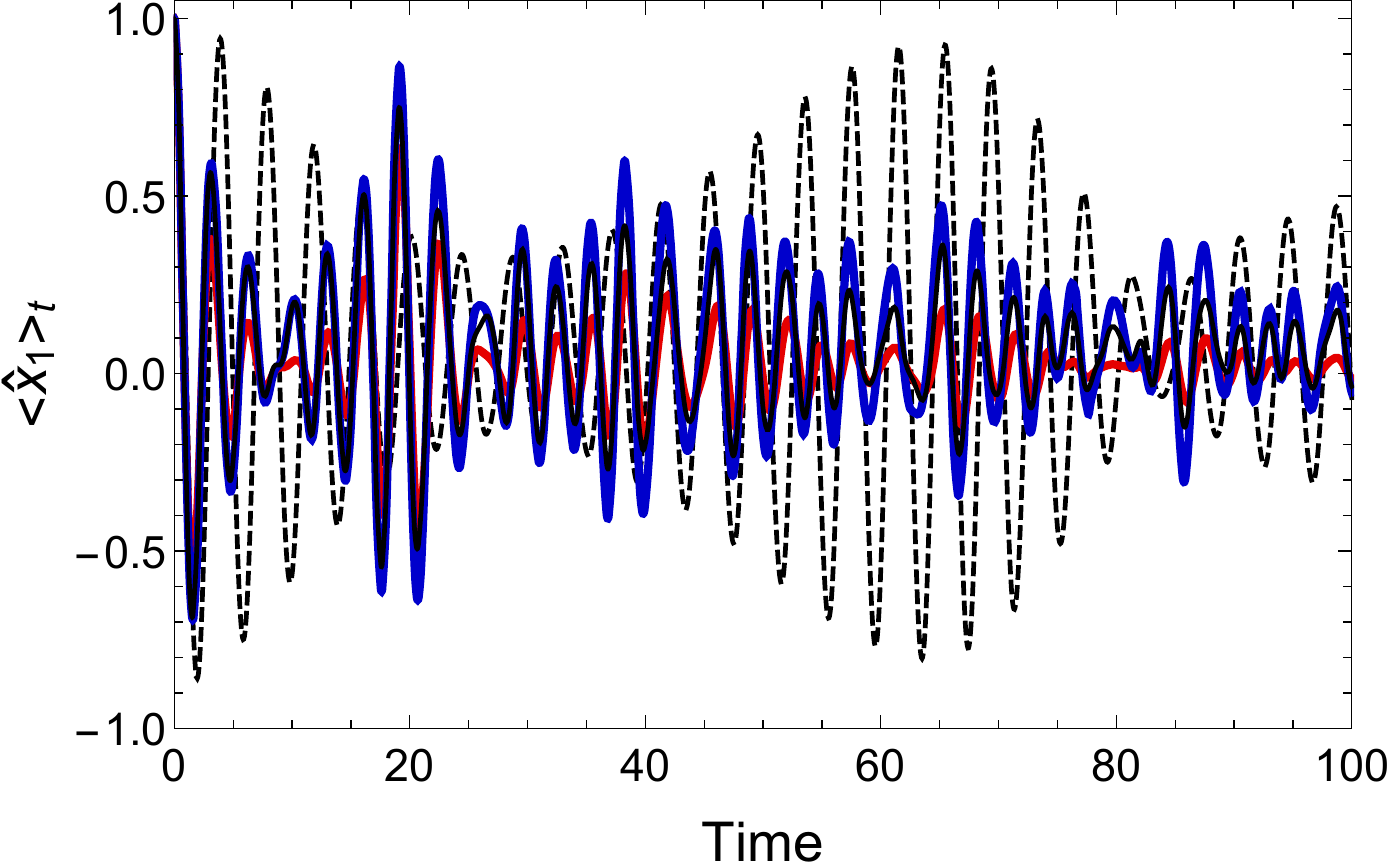}
    \caption{ The position correlation function, $<x_1>_t$ for a system-bath model with potential Eq.~\eqref{eq:sys-bath-pot},
         calculated using AMQC-IVR (blue), AMQC-IVR with SP approx. (red), 
    and Husimi-IVR (solid black). The coupling strength is  $\eta/m\omega_1 = 10^{-4}$ (top), 
    $\eta/m\omega_1 = 10^{-2}$ (middle), and $\eta/m\omega_1 = 1.0$ (bottom). 
    The dotted black line is the exact quantum average position of the 1d 
    anharmonic oscillator uncoupled from the bath. 
    Results adapted from Ref.~\citenum{Church2019a}.}
    \label{fig:amqc-2d}
\end{figure}

For very weak system-bath coupling, in Fig.~\ref{fig:amqc-2d},
we show that the AMQC-IVR results agree well with the exact 
quantum results for an uncoupled 1D anharmonic oscillator. 
As the coupling to the bath is increased, we see the expected 
damping in oscillatory structure in the AMQC-IVR results.~\cite{Church2019a}
We further investigate the applicability of the separable
prefactor approximation that is considerably less expensive,
requiring only the SC prefactor be computed for the single
system degrees of freedom in Fig.~\ref{fig:amqc-2d}.
We find that as the coupling to the bath is increased, the 
separable prefactor approximations begins to fail showing 
larger deviations from the AMQC-IVR result.~\cite{Church2019a}
Interestingly, we find that the classical limit Husimi 
IVR results agree better with AMQC-IVR than the separable 
prefactor approximation, suggesting again that an 
even-handed treatment of all modes is preferrable in 
the presence of strong coupling.~\cite{Church2019a}
\begin{figure}
    \centering
    \includegraphics[width=0.46\textwidth]{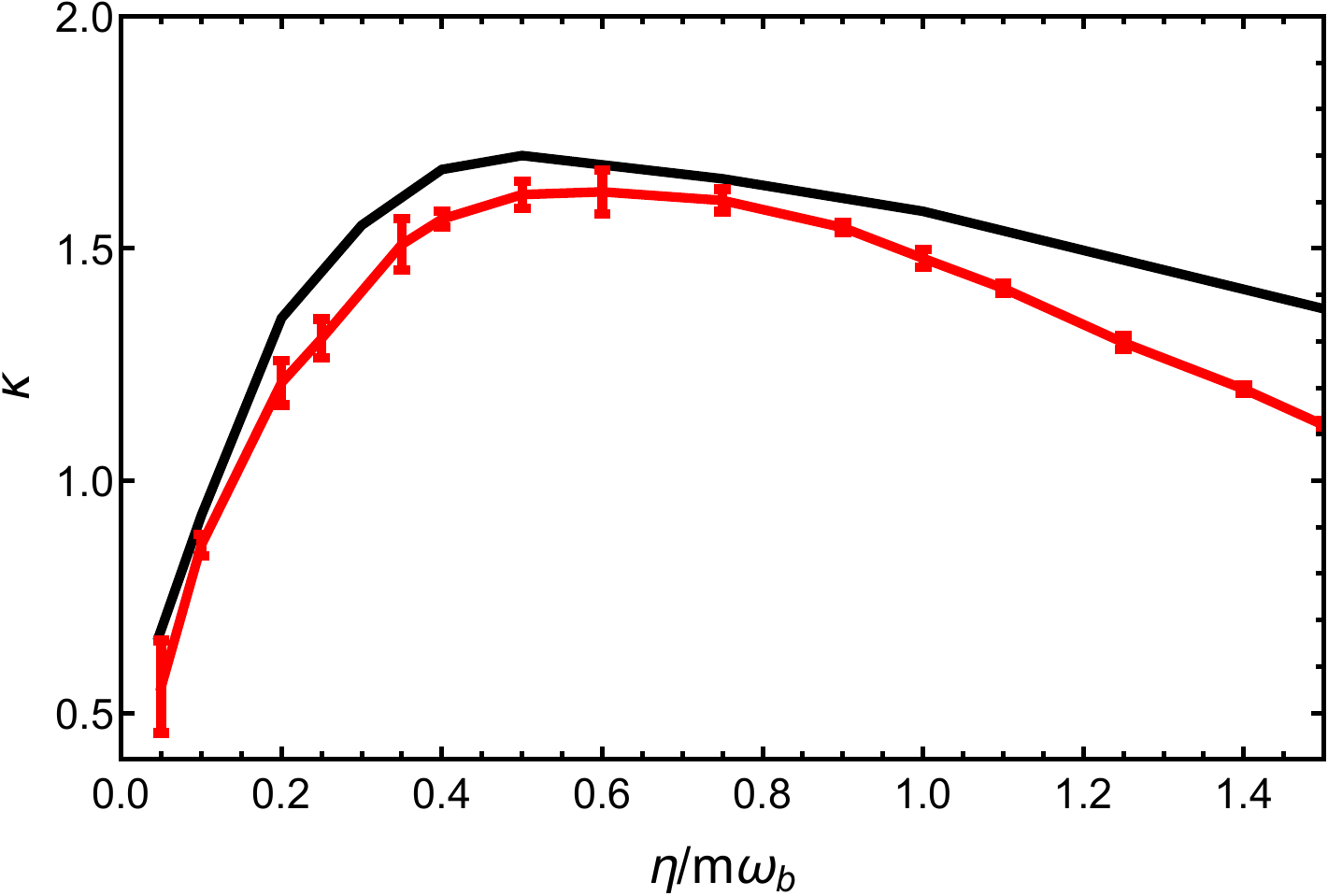}
    \caption{The thermal transmission coefficient at $T=300$K for a 
        proton-transfer model where a 1D double well potential (system) 
        is coupled to a bath of harmonic oscillators as a function of 
        system-bath coupling calculated using 
        AMQC-IVR (red) compared against exact path integral result (black) generated 
    using the Quasi-Adiabatic Path Integral method.~\cite{Topaler1994} 
Figure adapted from Ref.~\citenum{Church2019a}.}
    \label{fig:amqc-dw1}
\end{figure}

Finally, we calculate the thermal transmission coefficient, $\kappa(T)$,
from the flux-side correlation function for a proton transfer model 
system comprising a symmetric double well coupled to an bath of 
harmonic oscillators with an ohmic spectral density and exponential 
cut-off.~\cite{Topaler1994} 
The results from AMQC-IVR are compared against
exact real-time path integral simulations in Fig.~\ref{fig:amqc-dw1}
showing good agreement for a wide range of system-bath couplings. We 
note that this study also explored the applicability of the separable
prefactor approximation and found it to be valid only in the very 
weak-coupling regime.~\cite{Church2019a}

\subsection{A Better Path: Filinov Filtered Path Integrals }
It is evident that MQC-IVR and AMQC-IVR retain the accuracy of 
quantum limit SC methods like DHK-IVR while offering a significant
reduction in computational effort. The prefactor remains, perhaps,
the only limitation in extending these methods to truly complex 
system simulations: it is necessary to ensure continuity 
in evaluating the square root of the complex prefactor,
and time evolving the prefactor requires knowledge
of the Hessian.
In this section, we propose an alternate use of the 
modified Filinov filter to control the extent of oscillatory phase
in a real-time correlation function expression. Specifically,
rather than starting with a quantum limit SC expression with a prefactor
like the DHK-IVR, we derive an expression to filter the phase of an exact
real-time path integral correlation function. Note that this derivation 
takes its inspiration from previous work deriving LSC-IVR 
by linearizing the forward-backward paths in an 
exact real-time path integral expression for correlation functions.~\cite{Shi2003}

The path integral expression for a correlation function
is obtained by discretizing the forward and backward time evolution operators
by inserting identity in position space, and evaluating the resulting
short-time propagator matrix elements to obtain,~\cite{Feynman1965}
\begin{widetext}
\begin{align}
   C_{AB}(t) &= \text{Tr}\left[\hat{A}e^{i\hat{H}t/\hbar}\hat{B}e^{-i\hat{H}t/\hbar}\right] = \int dx_{0}^{+}\int dx_{0}^{-}\int dx_{N}^{+}\int dx_{N}^{-}\mel{x_{0}^{+}}{\hat{A}}{x_{0}^{-}}\mel{x_{0}^{-}}{e^{i\hat{H}t/\hbar}}{x_{N}^{-}} \mel{x_{N}^{-}}{\hat{B}}{x_{N}^{+}}\mel{x_{N}^{+}}{e^{-i\hat{H}t/\hbar}}{x_{0}^{+}} \notag \\
   &=\lim_{N\to\infty}\left(\frac{m}{2\pi\hbar\epsilon}\right)^N\int dx_{0}^{+}..\int dx_{N}^{+}\int dx_{0}^{-}..\int dx_{N}^{-}\mel{x_{0}^{+}}{\hat{A}}{x_{0}^{-}} \mel{x_{N}^{-}}{\hat{B}}{x_{N}^{+}}e^{\frac{i}{\hbar}\left(S_N^+-S_N^-\right)}.
   \label{eq:picf}
\end{align}
In Eq.~\ref{eq:picf}, Tr signifies the trace, $\hat{H}$ is the system Hamiltonian,
$N$ is the number of time slices, $\epsilon = t/N$, 
$x_0^\pm$ and $x_N^\pm$ are the initial and final positions for the forward/backward 
path, and $S_N^{\pm}$ represent the action corresponding to the forward/backward paths
respectively.
Rewriting Eq.~\eqref{eq:picf} in terms of mean ($\textbf{\textit{y}}$) 
and difference ($\textbf{\textit{z}}$) path variables we obtain
\begin{align}
    C_{AB}(t)&=\lim_{N\to\infty}\left(\frac{m}{2\pi\hbar\epsilon}\right)^N\int dy_{0}...\int dy_{N}\int dz_{0}...\int dz_{N}\mel{y_{0}+\frac{z_0}{2}}{\hat{A}}{y_{0}-\frac{z_0}{2}}     \mel{y_{N}-\frac{z_N}{2}}{\hat{B}}{y_{N}+\frac{z_N}{2}} e^{i\left(S_N^+-S_N^-\right)/\hbar},
    \label{eq:cf-yz}
\end{align}
where 
\begin{equation}
    y_j=\frac{1}{2}\left(x_j^++x_j^-\right) \qquad \text{and} \qquad z_j=x_j^+-x_j^-,
\end{equation}
and $x_j^\pm$ are positions corresponding to the $j^\text{th}$ time slice 
along the forward/backward paths. 
The phase due to action difference between the forward and backward paths can
also be expressed in mean and difference variables,
\begingroup
\allowdisplaybreaks
\begin{align}
    S_N^+-S_N^- &=\epsilon\left[\sum_{j=1}^{N-1}-\frac{m}{\epsilon^2}\left(y_{j+1}-2y_j+y_{j-1}\right)z_j-\left(V(y_{j}+\frac{z_j}{2})-V(y_{j}-\frac{z_j}{2})\right)\right.\notag \\
    &\left.-\frac{m}{\epsilon^2}\left(y_{1}-y_0\right)z_0-\frac{1}{2}\left(V(y_{0}+\frac{z_0}{2})-V(y_{0}-\frac{z_0}{2})\right)+\frac{m}{\epsilon^2}\left(y_{N}-y_{N-1}\right)z_N-\frac{1}{2}\left(V(y_{N}+\frac{z_N}{2})-V(y_{N}-\frac{z_N}{2})\right)\right]  \label{eq:Sfb} \\
    &=\epsilon\left[\sum_{j=1}^{N-1}z_j\left\{-\frac{m}{\epsilon^2}\left(y_{j+1}-2y_j+y_{j-1}\right)-(V'(y_{j})+V'''(y_{j})\frac{z_j^2}{24}...)\right\}\right.\notag\\
    & +\left.z_0\left\{-\frac{m}{\epsilon^2}\left(y_{1}-y_0\right)-\frac{1}{2}(V'(y_{0})+V'''(y_{0})\frac{z_0^2}{24}...)\right\}+ z_N\left\{\frac{m}{\epsilon^2}\left(y_{N}-y_{N-1}\right)-\frac{1}{2}(V'(y_{N})+V'''(y_{N})\frac{z_N^2}{24}...)\right\}\right], \label{eq:Sfb-Vexp}
\end{align}
\endgroup
\end{widetext}
where Eq.~\eqref{eq:Sfb-Vexp} is obtained by expanding the potential terms 
in Eq.~\eqref{eq:Sfb} around $y_j$. 
We note that the derivation up to this point follows earlier work 
establishing a direct derivation of LSC-IVR from an exact path integral 
expression for a correlation function.~\cite{Shi2003} 

We can now employ the MFF scheme to filter the oscillatory integral
in Eq.~\ref{eq:picf} with some care. In the limit of a large filter
strength, the MFF scheme is equivalent to a stationary phase approximation;
it is evident that if we choose to filter the phase due to 
the action difference , $S_N^+-S_N^-$, in Eq.~\eqref{eq:Sfb}
in the limit of $\bm{c}\rightarrow\infty$ we will obtain 
the semiclassical DVV-IVR expression.
The classical-limit LSC-IVR expression can be obtained in the 
$\bm{c}\rightarrow\infty$ limit, if we choose to filter
only a portion of the phase,
\begin{align}
    \phi \left(\textbf{r}\right) &= -\frac{1}{\hbar}\sum_{j=1}^{N-1}
    \left[\frac{m}{\epsilon}\left(y_{j+1}-2y_j+y_{j-1}\right)+
    \epsilon V'(y_j)\right]z_j \notag \\
    & = -\frac{1}{\hbar}\sum_{j=1}^{N-1} f_j z_j,
    \label{eq:fil-phase}
\end{align}
where $ f_j=\frac{m}{\epsilon}\left(y_{j+1}-2y_j+y_{j-1}\right)+\epsilon V'(y_j)$ and 
$\textbf{r}=\{f_1,z_1,f_2,z_2,...,f_{N-1},z_{N-1}\}$. 

Evaluating the first derivative of the phase and using the 
definition in Eq.~\eqref{eq:fil-sf}, we obtain the Filinov 
smoothing factor,
\begin{align}
    F(\textbf{r};\textbf{c}) &= 
    \prod_{j=1}^{N-1} \left(1+\frac{c_{z_j} c_{f_j}}{\hbar^2}\right)^{\frac{1}{2}} e^{-\frac{c_{z_j}z_j^2}{2\hbar^2}}e^{-\frac{c_{f_j}f_j^2}{2\hbar^2}}.
    \label{eq:fil-sf2}
\end{align}
and the Filinov Filtered Path Integral (FFPI) expression
for the correlation function,
\begin{align}
    C^\text{FFPI}_{AB}(t;{\bf c})&=\lim_{N\to\infty}\left(\frac{m}{2\pi\hbar\epsilon}\right)^N\int dy_{0}...\int dy_{N}\int dz_{0}...\int dz_{N}  \notag \\ 
    &\times 
    \mel{y_{0}+\frac{z_0}{2}}{\hat{A}}{y_{0}-\frac{z_0}{2}}     \mel{y_{N}-\frac{z_N}{2}}{\hat{B}}{y_{N}+\frac{z_N}{2}} \notag \\
    & \times e^{i\left(S_N^+-S_N^-\right)/\hbar}
    \prod_{j=1}^{N-1} \left(1+\frac{c_{z_j} c_{f_j}}{\hbar^2}\right)^{\frac{1}{2}} e^{-\frac{c_{z_j}z_j^2}{2\hbar^2}}e^{-\frac{c_{f_j}f_j^2}{2\hbar^2}}.
    \label{eq:ffpi_cf}
\end{align}
Additional derivation details are provided in Appendix B.

The FFPI correlation function in Eq.~\ref{eq:ffpi_cf} contains
no SC prefactor, however, the summation runs over {\it all} forward and 
backwards paths (expressed in mean and difference variables), unlike 
in the SC approximation where only classical and near-classical path 
contributions are included.

In the limit that the Filinov parameter goes to zero,
the FFPI correlation function in Eq.~\eqref{eq:ffpi_cf} 
becomes the exact path integral expression for the 
correlation function.
In the classical limit, $\textbf{c}\to\infty$, 
the smoothing factor simplifies to
\begin{align}
    \lim_{\textbf{c}\to\infty}F(\textbf{r};\textbf{c})&=
    \lim_{\textbf{c}\to\infty}\prod_{j=1}^{N-1} 
    \left(1+\frac{c_{z_j} c_{f_j}}{\hbar^2}\right)^{\frac{1}{2}} 
    e^{-\frac{c_{z_j}z_j^2}{2\hbar^2}}e^{-\frac{c_{f_j}f_j^2}{2\hbar^2}} \notag \\
    &= (2\pi\hbar)^{N-1}\prod_{j=1}^{N-1} \delta(z_j)~\delta(f_j),
\end{align}
and the delta function in $f_j$ corresponds to constraining
the mean variable to a classical path,
\begin{align}
    \lim_{\epsilon\to0}\prod_{j=1}^{N-1} \delta(f_j) \implies \frac{d^2}{dt^2}y(t)=-V'[y(t)],
\end{align}
where $\epsilon\to 0$ or alternatively, $N\to \infty$.
The delta function in $z_j$ ensures that the forward and backward paths coincide.
Physically, these two constraints describe the LSC-IVR approximation, and it 
can be shown that in the classical limit, the FFPI expression becomes 
identical to LSC-IVR,
\begin{widetext}
\begingroup
\allowdisplaybreaks
\begin{align}
    \lim_{ {\bf c}\to\infty}C_{AB}^\text{FFPI}(t)&=\lim_{N\to\infty}\left(\frac{m^N}{2\pi\hbar\epsilon^N}\right)
    \int dy_{0}...\int dy_{N}\int dz_{0}...\int dz_{N} 
   \mel{y_{0}+\frac{z_0}{2}}{\hat{A}}{y_{0}-\frac{z_0}{2}} 
    \mel{y_{N}-\frac{z_N}{2}}{\hat{B}}{y_{N}+\frac{z_N}{2}}  e^{\frac{i}{\hbar}\left(S_N^+-S_N^-\right)} \notag \\
    & \times \prod_{j=1}^{N-1}\delta(z_j)~\delta(f_j) \\
  & = \lim_{N\to\infty} \frac{m^N}{2\pi\hbar\epsilon^N} 
  \int dy_{0} \int dy_{N} \int dz_{0}...\int dz_{N} 
  \int df_{1}...\int df_{N-1}  
  \abs{\frac{\partial \textbf{\textit {y}}}{\partial \textbf{\textit{f}}}}
  \mel{y_{0}+\frac{z_0}{2}}{\hat{A}}{y_{0}-\frac{z_0}{2}} \notag \\
  & \times \mel{y_{N}-\frac{z_N}{2}}{\hat{B}}{y_{N}+\frac{z_N}{2}}  e^{\frac{i}{\hbar}\left(S_N^+-S_N^-\right)} \prod_{j=1}^{N-1}\delta(z_j)~\delta(f_j) 
  \label{eq:cf-cl1} \\
  &=\frac{1}{2\pi\hbar}\int dy_{0}\int dy_{N}\int dz_{0}\int dz_{N}\,
  \abs{\frac{\partial p_0}{\partial y_N}}
  \mel{y_{0}+\frac{z_0}{2}}{\hat{A}}{y_{0}-\frac{z_0}{2}}  
  \mel{y_{N}-\frac{z_N}{2}}{\hat{B}}{y_{N}+\frac{z_N}{2}}
  e^{-\frac{i}{\hbar}p_0z_0}e^{\frac{i}{\hbar}ip_N z_N} \label{eq:cf-cl2} \\
  &=(2\pi\hbar)^{-1}\int dy_{0}\int dp_{0}\,A_W(y_0,p_0)B_W(y_N,p_N). \label{eq:cf-lsc}
\end{align}
\endgroup
\end{widetext}
We obtain Eq.~\eqref{eq:cf-cl2} from Eq.~\eqref{eq:cf-cl1} by evaluating the delta
function integrals over $f_1\ldots f_N$ and $z_1\ldots z_N$,
simplifying the action difference,
\begin{align}
    \lim_{N\to\infty} (S_N^{+}-S_N^{-}) =  -p_0 z_0 + p_N z_N,
\end{align}
and using the derivative identity
\begin{align}
    \lim_{\epsilon\to 0}\left( \frac{m}{\epsilon} \right)^N
    \left|\frac{\partial \textbf{\textit {y}}}{\partial \textbf{\textit{f}}}\right|
  =\abs{\frac{\partial p_0}{\partial y_N}},
\end{align}
where we define the vectors $\textit{\textbf y}=\left\{ y_1,\ldots, y_{N-1} \right\}$
and $\textit{\textbf f}=\left\{ f_1,\ldots, f_{N-1} \right\}$, and we introduce
momentum variables $p_j$ conjugate to positions $y_j$.
Equation~\eqref{eq:cf-lsc} is the LSC-IVR correlation function, where 
we use standard notation $O_W$ to represent the Wigner transform of the 
corresponding operator $\hat{O}$. 
 
\begin{figure}
   \centering
    \includegraphics[width=0.45\textwidth]{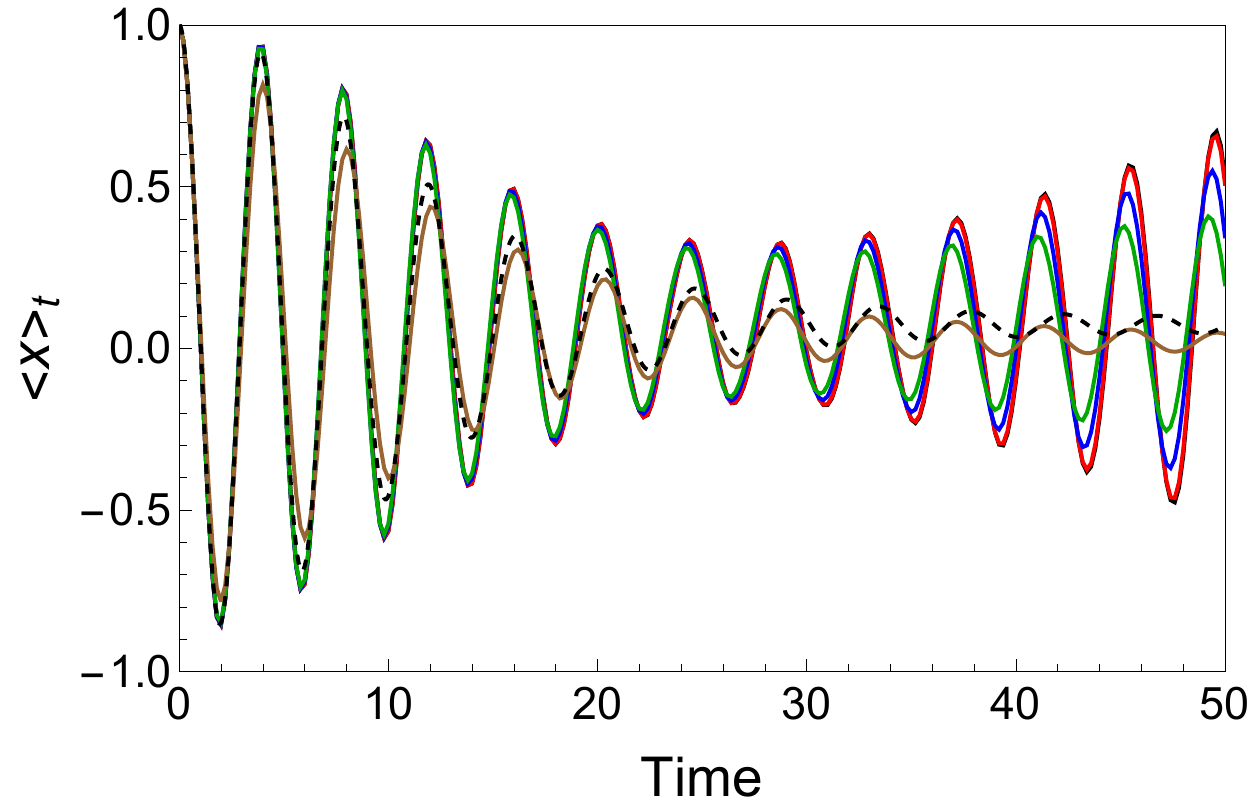}
    \caption{The position correlation function for a 1D anharmonic 
        oscillator obtained from exact real-time path integrals is 
        shown in black, LSC-IVR results are the black dashed line, and 
        FFPI correlation function results from our DVR implementation
        with $\bm{c}=10^{-4}$ shown in red, $\bm{c}=10^{-3}$ shown in blue, $\bm{c}=2\times 10^{-3}$ shown in green,
        and $\bm{c}=4\times10^{-3}$ shown in brown. 
        We find that while agreement
        in the quantum limit is good, for larger values of the Filinov 
        parameter there is significant deviation in the frequency of 
        oscillations from the LSC-IVR result.
    }
    \label{fig:ffpi_dvr}
\end{figure}

For finite, non-zero values of the filinov parameters, the FFPI correlation 
function in Eq.~\eqref{eq:ffpi_cf} must be evaluated by 
sampling \textit{all} paths that go from $(y_0,z_0)$ to $(y_N,z_N)$ in time $t$. 
Numerically demonstrating the systematic filtration achieved in the FFPI 
framework is challenging because in one limit, we must sample all paths 
and in the other limit we must sample only the classical paths associated
with the mean-variable. We start by diagonalizing the Hamiltonian matrix
on a standard DVR grid~\cite{Colbert1992a} to evaluate the short-time matrix elements
that comprise the exact real-time path integral correlation function. The 
exact path integral result is then obtained by multiplying these short-time 
matrix elements and summing over all possible intervening grid points. For
finite values of the Filinov parameter, we multiply by the Filinov smoothing
factor defined in Eq.~\ref{eq:fil-sf2}. 
In Fig.~\ref{fig:ffpi_dvr}, we plot the position correlation function 
for a 1D anharmonic oscillator previously studied using MQC-IVR. We show
that as the Filinov parameter is increased, we see quantum recurrence 
amplitudes decrease as expected. However, as expected this 
implementation cannot be used to obtain the numerical classical limit, 
with the results showing a mismatch in both frequency and amplitude 
compared with the LSC-IVR result. This is easily understood: the Filinov
smoothing factor in the classical limit corresponds to very narrow 
Gaussians, and correctly implementing this requires converging the grid to 
a density approaching the continuum. In addition, we recognize that although 
DVR matrix multiplication offers a way to numerically validate the FFPI 
correlation function expression, it is not scalable to realistic systems. 
This motivates the pursuit of a classical trajectory based approach that
reproduces the LSC-IVR limit and can perhaps, go beyond to capture some
quantum coherences.

In developing such a classical trajectory based implementation, we first 
classify the space of all paths into three types:
(i) Classical paths in the mean variable, $y(t)$, that have $f_j = z_j = 0$ for all $j$,
and that contribute in the $\textbf{c}\to\infty$ or LSC-IVR limit.
(ii) Pairs of classical forward and backward paths, $x^{+}(t)$ 
and $x^{-}(t)$ respectively. These are the paths that are explicitly 
generated when performing quantum-limit SC calculations. Re-writing 
the forward-backward paths in terms of the mean and difference 
variables, it can be shown that the corresponding $y(t)$ and $z(t)$ dynamics
correspond to classical dynamics under an effective potential
with $f_j \neq 0$ and $z_j \neq 0$.
(iii) All other `non-classical' paths that have, in general, $f_j \neq 0$ and $z_j \neq 0$. 
\begin{figure}[!htb]
    \centering
    \includegraphics[width=0.45\textwidth]{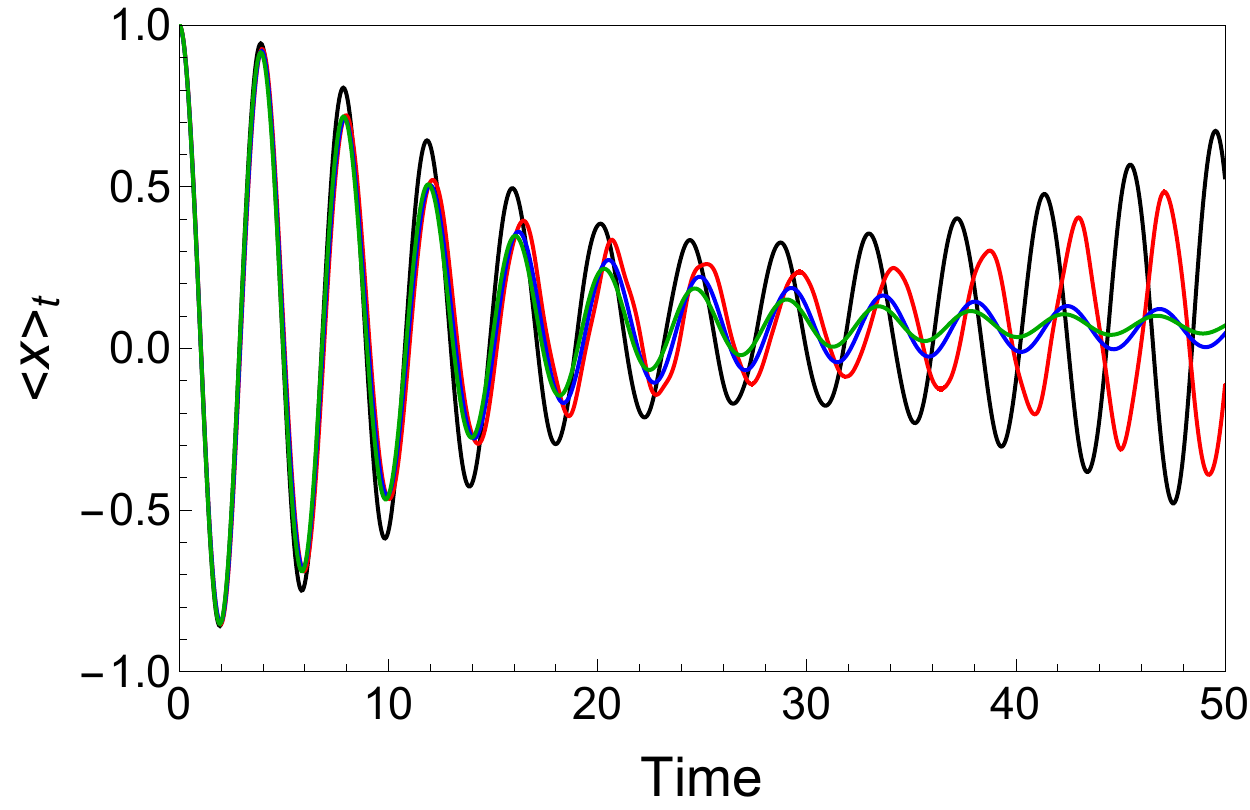}
    \caption{The FFPI position correlation function for a 1D anharmonic oscillator
        calculated using our approximate classical-trajectory based implementation
        is shown for $\bm{c}=0$ (red), $\bm{c}=10^{-2}$ (blue), 
     along with LSC-IVR results (green) and exact path integral results (black).}
    \label{fig:ffpi-1d}
\end{figure}

Numerical implementation challenges arise from the third type of path,
and indeed represent the challenge in undertaking any high-dimensional 
exact real-time path integral simulation. In Fig.~\ref{fig:ffpi-1d}, we 
show the position correlation function obtained when we approximately 
calculate the FFPI correlation function limiting ourselves to only
the first two types of classical paths. By design, this implementation does 
indeed correctly reproduce the classical limit LSC-IVR results, but 
while the short-time amplitudes are correctly described in the quantum
limit and some recurrence is observed, the amplitudes and frequencies
do not agree well in the quantum limit. 
An improved implementation is the subject of ongoing
work, but we note that devising a mixed-limit implementation where
a small number of system degrees of freedom are described at the DVR-level
of theory and the rest treated in the classical-limit SC level of theory
corresponds very closely to the Quantum-Classical
Path Integral method~\cite{Lambert2012a,Lambert2012b}.

\section{Conclusions}
We make the case that a rigorous SC framework for mixed quantum-classical
simulations offers a balanced approach to large system simulations,
retaining the ability to capture important quantum effects at a reduced 
cost by mitigating the sign problem. We derived the MQC-IVR expression
for real-time correlation functions using the MFF scheme to damp the 
phase of a quantum limit correlation function and verify that changing
the strength of Filinov parameter systematically changes the correlation
function from quantum-limit DHK-IVR to the classical-limit Husimi IVR
for linear operators. Further, we demonstrate the efficacy of mixed-limit
MQC-IVR implementations for high-dimensional systems where quantizing 
a few degrees of freedom is typically sufficient to capture quantum effects.
We show that several factors influence the choice of which modes to treat 
in the quantum limit including the nature of operator $\hat B$ and 
inter-mode coupling strength.
\begin{figure}
    \centering
    \includegraphics[width=0.45\textwidth]{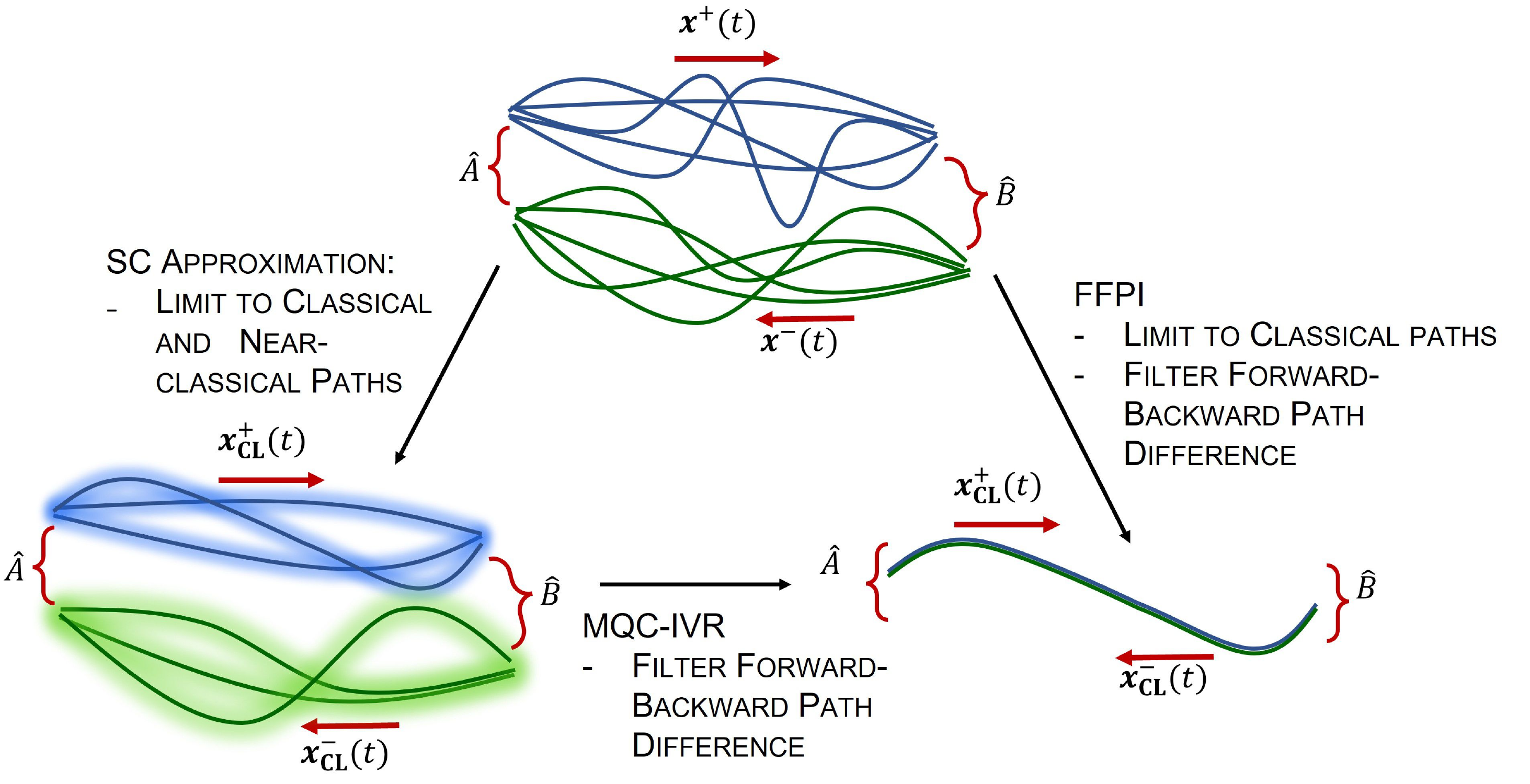}
    \caption{We sketch a cartoon representing {\it all} forward, 
        $\textbf{\textit{x}}^+ (t)$, and backward,
        $\textbf{\textit{x}}^- (t)$, paths that contribute to 
    the real-time correlation function $C_{AB}(t)$ in an exact 
    path integral calculation. As described in the introduction, using 
    the stationary phase approximation, it is possible to derive a quantum-limit
    SC approximation to the correlation function where only classical,
    $\textbf{\textit{x}}^{\pm}_\text{CL}(t)$, 
    and near-classical paths contribute. The MQC-IVR correlation function
    is obtained by Filinov filtering the difference in action 
    between the forward and backward paths; in the limit of a strong filter,
    we obtain an expression where the paths coincide leading to a classical-limit
    SC result, the Husimi-IVR. In order to eliminate the SC prefactor, we derive the FFPI 
    expression by simultaneously filtering contributions from non-classical
    paths and placing constraints on the difference in action between the 
    forward and backward paths. In the limit of a strong filter, the FFPI
    correlation function also leads to a classical-limit SC result, in 
    this case, LSC-IVR.}
    \label{fig:fb-paths}
\end{figure}

While MQC-IVR offers an accurate framework when $\hat B$ is a linear
operator, for non-linear operators, the classical-limit 
can be incorrect even at time zero.~\cite{LohoChoudhury2020} In a manuscript currently
under preparation we address this limitation and offer some strategies
to correct for the difference between the classical-limit MQC-IVR and 
the Husimi-IVR correlation function. In addition, we note that while
there is significant improvement over DHK-IVR, the presence of a 
full dimensional SC prefactor remains a computational challenge
to be overcome. In the mixed-limit AMQC-IVR, we offer a potential
solution through a separable prefactor approximation. In addition, we 
note that many advances in approximating the SC prefactor have been 
made~\cite{Gelabert2000d,Issack2005,Issack2007,Issack2007a,Issack2007b,Wong2011c,Guallar1999,Guallar2000,Tatchen2011,Liberto2016b} and several of these ideas
may be employed in the context of MQC-IVR.

Finally, we introduce the FFPI framework to achieve a prefactor-free
expression for real-time correlation functions. The connections between
the exact real-time path integral expression, MQC-IVR, FFPI, and LSC-IVR
are summarized in Fig.~\ref{fig:fb-paths}. Constructing a classical trajectory-based
approach for path sampling the FFPI correlation function remains an outstanding 
challenge and one that we intend to explore through both more traditional 
path sampling strategies and through data-driven strategies.


\begin{acknowledgments}
    The authors would like to acknowledge prior postdoctoral researchers, graduate
    and undergraduate students who contributed to the previously published work described
    in this feature, particularly Dr. Sergey Antipov. N. A. and S. M. acknowledge funding
    from NSF CAREER Grant No. CHE 1555205.
\end{acknowledgments}

\appendix
\section{MQC-IVR Prefactor}
The MQC-IVR prefactor derived for the Forward-Backward implementation for a general operator $\hat{B}$ is, 
\begin{align}
    D &= 2^{-\frac{N}{2}}\text{det}\left(\gamma_{0}^{-1}\textbf{G}\right)^{\frac{1}{2}} \notag \\
    & \times \text{det}\left[\frac{1}{2}\left(\textbf{M}_{pp}^{b}-i\gamma_{0}\textbf{M}_{qp}^{b}\right)\left(\textbf{G}^{-1}+\mathbb{1}\right)\left(\textbf{M}_{pp}^{f}\gamma_{0}+i\textbf{M}_{pq}^{f}\right)\right. \notag \\
    &+ \left. \left(\gamma_{0}\textbf{M}_{qq}^{b}+i\textbf{M}_{pq}^{b}\right) (\frac{1}{2}\gamma_{t}^{-1}+\textbf{c}_p)\textbf{G}^{-1}\left(\textbf{M}_{pp}^{f}\gamma_{0}+i\textbf{M}_{pq}^{f}\right)\right. \notag 
    \\
    &+ \left. \frac{1}{2}\left(\gamma_{0}\textbf{M}_{qq}^{b}+i\textbf{M}_{pq}^{b}\right)\left(\textbf{G}^{-1}+\mathbb{1}\right)\left(\textbf{M}_{qq}^{f}-i\textbf{M}_{qp}^{f}\gamma_{0}\right)\right. \notag \\ 
    &+ \left.\left(\textbf{M}_{pp}^{b}-i\gamma_{0}\textbf{M}_{qp}^{b}\right)(\frac{1}{2}\gamma_{t}+\textbf{c}_{q})\textbf{G}^{-1}\left(\textbf{M}_{qq}^{f}-i\textbf{M}_{qp}^{f}\gamma_{0}\right) \right]^{\frac{1}{2}},
\end{align}
where the diagonal matrix $\textbf{G}=(\textbf{c}_{q}+\gamma_{t})\textbf{c}_p+\textbf{c}_{q}(\gamma_{t}^{-1}+\textbf{c}_p)$.

\section{Deriving the FFPI correlation function-IVR Prefactor}
Once the choice of phase to be filtered is established [Eq.~\eqref{eq:fil-phase}], it's derivatives can be evaluated to be, 
\begin{align} 
    \frac{\partial \phi(\textbf{r})}{\partial\textbf{r}}= -\frac{1}{\hbar}\textbf{K}.\textbf{r} \qquad \& \qquad \frac{\partial^2 \phi(\textbf{r})}{\partial\textbf{r}^2}=-\frac{1}{\hbar}\textbf{K},
 \label{fil-der}
\end{align}
where the constant block diagonal matrix $\textbf{K}$  
is,
\begin{align}
    \textbf{K} = \left(\begin{array}{ccccccc}
    0 & 1 &\dots &\dots &\dots\\
    1 & 0 &\dots &\dots &\dots\\
     \vdots &\vdots & \ddots &\vdots &\vdots\\
     \dots &\dots &\dots & 0 & 1 \\
     \dots &\dots &\dots & 1 & 0 \\
    \end{array}\right)_{2(N-1)\times 2(N-1)}.
\end{align}
The diagonal matrix of Filinov parameters is defined as 
\begingroup
\allowdisplaybreaks
\begin{equation}
    \textbf{c}=\left(\begin{array}{ccccccc}
    c_{z_1} & \dots & \dots  & \dots & \dots & \dots & \dots \\
    \dots & c_{f_1} & \dots & \dots & \dots & \dots & \dots \\
     \dots & \dots & c_{z_2} & \dots & \dots & \dots & \dots\\
     \dots & \dots & \dots & c_{f_2} & \dots & \dots & \dots \\
    \vdots & \vdots& \vdots & \vdots & \ddots & \vdots &  \vdots \\
    \dots & \dots & \dots & \dots & \dots & c_{z_{N-1}} & \dots \\
    \dots & \dots & \dots & \dots & \dots & \dots & c_{f_{N-1}} \\ 
    \end{array}\right)_{2(N-1)\times 2(N-1)}.
\end{equation}
\endgroup
\bibliographystyle{apsrev4-2}
\bibliography{sc-review14.bib}

\providecommand{\noopsort}[1]{}\providecommand{\singleletter}[1]{#1}%
\begin{thebibliography}{166}%
\makeatletter
\providecommand \@ifxundefined [1]{%
 \@ifx{#1\undefined}
}%
\providecommand \@ifnum [1]{%
 \ifnum #1\expandafter \@firstoftwo
 \else \expandafter \@secondoftwo
 \fi
}%
\providecommand \@ifx [1]{%
 \ifx #1\expandafter \@firstoftwo
 \else \expandafter \@secondoftwo
 \fi
}%
\providecommand \natexlab [1]{#1}%
\providecommand \enquote  [1]{``#1''}%
\providecommand \bibnamefont  [1]{#1}%
\providecommand \bibfnamefont [1]{#1}%
\providecommand \citenamefont [1]{#1}%
\providecommand \href@noop [0]{\@secondoftwo}%
\providecommand \href [0]{\begingroup \@sanitize@url \@href}%
\providecommand \@href[1]{\@@startlink{#1}\@@href}%
\providecommand \@@href[1]{\endgroup#1\@@endlink}%
\providecommand \@sanitize@url [0]{\catcode `\\12\catcode `\$12\catcode
  `\&12\catcode `\#12\catcode `\^12\catcode `\_12\catcode `\%12\relax}%
\providecommand \@@startlink[1]{}%
\providecommand \@@endlink[0]{}%
\providecommand \url  [0]{\begingroup\@sanitize@url \@url }%
\providecommand \@url [1]{\endgroup\@href {#1}{\urlprefix }}%
\providecommand \urlprefix  [0]{URL }%
\providecommand \Eprint [0]{\href }%
\providecommand \doibase [0]{https://doi.org/}%
\providecommand \selectlanguage [0]{\@gobble}%
\providecommand \bibinfo  [0]{\@secondoftwo}%
\providecommand \bibfield  [0]{\@secondoftwo}%
\providecommand \translation [1]{[#1]}%
\providecommand \BibitemOpen [0]{}%
\providecommand \bibitemStop [0]{}%
\providecommand \bibitemNoStop [0]{.\EOS\space}%
\providecommand \EOS [0]{\spacefactor3000\relax}%
\providecommand \BibitemShut  [1]{\csname bibitem#1\endcsname}%
\let\auto@bib@innerbib\@empty
\bibitem [{\citenamefont {Miller}(2001)}]{Miller2001a}%
  \BibitemOpen
  \bibfield  {author} {\bibinfo {author} {\bibfnamefont {W.~H.}\ \bibnamefont
  {Miller}},\ }\href {https://doi.org/10.1021/JP003712K} {\bibfield  {journal}
  {\bibinfo  {journal} {Journal of Physical Chemistry A}\ }\textbf {\bibinfo
  {volume} {105}},\ \bibinfo {pages} {2942} (\bibinfo {year}
  {2001})}\BibitemShut {NoStop}%
\bibitem [{\citenamefont {Tannor}\ and\ \citenamefont
  {Garashchuk}(2000)}]{Tannor2000}%
  \BibitemOpen
  \bibfield  {author} {\bibinfo {author} {\bibfnamefont {D.~J.}\ \bibnamefont
  {Tannor}}\ and\ \bibinfo {author} {\bibfnamefont {S.}~\bibnamefont
  {Garashchuk}},\ }\href {https://doi.org/10.1146/annurev.physchem.51.1.553}
  {\bibfield  {journal} {\bibinfo  {journal} {Annual Review of Physical
  Chemistry}\ }\textbf {\bibinfo {volume} {51}},\ \bibinfo {pages} {553}
  (\bibinfo {year} {2000})}\BibitemShut {NoStop}%
\bibitem [{\citenamefont {Thoss}\ and\ \citenamefont {Wang}(2004)}]{Thoss2004}%
  \BibitemOpen
  \bibfield  {author} {\bibinfo {author} {\bibfnamefont {M.}~\bibnamefont
  {Thoss}}\ and\ \bibinfo {author} {\bibfnamefont {H.}~\bibnamefont {Wang}},\
  }\href {https://doi.org/10.1146/annurev.physchem.55.091602.094429} {\bibfield
   {journal} {\bibinfo  {journal} {Annual Review of Physical Chemistry}\
  }\textbf {\bibinfo {volume} {55}},\ \bibinfo {pages} {299} (\bibinfo {year}
  {2004})}\BibitemShut {NoStop}%
\bibitem [{\citenamefont {Kay}(2005)}]{Kay2005}%
  \BibitemOpen
  \bibfield  {author} {\bibinfo {author} {\bibfnamefont {K.~G.}\ \bibnamefont
  {Kay}},\ }\href {https://doi.org/10.1146/annurev.physchem.56.092503.141257}
  {\bibfield  {journal} {\bibinfo  {journal} {Annual Review of Physical
  Chemistry}\ }\textbf {\bibinfo {volume} {56}},\ \bibinfo {pages} {255}
  (\bibinfo {year} {2005})}\BibitemShut {NoStop}%
\bibitem [{\citenamefont {Stock}\ and\ \citenamefont
  {Thoss}(2005)}]{Stock2005}%
  \BibitemOpen
  \bibfield  {author} {\bibinfo {author} {\bibfnamefont {G.}~\bibnamefont
  {Stock}}\ and\ \bibinfo {author} {\bibfnamefont {M.}~\bibnamefont {Thoss}},\
  }in\ \href {https://doi.org/10.1002/0471739464.ch5} {\emph {\bibinfo
  {booktitle} {Advances in Chemical Physics}}},\ Vol.\ \bibinfo {volume} {131}\
  (\bibinfo  {publisher} {John Wiley \& Sons, Ltd},\ \bibinfo {year} {2005})\
  pp.\ \bibinfo {pages} {243--375}\BibitemShut {NoStop}%
\bibitem [{\citenamefont {Heller}(2006)}]{Heller2006}%
  \BibitemOpen
  \bibfield  {author} {\bibinfo {author} {\bibfnamefont {E.~J.}\ \bibnamefont
  {Heller}},\ }\href {https://doi.org/10.1021/ar040196y} {\bibfield  {journal}
  {\bibinfo  {journal} {Accounts of Chemical Research}\ }\textbf {\bibinfo
  {volume} {39}},\ \bibinfo {pages} {127} (\bibinfo {year} {2006})}\BibitemShut
  {NoStop}%
\bibitem [{\citenamefont {Kapral}(2006)}]{Kapral2006}%
  \BibitemOpen
  \bibfield  {author} {\bibinfo {author} {\bibfnamefont {R.}~\bibnamefont
  {Kapral}},\ }\href
  {https://doi.org/10.1146/annurev.physchem.57.032905.104702} {\bibfield
  {journal} {\bibinfo  {journal} {Annual Review of Physical Chemistry}\
  }\textbf {\bibinfo {volume} {57}},\ \bibinfo {pages} {129} (\bibinfo {year}
  {2006})}\BibitemShut {NoStop}%
\bibitem [{\citenamefont {Pollak}(2007)}]{Pollak2007}%
  \BibitemOpen
  \bibfield  {author} {\bibinfo {author} {\bibfnamefont {E.}~\bibnamefont
  {Pollak}},\ }\href {https://doi.org/10.1007/978-3-540-34460-5_11} {\bibfield
  {journal} {\bibinfo  {journal} {Springer Series in Chemical Physics}\
  }\textbf {\bibinfo {volume} {83}},\ \bibinfo {pages} {259} (\bibinfo {year}
  {2007})}\BibitemShut {NoStop}%
\bibitem [{\citenamefont {Miller}(2009)}]{Miller2009}%
  \BibitemOpen
  \bibfield  {author} {\bibinfo {author} {\bibfnamefont {W.~H.}\ \bibnamefont
  {Miller}},\ }\href {https://doi.org/10.1021/jp809907p} {\bibfield  {journal}
  {\bibinfo  {journal} {Journal of Physical Chemistry A}\ }\textbf {\bibinfo
  {volume} {113}},\ \bibinfo {pages} {1405} (\bibinfo {year}
  {2009})}\BibitemShut {NoStop}%
\bibitem [{\citenamefont {McRobbie}\ \emph {et~al.}(2009)\citenamefont
  {McRobbie}, \citenamefont {Hanna}, \citenamefont {Shi},\ and\ \citenamefont
  {Geva}}]{McRobbie2009b}%
  \BibitemOpen
  \bibfield  {author} {\bibinfo {author} {\bibfnamefont {P.~L.}\ \bibnamefont
  {McRobbie}}, \bibinfo {author} {\bibfnamefont {G.}~\bibnamefont {Hanna}},
  \bibinfo {author} {\bibfnamefont {Q.}~\bibnamefont {Shi}},\ and\ \bibinfo
  {author} {\bibfnamefont {E.}~\bibnamefont {Geva}},\ }\href
  {https://doi.org/10.1021/ar800280s} {\bibfield  {journal} {\bibinfo
  {journal} {Accounts of Chemical Research}\ }\textbf {\bibinfo {volume}
  {42}},\ \bibinfo {pages} {1299} (\bibinfo {year} {2009})}\BibitemShut
  {NoStop}%
\bibitem [{\citenamefont {Liu}(2015)}]{Liu2015}%
  \BibitemOpen
  \bibfield  {author} {\bibinfo {author} {\bibfnamefont {J.}~\bibnamefont
  {Liu}},\ }\href {https://doi.org/10.1002/qua.24872} {\bibfield  {journal}
  {\bibinfo  {journal} {International Journal of Quantum Chemistry}\ }\textbf
  {\bibinfo {volume} {115}},\ \bibinfo {pages} {657} (\bibinfo {year}
  {2015})}\BibitemShut {NoStop}%
\bibitem [{\citenamefont {Lee}\ \emph {et~al.}(2016)\citenamefont {Lee},
  \citenamefont {Huo},\ and\ \citenamefont {Coker}}]{Lee2016}%
  \BibitemOpen
  \bibfield  {author} {\bibinfo {author} {\bibfnamefont {M.~K.}\ \bibnamefont
  {Lee}}, \bibinfo {author} {\bibfnamefont {P.}~\bibnamefont {Huo}},\ and\
  \bibinfo {author} {\bibfnamefont {D.~F.}\ \bibnamefont {Coker}},\ }\href
  {https://doi.org/10.1146/ANNUREV-PHYSCHEM-040215-112252} {\bibfield
  {journal} {\bibinfo  {journal} {Annual Review of Physical Chemistry}\
  }\textbf {\bibinfo {volume} {67}},\ \bibinfo {pages} {639} (\bibinfo {year}
  {2016})}\BibitemShut {NoStop}%
\bibitem [{\citenamefont {L{\"{u}}}\ \emph {et~al.}(2019)\citenamefont
  {L{\"{u}}}, \citenamefont {Hu}, \citenamefont {Hedeg{\aa}rd},\ and\
  \citenamefont {Brandbyge}}]{Lu2019}%
  \BibitemOpen
  \bibfield  {author} {\bibinfo {author} {\bibfnamefont {J.-T.}\ \bibnamefont
  {L{\"{u}}}}, \bibinfo {author} {\bibfnamefont {B.-Z.}\ \bibnamefont {Hu}},
  \bibinfo {author} {\bibfnamefont {P.}~\bibnamefont {Hedeg{\aa}rd}},\ and\
  \bibinfo {author} {\bibfnamefont {M.}~\bibnamefont {Brandbyge}},\ }\href
  {https://doi.org/10.1016/j.progsurf.2018.07.002} {\bibfield  {journal}
  {\bibinfo  {journal} {Progress in Surface Science}\ }\textbf {\bibinfo
  {volume} {94}},\ \bibinfo {pages} {21} (\bibinfo {year} {2019})},\ \Eprint
  {https://arxiv.org/abs/1712.03863} {arXiv:1712.03863} \BibitemShut {NoStop}%
\bibitem [{\citenamefont {Conte}\ and\ \citenamefont
  {Ceotto}(2020)}]{Conte2020c}%
  \BibitemOpen
  \bibfield  {author} {\bibinfo {author} {\bibfnamefont {R.}~\bibnamefont
  {Conte}}\ and\ \bibinfo {author} {\bibfnamefont {M.}~\bibnamefont {Ceotto}},\
  }in\ \href {https://doi.org/10.1002/9781119417774.ch19} {\emph {\bibinfo
  {booktitle} {Quantum Chemistry and Dynamics of Excited States}}}\ (\bibinfo
  {publisher} {Wiley},\ \bibinfo {year} {2020})\ Chap.~\bibinfo {chapter} {19},
  pp.\ \bibinfo {pages} {595--628}\BibitemShut {NoStop}%
\bibitem [{\citenamefont {Bonfanti}\ \emph {et~al.}(2020)\citenamefont
  {Bonfanti}, \citenamefont {Worth},\ and\ \citenamefont
  {Burghardt}}]{Bonfanti2020}%
  \BibitemOpen
  \bibfield  {author} {\bibinfo {author} {\bibfnamefont {M.}~\bibnamefont
  {Bonfanti}}, \bibinfo {author} {\bibfnamefont {G.~A.}\ \bibnamefont
  {Worth}},\ and\ \bibinfo {author} {\bibfnamefont {I.}~\bibnamefont
  {Burghardt}},\ }in\ \href {https://doi.org/10.1002/9781119417774.ch12} {\emph
  {\bibinfo {booktitle} {Quantum Chemistry and Dynamics of Excited States}}}\
  (\bibinfo  {publisher} {Wiley},\ \bibinfo {year} {2020})\ pp.\ \bibinfo
  {pages} {383--411}\BibitemShut {NoStop}%
\bibitem [{\citenamefont {Van{\'{i}}{\v{c}}ek}\ and\ \citenamefont
  {Begu{\v{s}}i{\'{c}}}(2021)}]{Vanicek2021}%
  \BibitemOpen
  \bibfield  {author} {\bibinfo {author} {\bibfnamefont {J.}~\bibnamefont
  {Van{\'{i}}{\v{c}}ek}}\ and\ \bibinfo {author} {\bibfnamefont
  {T.}~\bibnamefont {Begu{\v{s}}i{\'{c}}}},\ }\href
  {https://doi.org/10.1016/B978-0-12-817234-6.00011-8} {\bibfield  {journal}
  {\bibinfo  {journal} {Molecular Spectroscopy and Quantum Dynamics}\ ,\
  \bibinfo {pages} {199}} (\bibinfo {year} {2021})}\BibitemShut {NoStop}%
\bibitem [{\citenamefont {Loring}(2022)}]{Loring2022}%
  \BibitemOpen
  \bibfield  {author} {\bibinfo {author} {\bibfnamefont {R.~F.}\ \bibnamefont
  {Loring}},\ }\href {https://doi.org/10.1146/ANNUREV-PHYSCHEM-082620-021302}
  {\bibfield  {journal} {\bibinfo  {journal} {Annual Review of Physical
  Chemistry}\ }\textbf {\bibinfo {volume} {73}},\ \bibinfo {pages} {273}
  (\bibinfo {year} {2022})}\BibitemShut {NoStop}%
\bibitem [{\citenamefont {Venkataraman}\ and\ \citenamefont
  {Miller}(2007)}]{Venkataraman2007}%
  \BibitemOpen
  \bibfield  {author} {\bibinfo {author} {\bibfnamefont {C.}~\bibnamefont
  {Venkataraman}}\ and\ \bibinfo {author} {\bibfnamefont {W.~H.}\ \bibnamefont
  {Miller}},\ }\href {https://doi.org/10.1063/1.2567200} {\bibfield  {journal}
  {\bibinfo  {journal} {The Journal of Chemical Physics}\ }\textbf {\bibinfo
  {volume} {126}},\ \bibinfo {pages} {094104} (\bibinfo {year}
  {2007})}\BibitemShut {NoStop}%
\bibitem [{\citenamefont {Sun}\ \emph {et~al.}(1998{\natexlab{a}})\citenamefont
  {Sun}, \citenamefont {Wang},\ and\ \citenamefont {Miller}}]{Sun1998}%
  \BibitemOpen
  \bibfield  {author} {\bibinfo {author} {\bibfnamefont {X.}~\bibnamefont
  {Sun}}, \bibinfo {author} {\bibfnamefont {H.}~\bibnamefont {Wang}},\ and\
  \bibinfo {author} {\bibfnamefont {W.~H.}\ \bibnamefont {Miller}},\ }\href
  {http://jcp.aip.org/jcp/copyright.jsp} {\bibfield  {journal} {\bibinfo
  {journal} {Journal of Chemical Physics}\ }\textbf {\bibinfo {volume} {109}},\
  \bibinfo {pages} {4190} (\bibinfo {year} {1998}{\natexlab{a}})}\BibitemShut
  {NoStop}%
\bibitem [{\citenamefont {Wang}\ \emph {et~al.}(1998)\citenamefont {Wang},
  \citenamefont {Sun},\ and\ \citenamefont {Miller}}]{Wang1998}%
  \BibitemOpen
  \bibfield  {author} {\bibinfo {author} {\bibfnamefont {H.}~\bibnamefont
  {Wang}}, \bibinfo {author} {\bibfnamefont {X.}~\bibnamefont {Sun}},\ and\
  \bibinfo {author} {\bibfnamefont {W.~H.}\ \bibnamefont {Miller}},\ }\href
  {https://doi.org/10.1063/1.476447} {\bibfield  {journal} {\bibinfo  {journal}
  {Journal of Chemical Physics}\ }\textbf {\bibinfo {volume} {108}},\ \bibinfo
  {pages} {9726} (\bibinfo {year} {1998})}\BibitemShut {NoStop}%
\bibitem [{\citenamefont {Shi}\ and\ \citenamefont
  {Geva}(2003{\natexlab{a}})}]{Shi2003c}%
  \BibitemOpen
  \bibfield  {author} {\bibinfo {author} {\bibfnamefont {Q.}~\bibnamefont
  {Shi}}\ and\ \bibinfo {author} {\bibfnamefont {E.}~\bibnamefont {Geva}},\
  }\href {https://doi.org/10.1021/jp030497} {\bibfield  {journal} {\bibinfo
  {journal} {Journal of Physical Chemistry A}\ }\textbf {\bibinfo {volume}
  {107}},\ \bibinfo {pages} {9059} (\bibinfo {year}
  {2003}{\natexlab{a}})}\BibitemShut {NoStop}%
\bibitem [{\citenamefont {Being}\ \emph {et~al.}(2005)\citenamefont {Being},
  \citenamefont {Shi},\ and\ \citenamefont {Geva}}]{BeingJ.Ka2005}%
  \BibitemOpen
  \bibfield  {author} {\bibinfo {author} {\bibfnamefont {K.~J.}\ \bibnamefont
  {Being}}, \bibinfo {author} {\bibfnamefont {Q.}~\bibnamefont {Shi}},\ and\
  \bibinfo {author} {\bibfnamefont {E.}~\bibnamefont {Geva}},\ }\href
  {https://doi.org/10.1021/JP051223K} {\bibfield  {journal} {\bibinfo
  {journal} {Journal of Physical Chemistry A}\ }\textbf {\bibinfo {volume}
  {109}},\ \bibinfo {pages} {5527} (\bibinfo {year} {2005})}\BibitemShut
  {NoStop}%
\bibitem [{\citenamefont {Navrotskaya}\ and\ \citenamefont
  {Geva}(2006)}]{Navrotskaya2006}%
  \BibitemOpen
  \bibfield  {author} {\bibinfo {author} {\bibfnamefont {I.}~\bibnamefont
  {Navrotskaya}}\ and\ \bibinfo {author} {\bibfnamefont {E.}~\bibnamefont
  {Geva}},\ }\href {https://doi.org/10.1021/JP066243G} {\bibfield  {journal}
  {\bibinfo  {journal} {Journal of Physical Chemistry A}\ }\textbf {\bibinfo
  {volume} {111}},\ \bibinfo {pages} {460} (\bibinfo {year}
  {2006})}\BibitemShut {NoStop}%
\bibitem [{\citenamefont {Being}\ and\ \citenamefont
  {Geva}(2006{\natexlab{a}})}]{Being2006}%
  \BibitemOpen
  \bibfield  {author} {\bibinfo {author} {\bibfnamefont {K.~J.}\ \bibnamefont
  {Being}}\ and\ \bibinfo {author} {\bibfnamefont {E.}~\bibnamefont {Geva}},\
  }\href {https://doi.org/10.1021/JP063907D} {\bibfield  {journal} {\bibinfo
  {journal} {Journal of Physical Chemistry A}\ }\textbf {\bibinfo {volume}
  {110}},\ \bibinfo {pages} {13131} (\bibinfo {year}
  {2006}{\natexlab{a}})}\BibitemShut {NoStop}%
\bibitem [{\citenamefont {Being}\ and\ \citenamefont
  {Geva}(2006{\natexlab{b}})}]{Being2006a}%
  \BibitemOpen
  \bibfield  {author} {\bibinfo {author} {\bibfnamefont {K.~J.}\ \bibnamefont
  {Being}}\ and\ \bibinfo {author} {\bibfnamefont {E.}~\bibnamefont {Geva}},\
  }\href {https://doi.org/10.1021/JP062363C} {\bibfield  {journal} {\bibinfo
  {journal} {Journal of Physical Chemistry A}\ }\textbf {\bibinfo {volume}
  {110}},\ \bibinfo {pages} {9555} (\bibinfo {year}
  {2006}{\natexlab{b}})}\BibitemShut {NoStop}%
\bibitem [{\citenamefont {V{\'{a}}zquez}\ \emph {et~al.}(2010)\citenamefont
  {V{\'{a}}zquez}, \citenamefont {Navrotskaya},\ and\ \citenamefont
  {Geva}}]{Vazquez2010}%
  \BibitemOpen
  \bibfield  {author} {\bibinfo {author} {\bibfnamefont {F.~X.}\ \bibnamefont
  {V{\'{a}}zquez}}, \bibinfo {author} {\bibfnamefont {I.}~\bibnamefont
  {Navrotskaya}},\ and\ \bibinfo {author} {\bibfnamefont {E.}~\bibnamefont
  {Geva}},\ }\href {https://doi.org/10.1021/JP1010499} {\bibfield  {journal}
  {\bibinfo  {journal} {Journal of Physical Chemistry A}\ }\textbf {\bibinfo
  {volume} {114}},\ \bibinfo {pages} {5682} (\bibinfo {year}
  {2010})}\BibitemShut {NoStop}%
\bibitem [{\citenamefont {Wang}\ \emph {et~al.}(2000)\citenamefont {Wang},
  \citenamefont {Thoss},\ and\ \citenamefont {Miller}}]{Wang2000a}%
  \BibitemOpen
  \bibfield  {author} {\bibinfo {author} {\bibfnamefont {H.}~\bibnamefont
  {Wang}}, \bibinfo {author} {\bibfnamefont {M.}~\bibnamefont {Thoss}},\ and\
  \bibinfo {author} {\bibfnamefont {W.~H.}\ \bibnamefont {Miller}},\ }\href
  {https://doi.org/10.1063/1.480560} {\bibfield  {journal} {\bibinfo  {journal}
  {Journal of Chemical Physics}\ }\textbf {\bibinfo {volume} {112}},\ \bibinfo
  {pages} {47} (\bibinfo {year} {2000})}\BibitemShut {NoStop}%
\bibitem [{\citenamefont {Grossmann}(2000)}]{Grossmann2000}%
  \BibitemOpen
  \bibfield  {author} {\bibinfo {author} {\bibfnamefont {F.}~\bibnamefont
  {Grossmann}},\ }\href@noop {} {\bibfield  {journal} {\bibinfo  {journal}
  {Physical Review Letters}\ }\textbf {\bibinfo {volume} {85}},\ \bibinfo
  {pages} {903} (\bibinfo {year} {2000})}\BibitemShut {NoStop}%
\bibitem [{\citenamefont {Burant}\ and\ \citenamefont
  {Batista}(2002)}]{Burant2002}%
  \BibitemOpen
  \bibfield  {author} {\bibinfo {author} {\bibfnamefont {J.~C.}\ \bibnamefont
  {Burant}}\ and\ \bibinfo {author} {\bibfnamefont {V.~S.}\ \bibnamefont
  {Batista}},\ }\href {https://doi.org/10.1063/1.1436306} {\bibfield  {journal}
  {\bibinfo  {journal} {The Journal of Chemical Physics}\ }\textbf {\bibinfo
  {volume} {116}},\ \bibinfo {pages} {2748} (\bibinfo {year}
  {2002})}\BibitemShut {NoStop}%
\bibitem [{\citenamefont {Sep{\'{u}}lveda}\ and\ \citenamefont
  {Grossmann}(1996)}]{Sepulveda1996}%
  \BibitemOpen
  \bibfield  {author} {\bibinfo {author} {\bibfnamefont {M.~A.}\ \bibnamefont
  {Sep{\'{u}}lveda}}\ and\ \bibinfo {author} {\bibfnamefont {F.}~\bibnamefont
  {Grossmann}},\ }in\ \href {https://doi.org/10.1002/9780470141557.ch4} {\emph
  {\bibinfo {booktitle} {Advances in Chemical Physics}}},\ Vol.~\bibinfo
  {volume} {96}\ (\bibinfo  {publisher} {John Wiley \& Sons, Ltd},\ \bibinfo
  {year} {1996})\ pp.\ \bibinfo {pages} {191--304}\BibitemShut {NoStop}%
\bibitem [{\citenamefont {Liu}\ \emph {et~al.}(2011{\natexlab{a}})\citenamefont
  {Liu}, \citenamefont {Miller}, \citenamefont {Fanourgakis}, \citenamefont
  {Xantheas}, \citenamefont {Imoto},\ and\ \citenamefont {Saito}}]{Liu2011a}%
  \BibitemOpen
  \bibfield  {author} {\bibinfo {author} {\bibfnamefont {J.}~\bibnamefont
  {Liu}}, \bibinfo {author} {\bibfnamefont {W.~H.}\ \bibnamefont {Miller}},
  \bibinfo {author} {\bibfnamefont {G.~S.}\ \bibnamefont {Fanourgakis}},
  \bibinfo {author} {\bibfnamefont {S.~S.}\ \bibnamefont {Xantheas}}, \bibinfo
  {author} {\bibfnamefont {S.}~\bibnamefont {Imoto}},\ and\ \bibinfo {author}
  {\bibfnamefont {S.}~\bibnamefont {Saito}},\ }\href
  {https://doi.org/10.1063/1.3670960} {\bibfield  {journal} {\bibinfo
  {journal} {The Journal of Chemical Physics}\ }\textbf {\bibinfo {volume}
  {135}},\ \bibinfo {pages} {244503} (\bibinfo {year}
  {2011}{\natexlab{a}})}\BibitemShut {NoStop}%
\bibitem [{\citenamefont {Poulsen}\ \emph {et~al.}(2005)\citenamefont
  {Poulsen}, \citenamefont {Nyman},\ and\ \citenamefont
  {Rossky}}]{Poulsen2005b}%
  \BibitemOpen
  \bibfield  {author} {\bibinfo {author} {\bibfnamefont {J.~A.}\ \bibnamefont
  {Poulsen}}, \bibinfo {author} {\bibfnamefont {G.}~\bibnamefont {Nyman}},\
  and\ \bibinfo {author} {\bibfnamefont {P.~J.}\ \bibnamefont {Rossky}},\
  }\href {www.pnas.orgcgidoi10.1073pnas.0408647102} {\bibfield  {journal}
  {\bibinfo  {journal} {Proceedings of the National Academy of Sciences}\
  }\textbf {\bibinfo {volume} {102}},\ \bibinfo {pages} {6709} (\bibinfo {year}
  {2005})}\BibitemShut {NoStop}%
\bibitem [{\citenamefont {Poulsen}\ \emph {et~al.}(2006)\citenamefont
  {Poulsen}, \citenamefont {Nyman},\ and\ \citenamefont
  {Rossky}}]{Poulsen2006b}%
  \BibitemOpen
  \bibfield  {author} {\bibinfo {author} {\bibfnamefont {J.~A.}\ \bibnamefont
  {Poulsen}}, \bibinfo {author} {\bibfnamefont {G.}~\bibnamefont {Nyman}},\
  and\ \bibinfo {author} {\bibfnamefont {P.~J.}\ \bibnamefont {Rossky}},\
  }\href {https://doi.org/10.1021/ct600167s} {\bibfield  {journal} {\bibinfo
  {journal} {Journal of Chemical Theory and Computation}\ }\textbf {\bibinfo
  {volume} {2}},\ \bibinfo {pages} {1482} (\bibinfo {year} {2006})}\BibitemShut
  {NoStop}%
\bibitem [{\citenamefont {Beutier}\ \emph {et~al.}(2015)\citenamefont
  {Beutier}, \citenamefont {Vuilleumier}, \citenamefont {Bonella},\ and\
  \citenamefont {Ciccotti}}]{Beutier2015}%
  \BibitemOpen
  \bibfield  {author} {\bibinfo {author} {\bibfnamefont {J.}~\bibnamefont
  {Beutier}}, \bibinfo {author} {\bibfnamefont {R.}~\bibnamefont
  {Vuilleumier}}, \bibinfo {author} {\bibfnamefont {S.}~\bibnamefont
  {Bonella}},\ and\ \bibinfo {author} {\bibfnamefont {G.}~\bibnamefont
  {Ciccotti}},\ }\href {https://doi.org/10.1080/00268976.2015.1064550}
  {\bibfield  {journal} {\bibinfo  {journal} {Molecular Physics}\ }\textbf
  {\bibinfo {volume} {113}},\ \bibinfo {pages} {2894} (\bibinfo {year}
  {2015})}\BibitemShut {NoStop}%
\bibitem [{\citenamefont {Kaledin}\ and\ \citenamefont
  {Miller}(2003{\natexlab{a}})}]{Kaledin2003c}%
  \BibitemOpen
  \bibfield  {author} {\bibinfo {author} {\bibfnamefont {A.~L.}\ \bibnamefont
  {Kaledin}}\ and\ \bibinfo {author} {\bibfnamefont {W.~H.}\ \bibnamefont
  {Miller}},\ }\href {https://doi.org/10.1063/1.1562158} {\bibfield  {journal}
  {\bibinfo  {journal} {The Journal of Chemical Physics}\ }\textbf {\bibinfo
  {volume} {118}},\ \bibinfo {pages} {7174} (\bibinfo {year}
  {2003}{\natexlab{a}})}\BibitemShut {NoStop}%
\bibitem [{\citenamefont {Kaledin}\ and\ \citenamefont
  {Miller}(2003{\natexlab{b}})}]{Kaledin2003d}%
  \BibitemOpen
  \bibfield  {author} {\bibinfo {author} {\bibfnamefont {A.~L.}\ \bibnamefont
  {Kaledin}}\ and\ \bibinfo {author} {\bibfnamefont {W.~H.}\ \bibnamefont
  {Miller}},\ }\href {https://doi.org/10.1063/1.1589477} {\bibfield  {journal}
  {\bibinfo  {journal} {The Journal of Chemical Physics}\ }\textbf {\bibinfo
  {volume} {119}},\ \bibinfo {pages} {3078} (\bibinfo {year}
  {2003}{\natexlab{b}})}\BibitemShut {NoStop}%
\bibitem [{\citenamefont {Kaledin}\ \emph {et~al.}(2004)\citenamefont
  {Kaledin}, \citenamefont {Huang},\ and\ \citenamefont
  {Bowman}}]{Kaledin2004}%
  \BibitemOpen
  \bibfield  {author} {\bibinfo {author} {\bibfnamefont {A.~L.}\ \bibnamefont
  {Kaledin}}, \bibinfo {author} {\bibfnamefont {X.}~\bibnamefont {Huang}},\
  and\ \bibinfo {author} {\bibfnamefont {J.~M.}\ \bibnamefont {Bowman}},\
  }\href {https://doi.org/10.1016/J.CPLETT.2003.12.013} {\bibfield  {journal}
  {\bibinfo  {journal} {Chemical Physics Letters}\ }\textbf {\bibinfo {volume}
  {384}},\ \bibinfo {pages} {80} (\bibinfo {year} {2004})}\BibitemShut
  {NoStop}%
\bibitem [{\citenamefont {K\"uhn}\ and\ \citenamefont
  {Makri}(1999)}]{Kuhn1999}%
  \BibitemOpen
  \bibfield  {author} {\bibinfo {author} {\bibfnamefont {O.}~\bibnamefont
  {K\"uhn}}\ and\ \bibinfo {author} {\bibfnamefont {N.}~\bibnamefont {Makri}},\
  }\href {https://doi.org/10.1021/jp991836v} {\bibfield  {journal} {\bibinfo
  {journal} {The Journal of Physical Chemistry A}\ }\textbf {\bibinfo {volume}
  {103}},\ \bibinfo {pages} {9487} (\bibinfo {year} {1999})}\BibitemShut
  {NoStop}%
\bibitem [{\citenamefont {Issack}\ and\ \citenamefont
  {Roy}(2007{\natexlab{a}})}]{Issack2007b}%
  \BibitemOpen
  \bibfield  {author} {\bibinfo {author} {\bibfnamefont {B.~B.}\ \bibnamefont
  {Issack}}\ and\ \bibinfo {author} {\bibfnamefont {P.-N.}\ \bibnamefont
  {Roy}},\ }\href {https://doi.org/10.1063/1.2755963} {\bibfield  {journal}
  {\bibinfo  {journal} {The Journal of Chemical Physics}\ }\textbf {\bibinfo
  {volume} {127}},\ \bibinfo {pages} {054105} (\bibinfo {year}
  {2007}{\natexlab{a}})}\BibitemShut {NoStop}%
\bibitem [{\citenamefont {Wong}\ \emph {et~al.}(2011)\citenamefont {Wong},
  \citenamefont {Benoit}, \citenamefont {Lewerenz}, \citenamefont {Brown},\
  and\ \citenamefont {Roy}}]{Wong2011c}%
  \BibitemOpen
  \bibfield  {author} {\bibinfo {author} {\bibfnamefont {S.~Y.~Y.}\
  \bibnamefont {Wong}}, \bibinfo {author} {\bibfnamefont {D.~M.}\ \bibnamefont
  {Benoit}}, \bibinfo {author} {\bibfnamefont {M.}~\bibnamefont {Lewerenz}},
  \bibinfo {author} {\bibfnamefont {A.}~\bibnamefont {Brown}},\ and\ \bibinfo
  {author} {\bibfnamefont {P.-N.}\ \bibnamefont {Roy}},\ }\href
  {https://doi.org/10.1063/1.3553179} {\bibfield  {journal} {\bibinfo
  {journal} {The Journal of Chemical Physics}\ }\textbf {\bibinfo {volume}
  {134}},\ \bibinfo {pages} {094110} (\bibinfo {year} {2011})}\BibitemShut
  {NoStop}%
\bibitem [{\citenamefont {Liberto}\ and\ \citenamefont
  {Ceotto}(2016)}]{Liberto2016b}%
  \BibitemOpen
  \bibfield  {author} {\bibinfo {author} {\bibfnamefont {G.~D.}\ \bibnamefont
  {Liberto}}\ and\ \bibinfo {author} {\bibfnamefont {M.}~\bibnamefont
  {Ceotto}},\ }\href {https://doi.org/10.1063/1.4964308} {\bibfield  {journal}
  {\bibinfo  {journal} {The Journal of Chemical Physics}\ }\textbf {\bibinfo
  {volume} {145}},\ \bibinfo {pages} {144107} (\bibinfo {year}
  {2016})}\BibitemShut {NoStop}%
\bibitem [{\citenamefont {Ceotto}\ \emph
  {et~al.}(2009{\natexlab{a}})\citenamefont {Ceotto}, \citenamefont {Atahan},
  \citenamefont {Shim}, \citenamefont {Tantardini},\ and\ \citenamefont
  {Aspuru-Guzik}}]{Ceotto2009c}%
  \BibitemOpen
  \bibfield  {author} {\bibinfo {author} {\bibfnamefont {M.}~\bibnamefont
  {Ceotto}}, \bibinfo {author} {\bibfnamefont {S.}~\bibnamefont {Atahan}},
  \bibinfo {author} {\bibfnamefont {S.}~\bibnamefont {Shim}}, \bibinfo {author}
  {\bibfnamefont {G.~F.}\ \bibnamefont {Tantardini}},\ and\ \bibinfo {author}
  {\bibfnamefont {A.}~\bibnamefont {Aspuru-Guzik}},\ }\href
  {https://doi.org/10.1039/B820785B} {\bibfield  {journal} {\bibinfo  {journal}
  {Physical Chemistry Chemical Physics}\ }\textbf {\bibinfo {volume} {11}},\
  \bibinfo {pages} {3861} (\bibinfo {year} {2009}{\natexlab{a}})}\BibitemShut
  {NoStop}%
\bibitem [{\citenamefont {Ceotto}\ \emph
  {et~al.}(2009{\natexlab{b}})\citenamefont {Ceotto}, \citenamefont {Atahan},
  \citenamefont {Tantardini},\ and\ \citenamefont
  {Aspuru-Guzik}}]{Ceotto2009a}%
  \BibitemOpen
  \bibfield  {author} {\bibinfo {author} {\bibfnamefont {M.}~\bibnamefont
  {Ceotto}}, \bibinfo {author} {\bibfnamefont {S.}~\bibnamefont {Atahan}},
  \bibinfo {author} {\bibfnamefont {G.~F.}\ \bibnamefont {Tantardini}},\ and\
  \bibinfo {author} {\bibfnamefont {A.}~\bibnamefont {Aspuru-Guzik}},\ }\href
  {https://doi.org/10.1063/1.3155062} {\bibfield  {journal} {\bibinfo
  {journal} {The Journal of Chemical Physics}\ }\textbf {\bibinfo {volume}
  {130}},\ \bibinfo {pages} {234113} (\bibinfo {year}
  {2009}{\natexlab{b}})}\BibitemShut {NoStop}%
\bibitem [{\citenamefont {Ceotto}\ \emph {et~al.}(2010)\citenamefont {Ceotto},
  \citenamefont {Dell'Angelo},\ and\ \citenamefont {Tantardini}}]{Ceotto2010c}%
  \BibitemOpen
  \bibfield  {author} {\bibinfo {author} {\bibfnamefont {M.}~\bibnamefont
  {Ceotto}}, \bibinfo {author} {\bibfnamefont {D.}~\bibnamefont
  {Dell'Angelo}},\ and\ \bibinfo {author} {\bibfnamefont {G.~F.}\ \bibnamefont
  {Tantardini}},\ }\href {https://doi.org/10.1063/1.3462242} {\bibfield
  {journal} {\bibinfo  {journal} {The Journal of Chemical Physics}\ }\textbf
  {\bibinfo {volume} {133}},\ \bibinfo {pages} {054701} (\bibinfo {year}
  {2010})}\BibitemShut {NoStop}%
\bibitem [{\citenamefont {Conte}\ \emph {et~al.}(2013)\citenamefont {Conte},
  \citenamefont {Aspuru-Guzik},\ and\ \citenamefont {Ceotto}}]{Conte2013c}%
  \BibitemOpen
  \bibfield  {author} {\bibinfo {author} {\bibfnamefont {R.}~\bibnamefont
  {Conte}}, \bibinfo {author} {\bibfnamefont {A.}~\bibnamefont
  {Aspuru-Guzik}},\ and\ \bibinfo {author} {\bibfnamefont {M.}~\bibnamefont
  {Ceotto}},\ }\href {https://doi.org/10.1021/JZ401603F} {\bibfield  {journal}
  {\bibinfo  {journal} {Journal of Physical Chemistry Letters}\ }\textbf
  {\bibinfo {volume} {4}},\ \bibinfo {pages} {3407} (\bibinfo {year}
  {2013})}\BibitemShut {NoStop}%
\bibitem [{\citenamefont {Gabas}\ \emph {et~al.}(2017)\citenamefont {Gabas},
  \citenamefont {Conte},\ and\ \citenamefont {Ceotto}}]{Gabas2017}%
  \BibitemOpen
  \bibfield  {author} {\bibinfo {author} {\bibfnamefont {F.}~\bibnamefont
  {Gabas}}, \bibinfo {author} {\bibfnamefont {R.}~\bibnamefont {Conte}},\ and\
  \bibinfo {author} {\bibfnamefont {M.}~\bibnamefont {Ceotto}},\ }\href
  {https://doi.org/10.1021/ACS.JCTC.6B01018} {\bibfield  {journal} {\bibinfo
  {journal} {Journal of Chemical Theory and Computation}\ }\textbf {\bibinfo
  {volume} {13}},\ \bibinfo {pages} {2378} (\bibinfo {year}
  {2017})}\BibitemShut {NoStop}%
\bibitem [{\citenamefont {Ceotto}\ \emph {et~al.}(2011)\citenamefont {Ceotto},
  \citenamefont {Valleau}, \citenamefont {Tantardini},\ and\ \citenamefont
  {Aspuru-Guzik}}]{Ceotto2011}%
  \BibitemOpen
  \bibfield  {author} {\bibinfo {author} {\bibfnamefont {M.}~\bibnamefont
  {Ceotto}}, \bibinfo {author} {\bibfnamefont {S.}~\bibnamefont {Valleau}},
  \bibinfo {author} {\bibfnamefont {G.~F.}\ \bibnamefont {Tantardini}},\ and\
  \bibinfo {author} {\bibfnamefont {A.}~\bibnamefont {Aspuru-Guzik}},\ }\href
  {https://doi.org/10.1063/1.3599469} {\bibfield  {journal} {\bibinfo
  {journal} {The Journal of Chemical Physics}\ }\textbf {\bibinfo {volume}
  {134}},\ \bibinfo {pages} {234103} (\bibinfo {year} {2011})}\BibitemShut
  {NoStop}%
\bibitem [{\citenamefont {Aieta}\ \emph
  {et~al.}(2020{\natexlab{a}})\citenamefont {Aieta}, \citenamefont {Bertaina},
  \citenamefont {Micciarelli},\ and\ \citenamefont {Ceotto}}]{Aieta2020}%
  \BibitemOpen
  \bibfield  {author} {\bibinfo {author} {\bibfnamefont {C.}~\bibnamefont
  {Aieta}}, \bibinfo {author} {\bibfnamefont {G.}~\bibnamefont {Bertaina}},
  \bibinfo {author} {\bibfnamefont {M.}~\bibnamefont {Micciarelli}},\ and\
  \bibinfo {author} {\bibfnamefont {M.}~\bibnamefont {Ceotto}},\ }\href
  {https://doi.org/10.1063/5.0031391} {\bibfield  {journal} {\bibinfo
  {journal} {The Journal of Chemical Physics}\ }\textbf {\bibinfo {volume}
  {153}},\ \bibinfo {pages} {214117} (\bibinfo {year}
  {2020}{\natexlab{a}})}\BibitemShut {NoStop}%
\bibitem [{\citenamefont {Aieta}\ \emph
  {et~al.}(2020{\natexlab{b}})\citenamefont {Aieta}, \citenamefont
  {Micciarelli}, \citenamefont {Bertaina},\ and\ \citenamefont
  {Ceotto}}]{Aieta2020a}%
  \BibitemOpen
  \bibfield  {author} {\bibinfo {author} {\bibfnamefont {C.}~\bibnamefont
  {Aieta}}, \bibinfo {author} {\bibfnamefont {M.}~\bibnamefont {Micciarelli}},
  \bibinfo {author} {\bibfnamefont {G.}~\bibnamefont {Bertaina}},\ and\
  \bibinfo {author} {\bibfnamefont {M.}~\bibnamefont {Ceotto}},\ }\href
  {https://doi.org/10.1038/s41467-020-18211-3} {\bibfield  {journal} {\bibinfo
  {journal} {Nature Communications}\ }\textbf {\bibinfo {volume} {11}},\
  \bibinfo {pages} {1} (\bibinfo {year} {2020}{\natexlab{b}})}\BibitemShut
  {NoStop}%
\bibitem [{\citenamefont {Micciarelli}\ \emph {et~al.}(2019)\citenamefont
  {Micciarelli}, \citenamefont {Gabas}, \citenamefont {Conte},\ and\
  \citenamefont {Ceotto}}]{Micciarelli2019}%
  \BibitemOpen
  \bibfield  {author} {\bibinfo {author} {\bibfnamefont {M.}~\bibnamefont
  {Micciarelli}}, \bibinfo {author} {\bibfnamefont {F.}~\bibnamefont {Gabas}},
  \bibinfo {author} {\bibfnamefont {R.}~\bibnamefont {Conte}},\ and\ \bibinfo
  {author} {\bibfnamefont {M.}~\bibnamefont {Ceotto}},\ }\href
  {https://doi.org/10.1063/1.5096968} {\bibfield  {journal} {\bibinfo
  {journal} {The Journal of Chemical Physics}\ }\textbf {\bibinfo {volume}
  {150}},\ \bibinfo {pages} {184113} (\bibinfo {year} {2019})}\BibitemShut
  {NoStop}%
\bibitem [{\citenamefont {Ceotto}\ \emph {et~al.}(2017)\citenamefont {Ceotto},
  \citenamefont {Liberto},\ and\ \citenamefont {Conte}}]{Ceotto2017c}%
  \BibitemOpen
  \bibfield  {author} {\bibinfo {author} {\bibfnamefont {M.}~\bibnamefont
  {Ceotto}}, \bibinfo {author} {\bibfnamefont {G.~D.}\ \bibnamefont
  {Liberto}},\ and\ \bibinfo {author} {\bibfnamefont {R.}~\bibnamefont
  {Conte}},\ }\href {https://doi.org/10.1103/PhysRevLett.119.010401} {\bibfield
   {journal} {\bibinfo  {journal} {Physical Review Letters}\ }\textbf {\bibinfo
  {volume} {119}},\ \bibinfo {pages} {010401} (\bibinfo {year}
  {2017})}\BibitemShut {NoStop}%
\bibitem [{\citenamefont {Liberto}\ \emph
  {et~al.}(2018{\natexlab{a}})\citenamefont {Liberto}, \citenamefont {Conte},\
  and\ \citenamefont {Ceotto}}]{Liberto2018}%
  \BibitemOpen
  \bibfield  {author} {\bibinfo {author} {\bibfnamefont {G.~D.}\ \bibnamefont
  {Liberto}}, \bibinfo {author} {\bibfnamefont {R.}~\bibnamefont {Conte}},\
  and\ \bibinfo {author} {\bibfnamefont {M.}~\bibnamefont {Ceotto}},\ }\href
  {https://doi.org/10.1063/1.5010388} {\bibfield  {journal} {\bibinfo
  {journal} {The Journal of Chemical Physics}\ }\textbf {\bibinfo {volume}
  {148}},\ \bibinfo {pages} {014307} (\bibinfo {year}
  {2018}{\natexlab{a}})}\BibitemShut {NoStop}%
\bibitem [{\citenamefont {Liberto}\ \emph
  {et~al.}(2018{\natexlab{b}})\citenamefont {Liberto}, \citenamefont {Conte},\
  and\ \citenamefont {Ceotto}}]{Liberto2018a}%
  \BibitemOpen
  \bibfield  {author} {\bibinfo {author} {\bibfnamefont {G.~D.}\ \bibnamefont
  {Liberto}}, \bibinfo {author} {\bibfnamefont {R.}~\bibnamefont {Conte}},\
  and\ \bibinfo {author} {\bibfnamefont {M.}~\bibnamefont {Ceotto}},\ }\href
  {https://doi.org/10.1063/1.5023155} {\bibfield  {journal} {\bibinfo
  {journal} {The Journal of Chemical Physics}\ }\textbf {\bibinfo {volume}
  {148}},\ \bibinfo {pages} {104302} (\bibinfo {year}
  {2018}{\natexlab{b}})}\BibitemShut {NoStop}%
\bibitem [{\citenamefont {Gabas}\ \emph {et~al.}(2018)\citenamefont {Gabas},
  \citenamefont {Liberto}, \citenamefont {Conte},\ and\ \citenamefont
  {Ceotto}}]{Gabas2018}%
  \BibitemOpen
  \bibfield  {author} {\bibinfo {author} {\bibfnamefont {F.}~\bibnamefont
  {Gabas}}, \bibinfo {author} {\bibfnamefont {G.~D.}\ \bibnamefont {Liberto}},
  \bibinfo {author} {\bibfnamefont {R.}~\bibnamefont {Conte}},\ and\ \bibinfo
  {author} {\bibfnamefont {M.}~\bibnamefont {Ceotto}},\ }\href
  {https://doi.org/10.1039/C8SC03041C} {\bibfield  {journal} {\bibinfo
  {journal} {Chemical Science}\ }\textbf {\bibinfo {volume} {9}},\ \bibinfo
  {pages} {7894} (\bibinfo {year} {2018})}\BibitemShut {NoStop}%
\bibitem [{\citenamefont {Gabas}\ \emph {et~al.}(2019)\citenamefont {Gabas},
  \citenamefont {Liberto},\ and\ \citenamefont {Ceotto}}]{Gabas2019}%
  \BibitemOpen
  \bibfield  {author} {\bibinfo {author} {\bibfnamefont {F.}~\bibnamefont
  {Gabas}}, \bibinfo {author} {\bibfnamefont {G.~D.}\ \bibnamefont {Liberto}},\
  and\ \bibinfo {author} {\bibfnamefont {M.}~\bibnamefont {Ceotto}},\ }\href
  {https://doi.org/10.1063/1.5100503} {\bibfield  {journal} {\bibinfo
  {journal} {The Journal of Chemical Physics}\ }\textbf {\bibinfo {volume}
  {150}},\ \bibinfo {pages} {224107} (\bibinfo {year} {2019})}\BibitemShut
  {NoStop}%
\bibitem [{\citenamefont {Bertaina}\ \emph {et~al.}(2019)\citenamefont
  {Bertaina}, \citenamefont {Liberto},\ and\ \citenamefont
  {Ceotto}}]{Bertaina2019c}%
  \BibitemOpen
  \bibfield  {author} {\bibinfo {author} {\bibfnamefont {G.}~\bibnamefont
  {Bertaina}}, \bibinfo {author} {\bibfnamefont {G.~D.}\ \bibnamefont
  {Liberto}},\ and\ \bibinfo {author} {\bibfnamefont {M.}~\bibnamefont
  {Ceotto}},\ }\href {https://doi.org/10.1063/1.5114616} {\bibfield  {journal}
  {\bibinfo  {journal} {The Journal of Chemical Physics}\ }\textbf {\bibinfo
  {volume} {151}},\ \bibinfo {pages} {114307} (\bibinfo {year}
  {2019})}\BibitemShut {NoStop}%
\bibitem [{\citenamefont {Gandolfi}\ \emph {et~al.}(2020)\citenamefont
  {Gandolfi}, \citenamefont {Rognoni}, \citenamefont {Aieta}, \citenamefont
  {Conte},\ and\ \citenamefont {Ceotto}}]{Gandolfi2020c}%
  \BibitemOpen
  \bibfield  {author} {\bibinfo {author} {\bibfnamefont {M.}~\bibnamefont
  {Gandolfi}}, \bibinfo {author} {\bibfnamefont {A.}~\bibnamefont {Rognoni}},
  \bibinfo {author} {\bibfnamefont {C.}~\bibnamefont {Aieta}}, \bibinfo
  {author} {\bibfnamefont {R.}~\bibnamefont {Conte}},\ and\ \bibinfo {author}
  {\bibfnamefont {M.}~\bibnamefont {Ceotto}},\ }\href
  {https://doi.org/10.1063/5.0031892} {\bibfield  {journal} {\bibinfo
  {journal} {The Journal of Chemical Physics}\ }\textbf {\bibinfo {volume}
  {153}},\ \bibinfo {pages} {204104} (\bibinfo {year} {2020})}\BibitemShut
  {NoStop}%
\bibitem [{\citenamefont {Ovchinnikov}\ and\ \citenamefont
  {Apkarian}(1996)}]{Ovchinnikov1996e}%
  \BibitemOpen
  \bibfield  {author} {\bibinfo {author} {\bibfnamefont {M.}~\bibnamefont
  {Ovchinnikov}}\ and\ \bibinfo {author} {\bibfnamefont {V.~A.}\ \bibnamefont
  {Apkarian}},\ }\href {https://doi.org/10.1063/1.472959} {\bibfield  {journal}
  {\bibinfo  {journal} {Journal of Chemical Physics}\ }\textbf {\bibinfo
  {volume} {105}},\ \bibinfo {pages} {10312} (\bibinfo {year}
  {1996})}\BibitemShut {NoStop}%
\bibitem [{\citenamefont {Ovchinnikov}\ and\ \citenamefont
  {Apkarian}(1998)}]{Ovchinnikov1998g}%
  \BibitemOpen
  \bibfield  {author} {\bibinfo {author} {\bibfnamefont {M.}~\bibnamefont
  {Ovchinnikov}}\ and\ \bibinfo {author} {\bibfnamefont {V.~A.}\ \bibnamefont
  {Apkarian}},\ }\href {https://doi.org/10.1063/1.473596} {\bibfield  {journal}
  {\bibinfo  {journal} {The Journal of Chemical Physics}\ }\textbf {\bibinfo
  {volume} {106}},\ \bibinfo {pages} {5775} (\bibinfo {year}
  {1998})}\BibitemShut {NoStop}%
\bibitem [{\citenamefont {Grossmann}(2016)}]{Grossmann2016}%
  \BibitemOpen
  \bibfield  {author} {\bibinfo {author} {\bibfnamefont {F.}~\bibnamefont
  {Grossmann}},\ }\href {https://doi.org/10.1088/0031-8949/91/4/044004}
  {\bibfield  {journal} {\bibinfo  {journal} {Physica Scripta}\ }\textbf
  {\bibinfo {volume} {91}},\ \bibinfo {pages} {044004} (\bibinfo {year}
  {2016})}\BibitemShut {NoStop}%
\bibitem [{\citenamefont {Buchholz}\ \emph {et~al.}(2012)\citenamefont
  {Buchholz}, \citenamefont {Goletz}, \citenamefont {Grossmann}, \citenamefont
  {Schmidt}, \citenamefont {Heyda},\ and\ \citenamefont
  {Jungwirth}}]{Buchholz2012}%
  \BibitemOpen
  \bibfield  {author} {\bibinfo {author} {\bibfnamefont {M.}~\bibnamefont
  {Buchholz}}, \bibinfo {author} {\bibfnamefont {C.-M.}\ \bibnamefont
  {Goletz}}, \bibinfo {author} {\bibfnamefont {F.}~\bibnamefont {Grossmann}},
  \bibinfo {author} {\bibfnamefont {B.}~\bibnamefont {Schmidt}}, \bibinfo
  {author} {\bibfnamefont {J.}~\bibnamefont {Heyda}},\ and\ \bibinfo {author}
  {\bibfnamefont {P.}~\bibnamefont {Jungwirth}},\ }\href
  {https://doi.org/10.1021/JP305084F} {\bibfield  {journal} {\bibinfo
  {journal} {Journal of Physical Chemistry A}\ }\textbf {\bibinfo {volume}
  {116}},\ \bibinfo {pages} {11199} (\bibinfo {year} {2012})}\BibitemShut
  {NoStop}%
\bibitem [{\citenamefont {Buchholz}\ \emph {et~al.}(2016)\citenamefont
  {Buchholz}, \citenamefont {Grossmann},\ and\ \citenamefont
  {Ceotto}}]{Buchholz2016}%
  \BibitemOpen
  \bibfield  {author} {\bibinfo {author} {\bibfnamefont {M.}~\bibnamefont
  {Buchholz}}, \bibinfo {author} {\bibfnamefont {F.}~\bibnamefont
  {Grossmann}},\ and\ \bibinfo {author} {\bibfnamefont {M.}~\bibnamefont
  {Ceotto}},\ }\href {https://doi.org/10.1063/1.4942536} {\bibfield  {journal}
  {\bibinfo  {journal} {The Journal of Chemical Physics}\ }\textbf {\bibinfo
  {volume} {144}},\ \bibinfo {pages} {094102} (\bibinfo {year}
  {2016})}\BibitemShut {NoStop}%
\bibitem [{\citenamefont {Buchholz}\ \emph {et~al.}(2017)\citenamefont
  {Buchholz}, \citenamefont {Grossmann},\ and\ \citenamefont
  {Ceotto}}]{Buchholz2017}%
  \BibitemOpen
  \bibfield  {author} {\bibinfo {author} {\bibfnamefont {M.}~\bibnamefont
  {Buchholz}}, \bibinfo {author} {\bibfnamefont {F.}~\bibnamefont
  {Grossmann}},\ and\ \bibinfo {author} {\bibfnamefont {M.}~\bibnamefont
  {Ceotto}},\ }\href {https://doi.org/10.1063/1.4998510} {\bibfield  {journal}
  {\bibinfo  {journal} {The Journal of Chemical Physics}\ }\textbf {\bibinfo
  {volume} {147}},\ \bibinfo {pages} {164110} (\bibinfo {year}
  {2017})}\BibitemShut {NoStop}%
\bibitem [{\citenamefont {Buchholz}\ \emph {et~al.}(2018)\citenamefont
  {Buchholz}, \citenamefont {Grossmann},\ and\ \citenamefont
  {Ceotto}}]{Buchholz2018}%
  \BibitemOpen
  \bibfield  {author} {\bibinfo {author} {\bibfnamefont {M.}~\bibnamefont
  {Buchholz}}, \bibinfo {author} {\bibfnamefont {F.}~\bibnamefont
  {Grossmann}},\ and\ \bibinfo {author} {\bibfnamefont {M.}~\bibnamefont
  {Ceotto}},\ }\href {https://doi.org/10.1063/1.5020144} {\bibfield  {journal}
  {\bibinfo  {journal} {The Journal of Chemical Physics}\ }\textbf {\bibinfo
  {volume} {148}},\ \bibinfo {pages} {114107} (\bibinfo {year}
  {2018})}\BibitemShut {NoStop}%
\bibitem [{\citenamefont {Sun}\ and\ \citenamefont
  {Miller}(1997{\natexlab{a}})}]{Sun1997}%
  \BibitemOpen
  \bibfield  {author} {\bibinfo {author} {\bibfnamefont {X.}~\bibnamefont
  {Sun}}\ and\ \bibinfo {author} {\bibfnamefont {W.~H.}\ \bibnamefont
  {Miller}},\ }\href {https://doi.org/10.1063/1.473624} {\bibfield  {journal}
  {\bibinfo  {journal} {The Journal of Chemical Physics}\ }\textbf {\bibinfo
  {volume} {106}},\ \bibinfo {pages} {6346} (\bibinfo {year}
  {1997}{\natexlab{a}})}\BibitemShut {NoStop}%
\bibitem [{\citenamefont {Sun}\ \emph {et~al.}(1998{\natexlab{b}})\citenamefont
  {Sun}, \citenamefont {Wang},\ and\ \citenamefont {Miller}}]{Sun1998b}%
  \BibitemOpen
  \bibfield  {author} {\bibinfo {author} {\bibfnamefont {X.}~\bibnamefont
  {Sun}}, \bibinfo {author} {\bibfnamefont {H.}~\bibnamefont {Wang}},\ and\
  \bibinfo {author} {\bibfnamefont {W.~H.}\ \bibnamefont {Miller}},\ }\href
  {https://doi.org/10.1063/1.477389} {\bibfield  {journal} {\bibinfo  {journal}
  {The Journal of Chemical Physics}\ }\textbf {\bibinfo {volume} {109}},\
  \bibinfo {pages} {7064} (\bibinfo {year} {1998}{\natexlab{b}})}\BibitemShut
  {NoStop}%
\bibitem [{\citenamefont {Batista}\ and\ \citenamefont
  {Miller}(1998)}]{Batista1998}%
  \BibitemOpen
  \bibfield  {author} {\bibinfo {author} {\bibfnamefont {V.~S.}\ \bibnamefont
  {Batista}}\ and\ \bibinfo {author} {\bibfnamefont {W.~H.}\ \bibnamefont
  {Miller}},\ }\href {https://doi.org/10.1063/1.475413} {\bibfield  {journal}
  {\bibinfo  {journal} {The Journal of Chemical Physics}\ }\textbf {\bibinfo
  {volume} {108}},\ \bibinfo {pages} {498} (\bibinfo {year}
  {1998})}\BibitemShut {NoStop}%
\bibitem [{\citenamefont {Rabani}\ \emph {et~al.}(1999)\citenamefont {Rabani},
  \citenamefont {Egorov},\ and\ \citenamefont {Berne}}]{Rabani1999}%
  \BibitemOpen
  \bibfield  {author} {\bibinfo {author} {\bibfnamefont {E.}~\bibnamefont
  {Rabani}}, \bibinfo {author} {\bibfnamefont {S.~A.}\ \bibnamefont {Egorov}},\
  and\ \bibinfo {author} {\bibfnamefont {B.~J.}\ \bibnamefont {Berne}},\ }\href
  {https://doi.org/10.1021/jp992189a} {\bibfield  {journal} {\bibinfo
  {journal} {The Journal of Physical Chemistry A}\ }\textbf {\bibinfo {volume}
  {103}},\ \bibinfo {pages} {9539} (\bibinfo {year} {1999})}\BibitemShut
  {NoStop}%
\bibitem [{\citenamefont {Wang}\ \emph {et~al.}(1999)\citenamefont {Wang},
  \citenamefont {Song}, \citenamefont {Chandler},\ and\ \citenamefont
  {Miller}}]{Wang1999}%
  \BibitemOpen
  \bibfield  {author} {\bibinfo {author} {\bibfnamefont {H.}~\bibnamefont
  {Wang}}, \bibinfo {author} {\bibfnamefont {X.}~\bibnamefont {Song}}, \bibinfo
  {author} {\bibfnamefont {D.}~\bibnamefont {Chandler}},\ and\ \bibinfo
  {author} {\bibfnamefont {W.~H.}\ \bibnamefont {Miller}},\ }\href
  {https://doi.org/10.1063/1.478388} {\bibfield  {journal} {\bibinfo  {journal}
  {The Journal of Chemical Physics}\ }\textbf {\bibinfo {volume} {110}},\
  \bibinfo {pages} {4828} (\bibinfo {year} {1999})}\BibitemShut {NoStop}%
\bibitem [{\citenamefont {Thoss}\ \emph {et~al.}(2000)\citenamefont {Thoss},
  \citenamefont {Miller},\ and\ \citenamefont {Stock}}]{Thoss2000}%
  \BibitemOpen
  \bibfield  {author} {\bibinfo {author} {\bibfnamefont {M.}~\bibnamefont
  {Thoss}}, \bibinfo {author} {\bibfnamefont {W.~H.}\ \bibnamefont {Miller}},\
  and\ \bibinfo {author} {\bibfnamefont {G.}~\bibnamefont {Stock}},\ }\href
  {https://doi.org/10.1063/1.481668} {\bibfield  {journal} {\bibinfo  {journal}
  {The Journal of Chemical Physics}\ }\textbf {\bibinfo {volume} {112}},\
  \bibinfo {pages} {10282} (\bibinfo {year} {2000})}\BibitemShut {NoStop}%
\bibitem [{\citenamefont {Shi}\ and\ \citenamefont {Geva}(2004)}]{Shi2004}%
  \BibitemOpen
  \bibfield  {author} {\bibinfo {author} {\bibfnamefont {Q.}~\bibnamefont
  {Shi}}\ and\ \bibinfo {author} {\bibfnamefont {E.}~\bibnamefont {Geva}},\
  }\href {https://doi.org/10.1021/JP049547G} {\bibfield  {journal} {\bibinfo
  {journal} {Journal of Physical Chemistry A}\ }\textbf {\bibinfo {volume}
  {108}},\ \bibinfo {pages} {6109} (\bibinfo {year} {2004})}\BibitemShut
  {NoStop}%
\bibitem [{\citenamefont {Shi}\ and\ \citenamefont {Geva}(2005)}]{Shi2005}%
  \BibitemOpen
  \bibfield  {author} {\bibinfo {author} {\bibfnamefont {Q.}~\bibnamefont
  {Shi}}\ and\ \bibinfo {author} {\bibfnamefont {E.}~\bibnamefont {Geva}},\
  }\href {https://doi.org/10.1063/1.1843813} {\bibfield  {journal} {\bibinfo
  {journal} {The Journal of Chemical Physics}\ }\textbf {\bibinfo {volume}
  {122}},\ \bibinfo {pages} {064506} (\bibinfo {year} {2005})}\BibitemShut
  {NoStop}%
\bibitem [{\citenamefont {Bonella}\ \emph {et~al.}(2005)\citenamefont
  {Bonella}, \citenamefont {Montemayor},\ and\ \citenamefont
  {Coker}}]{Bonella2005}%
  \BibitemOpen
  \bibfield  {author} {\bibinfo {author} {\bibfnamefont {S.}~\bibnamefont
  {Bonella}}, \bibinfo {author} {\bibfnamefont {D.}~\bibnamefont
  {Montemayor}},\ and\ \bibinfo {author} {\bibfnamefont {D.~F.}\ \bibnamefont
  {Coker}},\ }\href {https://doi.org/10.1073/PNAS.0408326102} {\bibfield
  {journal} {\bibinfo  {journal} {Proceedings of the National Academy of
  Sciences of the United States of America}\ }\textbf {\bibinfo {volume}
  {102}},\ \bibinfo {pages} {6715} (\bibinfo {year} {2005})}\BibitemShut
  {NoStop}%
\bibitem [{\citenamefont {Ananth}\ \emph {et~al.}(2007)\citenamefont {Ananth},
  \citenamefont {Venkataraman},\ and\ \citenamefont {Miller}}]{Ananth2007}%
  \BibitemOpen
  \bibfield  {author} {\bibinfo {author} {\bibfnamefont {N.}~\bibnamefont
  {Ananth}}, \bibinfo {author} {\bibfnamefont {C.}~\bibnamefont
  {Venkataraman}},\ and\ \bibinfo {author} {\bibfnamefont {W.~H.}\ \bibnamefont
  {Miller}},\ }\href {https://doi.org/10.1063/1.2759932} {\bibfield  {journal}
  {\bibinfo  {journal} {The Journal of Chemical Physics}\ }\textbf {\bibinfo
  {volume} {127}},\ \bibinfo {pages} {084114} (\bibinfo {year}
  {2007})}\BibitemShut {NoStop}%
\bibitem [{\citenamefont {Shi}\ and\ \citenamefont {Geva}(2008)}]{Shi2008}%
  \BibitemOpen
  \bibfield  {author} {\bibinfo {author} {\bibfnamefont {Q.}~\bibnamefont
  {Shi}}\ and\ \bibinfo {author} {\bibfnamefont {E.}~\bibnamefont {Geva}},\
  }\href {https://doi.org/10.1063/1.2981566} {\bibfield  {journal} {\bibinfo
  {journal} {The Journal of Chemical Physics}\ }\textbf {\bibinfo {volume}
  {129}},\ \bibinfo {pages} {124505} (\bibinfo {year} {2008})}\BibitemShut
  {NoStop}%
\bibitem [{\citenamefont {Miller}\ and\ \citenamefont
  {Tao}(2010)}]{Miller2010}%
  \BibitemOpen
  \bibfield  {author} {\bibinfo {author} {\bibfnamefont {W.~H.}\ \bibnamefont
  {Miller}}\ and\ \bibinfo {author} {\bibfnamefont {G.}~\bibnamefont {Tao}},\
  }\href
  {https://doi.org/10.1021/JZ1000825/ASSET/IMAGES/LARGE/JZ-2010-000825_0003.JPEG}
  {\bibfield  {journal} {\bibinfo  {journal} {Journal of Physical Chemistry
  Letters}\ }\textbf {\bibinfo {volume} {1}},\ \bibinfo {pages} {891} (\bibinfo
  {year} {2010})}\BibitemShut {NoStop}%
\bibitem [{\citenamefont {Venkataraman}(2011)}]{Venkataraman2011}%
  \BibitemOpen
  \bibfield  {author} {\bibinfo {author} {\bibfnamefont {C.}~\bibnamefont
  {Venkataraman}},\ }\href {https://doi.org/10.1063/1.3662095} {\bibfield
  {journal} {\bibinfo  {journal} {The Journal of Chemical Physics}\ }\textbf
  {\bibinfo {volume} {135}},\ \bibinfo {pages} {204503} (\bibinfo {year}
  {2011})}\BibitemShut {NoStop}%
\bibitem [{\citenamefont {Huo}\ and\ \citenamefont {Coker}(2011)}]{Huo2011}%
  \BibitemOpen
  \bibfield  {author} {\bibinfo {author} {\bibfnamefont {P.}~\bibnamefont
  {Huo}}\ and\ \bibinfo {author} {\bibfnamefont {D.~F.}\ \bibnamefont
  {Coker}},\ }\href {https://doi.org/10.1063/1.3664763} {\bibfield  {journal}
  {\bibinfo  {journal} {The Journal of Chemical Physics}\ }\textbf {\bibinfo
  {volume} {135}},\ \bibinfo {pages} {201101} (\bibinfo {year}
  {2011})}\BibitemShut {NoStop}%
\bibitem [{\citenamefont {Huo}\ and\ \citenamefont {Coker}(2012)}]{Huo2012}%
  \BibitemOpen
  \bibfield  {author} {\bibinfo {author} {\bibfnamefont {P.}~\bibnamefont
  {Huo}}\ and\ \bibinfo {author} {\bibfnamefont {D.~F.}\ \bibnamefont
  {Coker}},\ }\href {https://doi.org/10.1063/1.4748316} {\bibfield  {journal}
  {\bibinfo  {journal} {The Journal of Chemical Physics}\ }\textbf {\bibinfo
  {volume} {137}},\ \bibinfo {pages} {22A535} (\bibinfo {year}
  {2012})}\BibitemShut {NoStop}%
\bibitem [{\citenamefont {Cotton}\ and\ \citenamefont
  {Miller}(2013)}]{Cotton2013}%
  \BibitemOpen
  \bibfield  {author} {\bibinfo {author} {\bibfnamefont {S.~J.}\ \bibnamefont
  {Cotton}}\ and\ \bibinfo {author} {\bibfnamefont {W.~H.}\ \bibnamefont
  {Miller}},\ }\href {https://doi.org/10.1063/1.4845235} {\bibfield  {journal}
  {\bibinfo  {journal} {The Journal of Chemical Physics}\ }\textbf {\bibinfo
  {volume} {139}},\ \bibinfo {pages} {234112} (\bibinfo {year}
  {2013})}\BibitemShut {NoStop}%
\bibitem [{\citenamefont {Tao}(2013)}]{Tao2013b}%
  \BibitemOpen
  \bibfield  {author} {\bibinfo {author} {\bibfnamefont {G.}~\bibnamefont
  {Tao}},\ }\href
  {https://doi.org/10.1021/JP404856P/SUPPL_FILE/JP404856P_SI_001.PDF}
  {\bibfield  {journal} {\bibinfo  {journal} {Journal of Physical Chemistry A}\
  }\textbf {\bibinfo {volume} {117}},\ \bibinfo {pages} {5821} (\bibinfo {year}
  {2013})}\BibitemShut {NoStop}%
\bibitem [{\citenamefont {Sun}\ and\ \citenamefont {Geva}(2015)}]{Sun2015}%
  \BibitemOpen
  \bibfield  {author} {\bibinfo {author} {\bibfnamefont {X.}~\bibnamefont
  {Sun}}\ and\ \bibinfo {author} {\bibfnamefont {E.}~\bibnamefont {Geva}},\
  }\href {https://doi.org/10.1021/ACS.JPCA.5B08280} {\bibfield  {journal}
  {\bibinfo  {journal} {Journal of Physical Chemistry A}\ }\textbf {\bibinfo
  {volume} {120}},\ \bibinfo {pages} {2976} (\bibinfo {year}
  {2015})}\BibitemShut {NoStop}%
\bibitem [{\citenamefont {Sun}\ and\ \citenamefont
  {Geva}(2016{\natexlab{a}})}]{Sun2016}%
  \BibitemOpen
  \bibfield  {author} {\bibinfo {author} {\bibfnamefont {X.}~\bibnamefont
  {Sun}}\ and\ \bibinfo {author} {\bibfnamefont {E.}~\bibnamefont {Geva}},\
  }\href {https://doi.org/10.1021/ACS.JCTC.6B00236} {\bibfield  {journal}
  {\bibinfo  {journal} {Journal of Chemical Theory and Computation}\ }\textbf
  {\bibinfo {volume} {12}},\ \bibinfo {pages} {2926} (\bibinfo {year}
  {2016}{\natexlab{a}})}\BibitemShut {NoStop}%
\bibitem [{\citenamefont {Sun}\ and\ \citenamefont
  {Geva}(2016{\natexlab{b}})}]{Sun2016b}%
  \BibitemOpen
  \bibfield  {author} {\bibinfo {author} {\bibfnamefont {X.}~\bibnamefont
  {Sun}}\ and\ \bibinfo {author} {\bibfnamefont {E.}~\bibnamefont {Geva}},\
  }\href {https://doi.org/10.1063/1.4954509} {\bibfield  {journal} {\bibinfo
  {journal} {The Journal of Chemical Physics}\ }\textbf {\bibinfo {volume}
  {144}},\ \bibinfo {pages} {244105} (\bibinfo {year}
  {2016}{\natexlab{b}})}\BibitemShut {NoStop}%
\bibitem [{\citenamefont {Sun}\ and\ \citenamefont
  {Geva}(2016{\natexlab{c}})}]{Sun2016c}%
  \BibitemOpen
  \bibfield  {author} {\bibinfo {author} {\bibfnamefont {X.}~\bibnamefont
  {Sun}}\ and\ \bibinfo {author} {\bibfnamefont {E.}~\bibnamefont {Geva}},\
  }\href {https://doi.org/10.1063/1.4960337} {\bibfield  {journal} {\bibinfo
  {journal} {The Journal of Chemical Physics}\ }\textbf {\bibinfo {volume}
  {145}},\ \bibinfo {pages} {064109} (\bibinfo {year}
  {2016}{\natexlab{c}})}\BibitemShut {NoStop}%
\bibitem [{\citenamefont {Teh}\ and\ \citenamefont {Cheng}(2017)}]{Teh2017}%
  \BibitemOpen
  \bibfield  {author} {\bibinfo {author} {\bibfnamefont {H.-H.}\ \bibnamefont
  {Teh}}\ and\ \bibinfo {author} {\bibfnamefont {Y.-C.}\ \bibnamefont
  {Cheng}},\ }\href {https://doi.org/10.1063/1.4979894} {\bibfield  {journal}
  {\bibinfo  {journal} {The Journal of Chemical Physics}\ }\textbf {\bibinfo
  {volume} {146}},\ \bibinfo {pages} {144105} (\bibinfo {year}
  {2017})}\BibitemShut {NoStop}%
\bibitem [{\citenamefont {Kananenka}\ \emph {et~al.}(2017)\citenamefont
  {Kananenka}, \citenamefont {Sun}, \citenamefont {Schubert}, \citenamefont
  {Dunietz},\ and\ \citenamefont {Geva}}]{Kananenka2017}%
  \BibitemOpen
  \bibfield  {author} {\bibinfo {author} {\bibfnamefont {A.~A.}\ \bibnamefont
  {Kananenka}}, \bibinfo {author} {\bibfnamefont {X.}~\bibnamefont {Sun}},
  \bibinfo {author} {\bibfnamefont {A.}~\bibnamefont {Schubert}}, \bibinfo
  {author} {\bibfnamefont {B.~D.}\ \bibnamefont {Dunietz}},\ and\ \bibinfo
  {author} {\bibfnamefont {E.}~\bibnamefont {Geva}},\ }\href
  {https://doi.org/10.1063/1.4989509} {\bibfield  {journal} {\bibinfo
  {journal} {The Journal of Chemical Physics}\ }\textbf {\bibinfo {volume}
  {148}},\ \bibinfo {pages} {102304} (\bibinfo {year} {2017})}\BibitemShut
  {NoStop}%
\bibitem [{\citenamefont {Sun}\ \emph {et~al.}(2018)\citenamefont {Sun},
  \citenamefont {Zhang}, \citenamefont {Lai}, \citenamefont {Williams},
  \citenamefont {Cheung}, \citenamefont {Dunietz},\ and\ \citenamefont
  {Geva}}]{Sun2018}%
  \BibitemOpen
  \bibfield  {author} {\bibinfo {author} {\bibfnamefont {X.}~\bibnamefont
  {Sun}}, \bibinfo {author} {\bibfnamefont {P.}~\bibnamefont {Zhang}}, \bibinfo
  {author} {\bibfnamefont {Y.}~\bibnamefont {Lai}}, \bibinfo {author}
  {\bibfnamefont {K.~L.}\ \bibnamefont {Williams}}, \bibinfo {author}
  {\bibfnamefont {M.~S.}\ \bibnamefont {Cheung}}, \bibinfo {author}
  {\bibfnamefont {B.~D.}\ \bibnamefont {Dunietz}},\ and\ \bibinfo {author}
  {\bibfnamefont {E.}~\bibnamefont {Geva}},\ }\href
  {https://doi.org/10.1021/ACS.JPCC.8B02697} {\bibfield  {journal} {\bibinfo
  {journal} {The Journal of Physical Chemistry C}\ }\textbf {\bibinfo {volume}
  {122}},\ \bibinfo {pages} {11288} (\bibinfo {year} {2018})}\BibitemShut
  {NoStop}%
\bibitem [{\citenamefont {Provazza}\ \emph {et~al.}(2018)\citenamefont
  {Provazza}, \citenamefont {Segatta}, \citenamefont {Garavelli},\ and\
  \citenamefont {Coker}}]{Provazza2018}%
  \BibitemOpen
  \bibfield  {author} {\bibinfo {author} {\bibfnamefont {J.}~\bibnamefont
  {Provazza}}, \bibinfo {author} {\bibfnamefont {F.}~\bibnamefont {Segatta}},
  \bibinfo {author} {\bibfnamefont {M.}~\bibnamefont {Garavelli}},\ and\
  \bibinfo {author} {\bibfnamefont {D.~F.}\ \bibnamefont {Coker}},\ }\href
  {https://doi.org/10.1021/ACS.JCTC.7B01063/ASSET/IMAGES/CT-2017-01063N_M046.GIF}
  {\bibfield  {journal} {\bibinfo  {journal} {Journal of Chemical Theory and
  Computation}\ }\textbf {\bibinfo {volume} {14}},\ \bibinfo {pages} {856}
  (\bibinfo {year} {2018})}\BibitemShut {NoStop}%
\bibitem [{\citenamefont {Provazza}\ and\ \citenamefont
  {Coker}(2019)}]{Provazza2019}%
  \BibitemOpen
  \bibfield  {author} {\bibinfo {author} {\bibfnamefont {J.}~\bibnamefont
  {Provazza}}\ and\ \bibinfo {author} {\bibfnamefont {D.~F.}\ \bibnamefont
  {Coker}},\ }\href {https://doi.org/10.1063/1.5120253} {\bibfield  {journal}
  {\bibinfo  {journal} {The Journal of Chemical Physics}\ }\textbf {\bibinfo
  {volume} {151}},\ \bibinfo {pages} {154114} (\bibinfo {year}
  {2019})}\BibitemShut {NoStop}%
\bibitem [{\citenamefont {Mulvihill}\ \emph {et~al.}(2019)\citenamefont
  {Mulvihill}, \citenamefont {Gao}, \citenamefont {Liu}, \citenamefont
  {Schubert}, \citenamefont {Dunietz},\ and\ \citenamefont
  {Geva}}]{Mulvihill2019}%
  \BibitemOpen
  \bibfield  {author} {\bibinfo {author} {\bibfnamefont {E.}~\bibnamefont
  {Mulvihill}}, \bibinfo {author} {\bibfnamefont {X.}~\bibnamefont {Gao}},
  \bibinfo {author} {\bibfnamefont {Y.}~\bibnamefont {Liu}}, \bibinfo {author}
  {\bibfnamefont {A.}~\bibnamefont {Schubert}}, \bibinfo {author}
  {\bibfnamefont {B.~D.}\ \bibnamefont {Dunietz}},\ and\ \bibinfo {author}
  {\bibfnamefont {E.}~\bibnamefont {Geva}},\ }\href
  {https://doi.org/10.1063/1.5110891} {\bibfield  {journal} {\bibinfo
  {journal} {The Journal of Chemical Physics}\ }\textbf {\bibinfo {volume}
  {151}},\ \bibinfo {pages} {074103} (\bibinfo {year} {2019})}\BibitemShut
  {NoStop}%
\bibitem [{\citenamefont {Gao}\ \emph {et~al.}(2020)\citenamefont {Gao},
  \citenamefont {Saller}, \citenamefont {Liu}, \citenamefont {Kelly},
  \citenamefont {Richardson},\ and\ \citenamefont {Geva}}]{Gao2020}%
  \BibitemOpen
  \bibfield  {author} {\bibinfo {author} {\bibfnamefont {X.}~\bibnamefont
  {Gao}}, \bibinfo {author} {\bibfnamefont {M.~A.~C.}\ \bibnamefont {Saller}},
  \bibinfo {author} {\bibfnamefont {Y.}~\bibnamefont {Liu}}, \bibinfo {author}
  {\bibfnamefont {A.}~\bibnamefont {Kelly}}, \bibinfo {author} {\bibfnamefont
  {J.~O.}\ \bibnamefont {Richardson}},\ and\ \bibinfo {author} {\bibfnamefont
  {E.}~\bibnamefont {Geva}},\ }\href {https://doi.org/10.1021/ACS.JCTC.9B01267}
  {\bibfield  {journal} {\bibinfo  {journal} {Journal of Chemical Theory and
  Computation}\ }\textbf {\bibinfo {volume} {16}},\ \bibinfo {pages} {2883}
  (\bibinfo {year} {2020})}\BibitemShut {NoStop}%
\bibitem [{\citenamefont {Gao}\ and\ \citenamefont {Geva}(2020)}]{Gao2020b}%
  \BibitemOpen
  \bibfield  {author} {\bibinfo {author} {\bibfnamefont {X.}~\bibnamefont
  {Gao}}\ and\ \bibinfo {author} {\bibfnamefont {E.}~\bibnamefont {Geva}},\
  }\href
  {https://doi.org/10.1021/ACS.JPCA.0C09750/ASSET/IMAGES/LARGE/JP0C09750_0006.JPEG}
  {\bibfield  {journal} {\bibinfo  {journal} {Journal of Physical Chemistry A}\
  }\textbf {\bibinfo {volume} {124}},\ \bibinfo {pages} {11006} (\bibinfo
  {year} {2020})}\BibitemShut {NoStop}%
\bibitem [{\citenamefont {Dodin}\ \emph {et~al.}(2022)\citenamefont {Dodin},
  \citenamefont {Provazza}, \citenamefont {Coker},\ and\ \citenamefont
  {Willard}}]{Dodin2022}%
  \BibitemOpen
  \bibfield  {author} {\bibinfo {author} {\bibfnamefont {A.}~\bibnamefont
  {Dodin}}, \bibinfo {author} {\bibfnamefont {J.}~\bibnamefont {Provazza}},
  \bibinfo {author} {\bibfnamefont {D.~F.}\ \bibnamefont {Coker}},\ and\
  \bibinfo {author} {\bibfnamefont {A.~P.}\ \bibnamefont {Willard}},\ }\href
  {https://doi.org/10.1021/acs.jctc.1c00477} {\bibfield  {journal} {\bibinfo
  {journal} {Journal of Chemical Theory and Computation}\ }\textbf {\bibinfo
  {volume} {18}},\ \bibinfo {pages} {2047} (\bibinfo {year}
  {2022})}\BibitemShut {NoStop}%
\bibitem [{\citenamefont {Kumar}\ \emph {et~al.}(2021)\citenamefont {Kumar},
  \citenamefont {Provazza},\ and\ \citenamefont {Coker}}]{Kumar2021}%
  \BibitemOpen
  \bibfield  {author} {\bibinfo {author} {\bibfnamefont {M.}~\bibnamefont
  {Kumar}}, \bibinfo {author} {\bibfnamefont {J.}~\bibnamefont {Provazza}},\
  and\ \bibinfo {author} {\bibfnamefont {D.~F.}\ \bibnamefont {Coker}},\ }\href
  {https://doi.org/10.1063/5.0054377} {\bibfield  {journal} {\bibinfo
  {journal} {The Journal of Chemical Physics}\ }\textbf {\bibinfo {volume}
  {154}},\ \bibinfo {pages} {224109} (\bibinfo {year} {2021})}\BibitemShut
  {NoStop}%
\bibitem [{\citenamefont {Feynman}\ and\ \citenamefont
  {Hibbs}(1965)}]{Feynman1965}%
  \BibitemOpen
  \bibfield  {author} {\bibinfo {author} {\bibfnamefont {R.~P.}\ \bibnamefont
  {Feynman}}\ and\ \bibinfo {author} {\bibfnamefont {A.~R.}\ \bibnamefont
  {Hibbs}},\ }\href@noop {} {\emph {\bibinfo {title} {{Quantum Mechanics and
  Path Integrals}}}}\ (\bibinfo  {publisher} {McGraw-Hill},\ \bibinfo {year}
  {1965})\BibitemShut {NoStop}%
\bibitem [{\citenamefont {{Van Vleck}}(1928)}]{VanVleck1928}%
  \BibitemOpen
  \bibfield  {author} {\bibinfo {author} {\bibfnamefont {J.~H.}\ \bibnamefont
  {{Van Vleck}}},\ }\href {https://doi.org/10.1073/pnas.14.2.178} {\bibfield
  {journal} {\bibinfo  {journal} {Proceedings of the National Academy of
  Sciences}\ }\textbf {\bibinfo {volume} {14}},\ \bibinfo {pages} {178}
  (\bibinfo {year} {1928})}\BibitemShut {NoStop}%
\bibitem [{\citenamefont {Miller}(1970)}]{Miller1970}%
  \BibitemOpen
  \bibfield  {author} {\bibinfo {author} {\bibfnamefont {W.~H.}\ \bibnamefont
  {Miller}},\ }\href {https://doi.org/10.1063/1.1674535} {\bibfield  {journal}
  {\bibinfo  {journal} {The Journal of Chemical Physics}\ }\textbf {\bibinfo
  {volume} {53}},\ \bibinfo {pages} {3578} (\bibinfo {year}
  {1970})}\BibitemShut {NoStop}%
\bibitem [{\citenamefont {Heller}(1991{\natexlab{a}})}]{Heller1991}%
  \BibitemOpen
  \bibfield  {author} {\bibinfo {author} {\bibfnamefont {E.~J.}\ \bibnamefont
  {Heller}},\ }\href {https://doi.org/10.1063/1.459848} {\bibfield  {journal}
  {\bibinfo  {journal} {Journal of Chemical Physics}\ }\textbf {\bibinfo
  {volume} {94}},\ \bibinfo {pages} {2723} (\bibinfo {year}
  {1991}{\natexlab{a}})}\BibitemShut {NoStop}%
\bibitem [{\citenamefont {Heller}(1991{\natexlab{b}})}]{Heller1991a}%
  \BibitemOpen
  \bibfield  {author} {\bibinfo {author} {\bibfnamefont {E.~J.}\ \bibnamefont
  {Heller}},\ }\href {https://doi.org/10.1063/1.461178} {\bibfield  {journal}
  {\bibinfo  {journal} {Journal of Chemical Physics}\ }\textbf {\bibinfo
  {volume} {95}},\ \bibinfo {pages} {9431} (\bibinfo {year}
  {1991}{\natexlab{b}})}\BibitemShut {NoStop}%
\bibitem [{\citenamefont {Miller}(1991)}]{Miller1991}%
  \BibitemOpen
  \bibfield  {author} {\bibinfo {author} {\bibfnamefont {W.~H.}\ \bibnamefont
  {Miller}},\ }\href {https://doi.org/10.1063/1.461177} {\bibfield  {journal}
  {\bibinfo  {journal} {The Journal of Chemical Physics}\ }\textbf {\bibinfo
  {volume} {95}},\ \bibinfo {pages} {9428} (\bibinfo {year}
  {1991})}\BibitemShut {NoStop}%
\bibitem [{\citenamefont {Herman}\ and\ \citenamefont
  {Kluk}(1984)}]{Herman1984}%
  \BibitemOpen
  \bibfield  {author} {\bibinfo {author} {\bibfnamefont {M.~F.}\ \bibnamefont
  {Herman}}\ and\ \bibinfo {author} {\bibfnamefont {E.}~\bibnamefont {Kluk}},\
  }\href {https://doi.org/10.1016/0301-0104(84)80039-7} {\bibfield  {journal}
  {\bibinfo  {journal} {Chemical Physics}\ }\textbf {\bibinfo {volume} {91}},\
  \bibinfo {pages} {27} (\bibinfo {year} {1984})}\BibitemShut {NoStop}%
\bibitem [{\citenamefont {Sun}\ and\ \citenamefont
  {Miller}(1997{\natexlab{b}})}]{Sun1997c}%
  \BibitemOpen
  \bibfield  {author} {\bibinfo {author} {\bibfnamefont {X.}~\bibnamefont
  {Sun}}\ and\ \bibinfo {author} {\bibfnamefont {W.~H.}\ \bibnamefont
  {Miller}},\ }\href {https://doi.org/10.1063/1.473171} {\bibfield  {journal}
  {\bibinfo  {journal} {Journal of Chemical Physics}\ }\textbf {\bibinfo
  {volume} {106}},\ \bibinfo {pages} {916} (\bibinfo {year}
  {1997}{\natexlab{b}})}\BibitemShut {NoStop}%
\bibitem [{\citenamefont {Shi}\ and\ \citenamefont
  {Geva}(2003{\natexlab{b}})}]{Shi2003}%
  \BibitemOpen
  \bibfield  {author} {\bibinfo {author} {\bibfnamefont {Q.}~\bibnamefont
  {Shi}}\ and\ \bibinfo {author} {\bibfnamefont {E.}~\bibnamefont {Geva}},\
  }\href {https://doi.org/10.1063/1.1564814} {\bibfield  {journal} {\bibinfo
  {journal} {The Journal of Chemical Physics}\ }\textbf {\bibinfo {volume}
  {118}},\ \bibinfo {pages} {8173} (\bibinfo {year}
  {2003}{\natexlab{b}})}\BibitemShut {NoStop}%
\bibitem [{\citenamefont {Zhao}\ and\ \citenamefont
  {Makri}(2002)}]{YiZhao2002}%
  \BibitemOpen
  \bibfield  {author} {\bibinfo {author} {\bibfnamefont {Y.}~\bibnamefont
  {Zhao}}\ and\ \bibinfo {author} {\bibfnamefont {N.}~\bibnamefont {Makri}},\
  }\href@noop {} {\bibfield  {journal} {\bibinfo  {journal} {Chemical Physics}\
  }\textbf {\bibinfo {volume} {280}},\ \bibinfo {pages} {135} (\bibinfo {year}
  {2002})}\BibitemShut {NoStop}%
\bibitem [{\citenamefont {Wright}\ and\ \citenamefont
  {Makri}(2004)}]{Wright2004}%
  \BibitemOpen
  \bibfield  {author} {\bibinfo {author} {\bibfnamefont {N.~J.}\ \bibnamefont
  {Wright}}\ and\ \bibinfo {author} {\bibfnamefont {N.}~\bibnamefont {Makri}},\
  }\href {https://doi.org/10.1021/jp037600f} {\bibfield  {journal} {\bibinfo
  {journal} {Journal of Physical Chemistry B}\ }\textbf {\bibinfo {volume}
  {108}},\ \bibinfo {pages} {6816} (\bibinfo {year} {2004})}\BibitemShut
  {NoStop}%
\bibitem [{\citenamefont {Gelabert}\ \emph {et~al.}(2001)\citenamefont
  {Gelabert}, \citenamefont {Gim{\'{e}}nez}, \citenamefont {Thoss},
  \citenamefont {Wang},\ and\ \citenamefont {Miller}}]{Gelabert2001}%
  \BibitemOpen
  \bibfield  {author} {\bibinfo {author} {\bibfnamefont {R.}~\bibnamefont
  {Gelabert}}, \bibinfo {author} {\bibfnamefont {X.}~\bibnamefont
  {Gim{\'{e}}nez}}, \bibinfo {author} {\bibfnamefont {M.}~\bibnamefont
  {Thoss}}, \bibinfo {author} {\bibfnamefont {H.}~\bibnamefont {Wang}},\ and\
  \bibinfo {author} {\bibfnamefont {W.~H.}\ \bibnamefont {Miller}},\ }\href
  {https://doi.org/10.1063/1.1337803} {\bibfield  {journal} {\bibinfo
  {journal} {Journal of Chemical Physics}\ }\textbf {\bibinfo {volume} {114}},\
  \bibinfo {pages} {2572} (\bibinfo {year} {2001})}\BibitemShut {NoStop}%
\bibitem [{\citenamefont {Liu}\ and\ \citenamefont {Miller}(2006)}]{Liu2006}%
  \BibitemOpen
  \bibfield  {author} {\bibinfo {author} {\bibfnamefont {J.}~\bibnamefont
  {Liu}}\ and\ \bibinfo {author} {\bibfnamefont {W.~H.}\ \bibnamefont
  {Miller}},\ }\href {https://doi.org/10.1063/1.2395941} {\bibfield  {journal}
  {\bibinfo  {journal} {The Journal of Chemical Physics}\ }\textbf {\bibinfo
  {volume} {125}},\ \bibinfo {pages} {224104} (\bibinfo {year}
  {2006})}\BibitemShut {NoStop}%
\bibitem [{\citenamefont {Liu}\ and\ \citenamefont {Miller}(2007)}]{Liu2007}%
  \BibitemOpen
  \bibfield  {author} {\bibinfo {author} {\bibfnamefont {J.}~\bibnamefont
  {Liu}}\ and\ \bibinfo {author} {\bibfnamefont {W.~H.}\ \bibnamefont
  {Miller}},\ }\href {https://doi.org/10.1063/1.2774990} {\bibfield  {journal}
  {\bibinfo  {journal} {The Journal of Chemical Physics}\ }\textbf {\bibinfo
  {volume} {127}},\ \bibinfo {pages} {114506} (\bibinfo {year}
  {2007})}\BibitemShut {NoStop}%
\bibitem [{\citenamefont {Liu}\ and\ \citenamefont {Miller}(2008)}]{Liu2008}%
  \BibitemOpen
  \bibfield  {author} {\bibinfo {author} {\bibfnamefont {J.}~\bibnamefont
  {Liu}}\ and\ \bibinfo {author} {\bibfnamefont {W.~H.}\ \bibnamefont
  {Miller}},\ }\href {https://doi.org/10.1063/1.2889945} {\bibfield  {journal}
  {\bibinfo  {journal} {The Journal of Chemical Physics}\ }\textbf {\bibinfo
  {volume} {128}},\ \bibinfo {pages} {144511} (\bibinfo {year}
  {2008})}\BibitemShut {NoStop}%
\bibitem [{\citenamefont {Liu}\ \emph {et~al.}(2011{\natexlab{b}})\citenamefont
  {Liu}, \citenamefont {Alder},\ and\ \citenamefont {Miller}}]{Liu2011}%
  \BibitemOpen
  \bibfield  {author} {\bibinfo {author} {\bibfnamefont {J.}~\bibnamefont
  {Liu}}, \bibinfo {author} {\bibfnamefont {B.~J.}\ \bibnamefont {Alder}},\
  and\ \bibinfo {author} {\bibfnamefont {W.~H.}\ \bibnamefont {Miller}},\
  }\href {https://doi.org/10.1063/1.3639107} {\bibfield  {journal} {\bibinfo
  {journal} {The Journal of Chemical Physics}\ }\textbf {\bibinfo {volume}
  {135}},\ \bibinfo {pages} {114105} (\bibinfo {year}
  {2011}{\natexlab{b}})}\BibitemShut {NoStop}%
\bibitem [{\citenamefont {Monteferrante}\ \emph {et~al.}(2013)\citenamefont
  {Monteferrante}, \citenamefont {Bonella},\ and\ \citenamefont
  {Ciccotti}}]{Monteferrante2013}%
  \BibitemOpen
  \bibfield  {author} {\bibinfo {author} {\bibfnamefont {M.}~\bibnamefont
  {Monteferrante}}, \bibinfo {author} {\bibfnamefont {S.}~\bibnamefont
  {Bonella}},\ and\ \bibinfo {author} {\bibfnamefont {G.}~\bibnamefont
  {Ciccotti}},\ }\href {https://doi.org/10.1063/1.4789760} {\bibfield
  {journal} {\bibinfo  {journal} {The Journal of Chemical Physics}\ }\textbf
  {\bibinfo {volume} {138}},\ \bibinfo {pages} {054118} (\bibinfo {year}
  {2013})}\BibitemShut {NoStop}%
\bibitem [{\citenamefont {Poulsen}\ \emph {et~al.}(2003)\citenamefont
  {Poulsen}, \citenamefont {Nyman},\ and\ \citenamefont
  {Rossky}}]{Poulsen2003}%
  \BibitemOpen
  \bibfield  {author} {\bibinfo {author} {\bibfnamefont {J.~A.}\ \bibnamefont
  {Poulsen}}, \bibinfo {author} {\bibfnamefont {G.}~\bibnamefont {Nyman}},\
  and\ \bibinfo {author} {\bibfnamefont {P.~J.}\ \bibnamefont {Rossky}},\
  }\href {https://doi.org/10.1063/1.1626631} {\bibfield  {journal} {\bibinfo
  {journal} {Journal of Chemical Physics}\ }\textbf {\bibinfo {volume} {119}},\
  \bibinfo {pages} {12179} (\bibinfo {year} {2003})}\BibitemShut {NoStop}%
\bibitem [{\citenamefont {Poulsen}\ \emph
  {et~al.}(2004{\natexlab{a}})\citenamefont {Poulsen}, \citenamefont {Nyman},\
  and\ \citenamefont {Rossky}}]{Poulsen2004}%
  \BibitemOpen
  \bibfield  {author} {\bibinfo {author} {\bibfnamefont {J.~A.}\ \bibnamefont
  {Poulsen}}, \bibinfo {author} {\bibfnamefont {G.}~\bibnamefont {Nyman}},\
  and\ \bibinfo {author} {\bibfnamefont {P.~J.}\ \bibnamefont {Rossky}},\
  }\href {https://doi.org/10.1021/JP040425Y} {\bibfield  {journal} {\bibinfo
  {journal} {Journal of Physical Chemistry B}\ }\textbf {\bibinfo {volume}
  {108}},\ \bibinfo {pages} {19799} (\bibinfo {year}
  {2004}{\natexlab{a}})}\BibitemShut {NoStop}%
\bibitem [{\citenamefont {Poulsen}\ \emph
  {et~al.}(2004{\natexlab{b}})\citenamefont {Poulsen}, \citenamefont {Nyman},\
  and\ \citenamefont {Rossky}}]{Poulsen2004b}%
  \BibitemOpen
  \bibfield  {author} {\bibinfo {author} {\bibfnamefont {J.~A.}\ \bibnamefont
  {Poulsen}}, \bibinfo {author} {\bibfnamefont {G.}~\bibnamefont {Nyman}},\
  and\ \bibinfo {author} {\bibfnamefont {P.~J.}\ \bibnamefont {Rossky}},\
  }\href {https://doi.org/10.1021/jp049281d} {\bibfield  {journal} {\bibinfo
  {journal} {Journal of Physical Chemistry A}\ }\textbf {\bibinfo {volume}
  {108}},\ \bibinfo {pages} {8743} (\bibinfo {year}
  {2004}{\natexlab{b}})}\BibitemShut {NoStop}%
\bibitem [{\citenamefont {Poulsen}\ \emph {et~al.}(2007)\citenamefont
  {Poulsen}, \citenamefont {Scheers}, \citenamefont {Nyman},\ and\
  \citenamefont {Rossky}}]{Poulsen2007}%
  \BibitemOpen
  \bibfield  {author} {\bibinfo {author} {\bibfnamefont {J.~A.}\ \bibnamefont
  {Poulsen}}, \bibinfo {author} {\bibfnamefont {J.}~\bibnamefont {Scheers}},
  \bibinfo {author} {\bibfnamefont {G.}~\bibnamefont {Nyman}},\ and\ \bibinfo
  {author} {\bibfnamefont {P.~J.}\ \bibnamefont {Rossky}},\ }\href
  {https://doi.org/10.1103/PhysRevB.75.224505} {\bibfield  {journal} {\bibinfo
  {journal} {Physical Review B}\ }\textbf {\bibinfo {volume} {75}},\ \bibinfo
  {pages} {224505} (\bibinfo {year} {2007})}\BibitemShut {NoStop}%
\bibitem [{\citenamefont {Pan}\ and\ \citenamefont {Tao}(2013)}]{Pan2013}%
  \BibitemOpen
  \bibfield  {author} {\bibinfo {author} {\bibfnamefont {F.}~\bibnamefont
  {Pan}}\ and\ \bibinfo {author} {\bibfnamefont {G.}~\bibnamefont {Tao}},\
  }\href {https://doi.org/10.1063/1.4794191} {\bibfield  {journal} {\bibinfo
  {journal} {The Journal of Chemical Physics}\ }\textbf {\bibinfo {volume}
  {138}},\ \bibinfo {pages} {091101} (\bibinfo {year} {2013})}\BibitemShut
  {NoStop}%
\bibitem [{\citenamefont {Tao}\ and\ \citenamefont {Miller}(2011)}]{Tao2011}%
  \BibitemOpen
  \bibfield  {author} {\bibinfo {author} {\bibfnamefont {G.}~\bibnamefont
  {Tao}}\ and\ \bibinfo {author} {\bibfnamefont {W.~H.}\ \bibnamefont
  {Miller}},\ }\href {https://doi.org/10.1063/1.3600656} {\bibfield  {journal}
  {\bibinfo  {journal} {The Journal of Chemical Physics}\ }\textbf {\bibinfo
  {volume} {135}},\ \bibinfo {pages} {024104} (\bibinfo {year}
  {2011})}\BibitemShut {NoStop}%
\bibitem [{\citenamefont {Tao}\ and\ \citenamefont {Miller}(2012)}]{Tao2012}%
  \BibitemOpen
  \bibfield  {author} {\bibinfo {author} {\bibfnamefont {G.}~\bibnamefont
  {Tao}}\ and\ \bibinfo {author} {\bibfnamefont {W.~H.}\ \bibnamefont
  {Miller}},\ }\href {https://doi.org/10.1063/1.4752206} {\bibfield  {journal}
  {\bibinfo  {journal} {The Journal of Chemical Physics}\ }\textbf {\bibinfo
  {volume} {137}},\ \bibinfo {pages} {124105} (\bibinfo {year}
  {2012})}\BibitemShut {NoStop}%
\bibitem [{\citenamefont {Tao}\ and\ \citenamefont {Miller}(2013)}]{Tao2013}%
  \BibitemOpen
  \bibfield  {author} {\bibinfo {author} {\bibfnamefont {G.}~\bibnamefont
  {Tao}}\ and\ \bibinfo {author} {\bibfnamefont {W.~H.}\ \bibnamefont
  {Miller}},\ }\href {https://doi.org/10.1080/00268976.2013.776712} {\bibfield
  {journal} {\bibinfo  {journal} {Molecular Physics}\ }\textbf {\bibinfo
  {volume} {111}},\ \bibinfo {pages} {1987} (\bibinfo {year}
  {2013})}\BibitemShut {NoStop}%
\bibitem [{\citenamefont {Tao}(2014)}]{Tao2014c}%
  \BibitemOpen
  \bibfield  {author} {\bibinfo {author} {\bibfnamefont {G.}~\bibnamefont
  {Tao}},\ }\href {https://doi.org/10.1007/s00214-014-1448-y} {\bibfield
  {journal} {\bibinfo  {journal} {Theoretical Chemistry Accounts}\ }\textbf
  {\bibinfo {volume} {133}},\ \bibinfo {pages} {1448} (\bibinfo {year}
  {2014})}\BibitemShut {NoStop}%
\bibitem [{\citenamefont {Elran}\ and\ \citenamefont
  {Kay}(1999{\natexlab{a}})}]{Elran1999}%
  \BibitemOpen
  \bibfield  {author} {\bibinfo {author} {\bibfnamefont {Y.}~\bibnamefont
  {Elran}}\ and\ \bibinfo {author} {\bibfnamefont {K.~G.}\ \bibnamefont
  {Kay}},\ }\href {https://doi.org/10.1063/1.478255} {\bibfield  {journal}
  {\bibinfo  {journal} {The Journal of Chemical Physics}\ }\textbf {\bibinfo
  {volume} {110}},\ \bibinfo {pages} {3653} (\bibinfo {year}
  {1999}{\natexlab{a}})}\BibitemShut {NoStop}%
\bibitem [{\citenamefont {Elran}\ and\ \citenamefont
  {Kay}(1999{\natexlab{b}})}]{Elran1999a}%
  \BibitemOpen
  \bibfield  {author} {\bibinfo {author} {\bibfnamefont {Y.}~\bibnamefont
  {Elran}}\ and\ \bibinfo {author} {\bibfnamefont {K.~G.}\ \bibnamefont
  {Kay}},\ }\href {https://doi.org/10.1063/1.478810} {\bibfield  {journal}
  {\bibinfo  {journal} {The Journal of Chemical Physics}\ }\textbf {\bibinfo
  {volume} {110}},\ \bibinfo {pages} {8912} (\bibinfo {year}
  {1999}{\natexlab{b}})}\BibitemShut {NoStop}%
\bibitem [{\citenamefont {Issack}\ and\ \citenamefont
  {Roy}(2007{\natexlab{b}})}]{Issack2007}%
  \BibitemOpen
  \bibfield  {author} {\bibinfo {author} {\bibfnamefont {B.~B.}\ \bibnamefont
  {Issack}}\ and\ \bibinfo {author} {\bibfnamefont {P.-N.}\ \bibnamefont
  {Roy}},\ }\href {https://doi.org/10.1063/1.2786456} {\bibfield  {journal}
  {\bibinfo  {journal} {The Journal of Chemical Physics}\ }\textbf {\bibinfo
  {volume} {127}},\ \bibinfo {pages} {144306} (\bibinfo {year}
  {2007}{\natexlab{b}})}\BibitemShut {NoStop}%
\bibitem [{\citenamefont {Makri}\ and\ \citenamefont
  {Thompson}(1998)}]{Makri1998a}%
  \BibitemOpen
  \bibfield  {author} {\bibinfo {author} {\bibfnamefont {N.}~\bibnamefont
  {Makri}}\ and\ \bibinfo {author} {\bibfnamefont {K.}~\bibnamefont
  {Thompson}},\ }\href {https://doi.org/10.1016/S0009-2614(98)00590-9}
  {\bibfield  {journal} {\bibinfo  {journal} {Chemical Physics Letters}\
  }\textbf {\bibinfo {volume} {291}},\ \bibinfo {pages} {101} (\bibinfo {year}
  {1998})}\BibitemShut {NoStop}%
\bibitem [{\citenamefont {Thompson}\ and\ \citenamefont
  {Makri}(1999{\natexlab{a}})}]{Thompson1999a}%
  \BibitemOpen
  \bibfield  {author} {\bibinfo {author} {\bibfnamefont {K.}~\bibnamefont
  {Thompson}}\ and\ \bibinfo {author} {\bibfnamefont {N.}~\bibnamefont
  {Makri}},\ }\href@noop {} {\bibfield  {journal} {\bibinfo  {journal}
  {Physical Review E}\ }\textbf {\bibinfo {volume} {59}},\ \bibinfo {pages}
  {R4729} (\bibinfo {year} {1999}{\natexlab{a}})}\BibitemShut {NoStop}%
\bibitem [{\citenamefont {Thompson}\ and\ \citenamefont
  {Makri}(1999{\natexlab{b}})}]{Thompson1999b}%
  \BibitemOpen
  \bibfield  {author} {\bibinfo {author} {\bibfnamefont {K.}~\bibnamefont
  {Thompson}}\ and\ \bibinfo {author} {\bibfnamefont {N.}~\bibnamefont
  {Makri}},\ }\href {https://doi.org/10.1063/1.478011} {\bibfield  {journal}
  {\bibinfo  {journal} {The Journal of Chemical Physics}\ }\textbf {\bibinfo
  {volume} {110}},\ \bibinfo {pages} {64112} (\bibinfo {year}
  {1999}{\natexlab{b}})}\BibitemShut {NoStop}%
\bibitem [{\citenamefont {Nakayama}\ and\ \citenamefont
  {Makri}(2003)}]{Nakayama2003}%
  \BibitemOpen
  \bibfield  {author} {\bibinfo {author} {\bibfnamefont {A.}~\bibnamefont
  {Nakayama}}\ and\ \bibinfo {author} {\bibfnamefont {N.}~\bibnamefont
  {Makri}},\ }\href {https://doi.org/10.1063/1.1611473} {\bibfield  {journal}
  {\bibinfo  {journal} {The Journal of Chemical Physics}\ }\textbf {\bibinfo
  {volume} {119}},\ \bibinfo {pages} {8592} (\bibinfo {year}
  {2003})}\BibitemShut {NoStop}%
\bibitem [{\citenamefont {Wright}\ and\ \citenamefont
  {Makri}(2003)}]{Wright2003}%
  \BibitemOpen
  \bibfield  {author} {\bibinfo {author} {\bibfnamefont {N.~J.}\ \bibnamefont
  {Wright}}\ and\ \bibinfo {author} {\bibfnamefont {N.}~\bibnamefont {Makri}},\
  }\href {https://doi.org/10.1063/1.1580472} {\bibfield  {journal} {\bibinfo
  {journal} {The Journal of Chemical Physics}\ }\textbf {\bibinfo {volume}
  {119}},\ \bibinfo {pages} {1634} (\bibinfo {year} {2003})}\BibitemShut
  {NoStop}%
\bibitem [{\citenamefont {Makri}\ \emph {et~al.}(2004)\citenamefont {Makri},
  \citenamefont {Nakayama},\ and\ \citenamefont {Wright}}]{Makri2004}%
  \BibitemOpen
  \bibfield  {author} {\bibinfo {author} {\bibfnamefont {N.}~\bibnamefont
  {Makri}}, \bibinfo {author} {\bibfnamefont {A.}~\bibnamefont {Nakayama}},\
  and\ \bibinfo {author} {\bibfnamefont {N.~J.}\ \bibnamefont {Wright}},\
  }\href {https://doi.org/10.1142/S0219633604001112} {\bibfield  {journal}
  {\bibinfo  {journal} {Journal of Theoretical and Computational Chemistry}\
  }\textbf {\bibinfo {volume} {3}},\ \bibinfo {pages} {391} (\bibinfo {year}
  {2004})}\BibitemShut {NoStop}%
\bibitem [{\citenamefont {Lawrence}\ \emph {et~al.}(2004)\citenamefont
  {Lawrence}, \citenamefont {Nakayama}, \citenamefont {Makri},\ and\
  \citenamefont {Skinner}}]{Lawrence2004}%
  \BibitemOpen
  \bibfield  {author} {\bibinfo {author} {\bibfnamefont {C.~P.}\ \bibnamefont
  {Lawrence}}, \bibinfo {author} {\bibfnamefont {A.}~\bibnamefont {Nakayama}},
  \bibinfo {author} {\bibfnamefont {N.}~\bibnamefont {Makri}},\ and\ \bibinfo
  {author} {\bibfnamefont {J.~L.}\ \bibnamefont {Skinner}},\ }\href
  {https://doi.org/10.1063/1.1645783} {\bibfield  {journal} {\bibinfo
  {journal} {The Journal of Chemical Physics}\ }\textbf {\bibinfo {volume}
  {120}},\ \bibinfo {pages} {6621} (\bibinfo {year} {2004})}\BibitemShut
  {NoStop}%
\bibitem [{\citenamefont {Makri}(2011)}]{Makri2011}%
  \BibitemOpen
  \bibfield  {author} {\bibinfo {author} {\bibfnamefont {N.}~\bibnamefont
  {Makri}},\ }\href {https://doi.org/10.1039/c0cp02374d} {\bibfield  {journal}
  {\bibinfo  {journal} {Phys. Chem. Chem. Phys}\ }\textbf {\bibinfo {volume}
  {13}},\ \bibinfo {pages} {14442} (\bibinfo {year} {2011})}\BibitemShut
  {NoStop}%
\bibitem [{\citenamefont {Tao}\ and\ \citenamefont {Miller}(2009)}]{Tao2009}%
  \BibitemOpen
  \bibfield  {author} {\bibinfo {author} {\bibfnamefont {G.}~\bibnamefont
  {Tao}}\ and\ \bibinfo {author} {\bibfnamefont {W.~H.}\ \bibnamefont
  {Miller}},\ }\href {https://doi.org/10.1063/1.3132224} {\bibfield  {journal}
  {\bibinfo  {journal} {The Journal of Chemical Physics}\ }\textbf {\bibinfo
  {volume} {130}},\ \bibinfo {pages} {184108} (\bibinfo {year}
  {2009})}\BibitemShut {NoStop}%
\bibitem [{\citenamefont {Grossmann}(2006)}]{Grossmann2006}%
  \BibitemOpen
  \bibfield  {author} {\bibinfo {author} {\bibfnamefont {F.}~\bibnamefont
  {Grossmann}},\ }\href {https://doi.org/10.1063/1.2213255} {\bibfield
  {journal} {\bibinfo  {journal} {The Journal of Chemical Physics}\ }\textbf
  {\bibinfo {volume} {125}},\ \bibinfo {pages} {014111} (\bibinfo {year}
  {2006})}\BibitemShut {NoStop}%
\bibitem [{\citenamefont {Goletz}\ and\ \citenamefont
  {Grossmann}(2009)}]{Goletz2009}%
  \BibitemOpen
  \bibfield  {author} {\bibinfo {author} {\bibfnamefont {C.-M.}\ \bibnamefont
  {Goletz}}\ and\ \bibinfo {author} {\bibfnamefont {F.}~\bibnamefont
  {Grossmann}},\ }\href {https://doi.org/10.1063/1.3157162} {\bibfield
  {journal} {\bibinfo  {journal} {The Journal of Chemical Physics}\ }\textbf
  {\bibinfo {volume} {130}},\ \bibinfo {pages} {244107} (\bibinfo {year}
  {2009})}\BibitemShut {NoStop}%
\bibitem [{\citenamefont {Goletz}\ \emph {et~al.}(2010)\citenamefont {Goletz},
  \citenamefont {Koch},\ and\ \citenamefont {Grossmann}}]{Goletz2010}%
  \BibitemOpen
  \bibfield  {author} {\bibinfo {author} {\bibfnamefont {C.~M.}\ \bibnamefont
  {Goletz}}, \bibinfo {author} {\bibfnamefont {W.}~\bibnamefont {Koch}},\ and\
  \bibinfo {author} {\bibfnamefont {F.}~\bibnamefont {Grossmann}},\ }\href
  {https://doi.org/10.1016/J.CHEMPHYS.2010.06.019} {\bibfield  {journal}
  {\bibinfo  {journal} {Chemical Physics}\ }\textbf {\bibinfo {volume} {375}},\
  \bibinfo {pages} {227} (\bibinfo {year} {2010})}\BibitemShut {NoStop}%
\bibitem [{\citenamefont {Heller}(1975)}]{Heller1975}%
  \BibitemOpen
  \bibfield  {author} {\bibinfo {author} {\bibfnamefont {E.~J.}\ \bibnamefont
  {Heller}},\ }\href {https://doi.org/10.1063/1.430620} {\bibfield  {journal}
  {\bibinfo  {journal} {The Journal of Chemical Physics}\ }\textbf {\bibinfo
  {volume} {62}},\ \bibinfo {pages} {1544} (\bibinfo {year}
  {1975})}\BibitemShut {NoStop}%
\bibitem [{\citenamefont {Littlejohn}(1986)}]{Littlejohn1986}%
  \BibitemOpen
  \bibfield  {author} {\bibinfo {author} {\bibfnamefont {R.~G.}\ \bibnamefont
  {Littlejohn}},\ }\href {https://doi.org/10.1016/0370-1573(86)90103-1}
  {\bibfield  {journal} {\bibinfo  {journal} {Physics Reports}\ }\textbf
  {\bibinfo {volume} {138}},\ \bibinfo {pages} {193} (\bibinfo {year}
  {1986})}\BibitemShut {NoStop}%
\bibitem [{\citenamefont {Deshpande}\ and\ \citenamefont
  {Ezra}(2006)}]{Deshpande2006}%
  \BibitemOpen
  \bibfield  {author} {\bibinfo {author} {\bibfnamefont {S.~A.}\ \bibnamefont
  {Deshpande}}\ and\ \bibinfo {author} {\bibfnamefont {G.~S.}\ \bibnamefont
  {Ezra}},\ }\href {https://doi.org/10.1088/0305-4470/39/18/020} {\bibfield
  {journal} {\bibinfo  {journal} {Journal of Physics A: Mathematical and
  General}\ }\textbf {\bibinfo {volume} {39}},\ \bibinfo {pages} {5067}
  (\bibinfo {year} {2006})}\BibitemShut {NoStop}%
\bibitem [{\citenamefont {Gelabert}\ \emph {et~al.}(2000)\citenamefont
  {Gelabert}, \citenamefont {Gim{\'{e}}nez}, \citenamefont {Thoss},
  \citenamefont {Wang},\ and\ \citenamefont {Miller}}]{Gelabert2000d}%
  \BibitemOpen
  \bibfield  {author} {\bibinfo {author} {\bibfnamefont {R.}~\bibnamefont
  {Gelabert}}, \bibinfo {author} {\bibfnamefont {X.}~\bibnamefont
  {Gim{\'{e}}nez}}, \bibinfo {author} {\bibfnamefont {M.}~\bibnamefont
  {Thoss}}, \bibinfo {author} {\bibfnamefont {H.}~\bibnamefont {Wang}},\ and\
  \bibinfo {author} {\bibfnamefont {W.~H.}\ \bibnamefont {Miller}},\ }\href
  {https://doi.org/10.1021/jp0012451} {\bibfield  {journal} {\bibinfo
  {journal} {Journal of Physical Chemistry A}\ }\textbf {\bibinfo {volume}
  {104}},\ \bibinfo {pages} {10321} (\bibinfo {year} {2000})}\BibitemShut
  {NoStop}%
\bibitem [{\citenamefont {Issack}\ and\ \citenamefont
  {Roy}(2005)}]{Issack2005}%
  \BibitemOpen
  \bibfield  {author} {\bibinfo {author} {\bibfnamefont {B.~B.}\ \bibnamefont
  {Issack}}\ and\ \bibinfo {author} {\bibfnamefont {P.-N.}\ \bibnamefont
  {Roy}},\ }\href {https://doi.org/10.1063/1.2004947} {\bibfield  {journal}
  {\bibinfo  {journal} {The Journal of Chemical Physics}\ }\textbf {\bibinfo
  {volume} {123}},\ \bibinfo {pages} {084103} (\bibinfo {year}
  {2005})}\BibitemShut {NoStop}%
\bibitem [{\citenamefont {Issack}\ and\ \citenamefont
  {Roy}(2007{\natexlab{c}})}]{Issack2007a}%
  \BibitemOpen
  \bibfield  {author} {\bibinfo {author} {\bibfnamefont {B.~B.}\ \bibnamefont
  {Issack}}\ and\ \bibinfo {author} {\bibfnamefont {P.-N.}\ \bibnamefont
  {Roy}},\ }\href {https://doi.org/10.1063/1.2423019} {\bibfield  {journal}
  {\bibinfo  {journal} {The Journal of Chemical Physics}\ }\textbf {\bibinfo
  {volume} {126}},\ \bibinfo {pages} {024111} (\bibinfo {year}
  {2007}{\natexlab{c}})}\BibitemShut {NoStop}%
\bibitem [{\citenamefont {Guallar}\ \emph {et~al.}(1999)\citenamefont
  {Guallar}, \citenamefont {Batista},\ and\ \citenamefont
  {Miller}}]{Guallar1999}%
  \BibitemOpen
  \bibfield  {author} {\bibinfo {author} {\bibfnamefont {V.}~\bibnamefont
  {Guallar}}, \bibinfo {author} {\bibfnamefont {V.~S.}\ \bibnamefont
  {Batista}},\ and\ \bibinfo {author} {\bibfnamefont {W.~H.}\ \bibnamefont
  {Miller}},\ }\href {https://doi.org/10.1063/1.478866} {\bibfield  {journal}
  {\bibinfo  {journal} {Journal of Chemical Physics}\ }\textbf {\bibinfo
  {volume} {110}},\ \bibinfo {pages} {9922} (\bibinfo {year}
  {1999})}\BibitemShut {NoStop}%
\bibitem [{\citenamefont {Guallar}\ \emph {et~al.}(2000)\citenamefont
  {Guallar}, \citenamefont {Batista},\ and\ \citenamefont
  {Miller}}]{Guallar2000}%
  \BibitemOpen
  \bibfield  {author} {\bibinfo {author} {\bibfnamefont {V.}~\bibnamefont
  {Guallar}}, \bibinfo {author} {\bibfnamefont {V.~S.}\ \bibnamefont
  {Batista}},\ and\ \bibinfo {author} {\bibfnamefont {W.~H.}\ \bibnamefont
  {Miller}},\ }\href {https://doi.org/10.1063/1.1321049} {\bibfield  {journal}
  {\bibinfo  {journal} {The Journal of Chemical Physics}\ }\textbf {\bibinfo
  {volume} {113}},\ \bibinfo {pages} {9510} (\bibinfo {year}
  {2000})}\BibitemShut {NoStop}%
\bibitem [{\citenamefont {Tatchen}\ \emph {et~al.}(2011)\citenamefont
  {Tatchen}, \citenamefont {Pollak}, \citenamefont {Tao},\ and\ \citenamefont
  {Miller}}]{Tatchen2011}%
  \BibitemOpen
  \bibfield  {author} {\bibinfo {author} {\bibfnamefont {J.}~\bibnamefont
  {Tatchen}}, \bibinfo {author} {\bibfnamefont {E.}~\bibnamefont {Pollak}},
  \bibinfo {author} {\bibfnamefont {G.}~\bibnamefont {Tao}},\ and\ \bibinfo
  {author} {\bibfnamefont {W.~H.}\ \bibnamefont {Miller}},\ }\href
  {https://doi.org/10.1063/1.3573566} {\bibfield  {journal} {\bibinfo
  {journal} {The Journal of Chemical Physics}\ }\textbf {\bibinfo {volume}
  {134}},\ \bibinfo {pages} {134104} (\bibinfo {year} {2011})}\BibitemShut
  {NoStop}%
\bibitem [{\citenamefont {Zhang}\ and\ \citenamefont
  {Pollak}(2004)}]{Zhang2004}%
  \BibitemOpen
  \bibfield  {author} {\bibinfo {author} {\bibfnamefont {S.}~\bibnamefont
  {Zhang}}\ and\ \bibinfo {author} {\bibfnamefont {E.}~\bibnamefont {Pollak}},\
  }\href {https://doi.org/10.1063/1.1772361} {\bibfield  {journal} {\bibinfo
  {journal} {Journal of Chemical Physics}\ }\textbf {\bibinfo {volume} {121}},\
  \bibinfo {pages} {3384} (\bibinfo {year} {2004})}\BibitemShut {NoStop}%
\bibitem [{\citenamefont {Pollak}\ and\ \citenamefont
  {Shao}(2003)}]{Pollak2003}%
  \BibitemOpen
  \bibfield  {author} {\bibinfo {author} {\bibfnamefont {E.}~\bibnamefont
  {Pollak}}\ and\ \bibinfo {author} {\bibfnamefont {J.}~\bibnamefont {Shao}},\
  }\href {https://doi.org/10.1021/JP030098E} {\bibfield  {journal} {\bibinfo
  {journal} {Journal of Physical Chemistry A}\ }\textbf {\bibinfo {volume}
  {107}},\ \bibinfo {pages} {7112} (\bibinfo {year} {2003})}\BibitemShut
  {NoStop}%
\bibitem [{\citenamefont {Zhang}\ and\ \citenamefont
  {Pollak}(2003{\natexlab{a}})}]{Zhang2003}%
  \BibitemOpen
  \bibfield  {author} {\bibinfo {author} {\bibfnamefont {S.}~\bibnamefont
  {Zhang}}\ and\ \bibinfo {author} {\bibfnamefont {E.}~\bibnamefont {Pollak}},\
  }\href {https://doi.org/10.1103/PhysRevLett.91.190201} {\bibfield  {journal}
  {\bibinfo  {journal} {Physical Review Letters}\ }\textbf {\bibinfo {volume}
  {91}},\ \bibinfo {pages} {190201} (\bibinfo {year}
  {2003}{\natexlab{a}})}\BibitemShut {NoStop}%
\bibitem [{\citenamefont {Zhang}\ and\ \citenamefont
  {Pollak}(2003{\natexlab{b}})}]{Zhang2003a}%
  \BibitemOpen
  \bibfield  {author} {\bibinfo {author} {\bibfnamefont {S.}~\bibnamefont
  {Zhang}}\ and\ \bibinfo {author} {\bibfnamefont {E.}~\bibnamefont {Pollak}},\
  }\href {https://doi.org/10.1063/1.1622931} {\bibfield  {journal} {\bibinfo
  {journal} {The Journal of Chemical Physics}\ }\textbf {\bibinfo {volume}
  {119}},\ \bibinfo {pages} {11058} (\bibinfo {year}
  {2003}{\natexlab{b}})}\BibitemShut {NoStop}%
\bibitem [{\citenamefont {Antipov}\ \emph {et~al.}(2015)\citenamefont
  {Antipov}, \citenamefont {Ye},\ and\ \citenamefont {Ananth}}]{Antipov2015}%
  \BibitemOpen
  \bibfield  {author} {\bibinfo {author} {\bibfnamefont {S.~V.}\ \bibnamefont
  {Antipov}}, \bibinfo {author} {\bibfnamefont {Z.}~\bibnamefont {Ye}},\ and\
  \bibinfo {author} {\bibfnamefont {N.}~\bibnamefont {Ananth}},\ }\href
  {https://doi.org/10.1063/1.4919667} {\bibfield  {journal} {\bibinfo
  {journal} {The Journal of Chemical Physics}\ }\textbf {\bibinfo {volume}
  {142}},\ \bibinfo {pages} {184102} (\bibinfo {year} {2015})}\BibitemShut
  {NoStop}%
\bibitem [{\citenamefont {Church}\ \emph {et~al.}(2017)\citenamefont {Church},
  \citenamefont {Antipov},\ and\ \citenamefont {Ananth}}]{Church2017}%
  \BibitemOpen
  \bibfield  {author} {\bibinfo {author} {\bibfnamefont {M.~S.}\ \bibnamefont
  {Church}}, \bibinfo {author} {\bibfnamefont {S.~V.}\ \bibnamefont
  {Antipov}},\ and\ \bibinfo {author} {\bibfnamefont {N.}~\bibnamefont
  {Ananth}},\ }\href {https://doi.org/10.1063/1.4986645} {\bibfield  {journal}
  {\bibinfo  {journal} {The Journal of Chemical Physics}\ }\textbf {\bibinfo
  {volume} {146}},\ \bibinfo {pages} {234104} (\bibinfo {year}
  {2017})}\BibitemShut {NoStop}%
\bibitem [{\citenamefont {Church}\ and\ \citenamefont
  {Ananth}(2019)}]{Church2019a}%
  \BibitemOpen
  \bibfield  {author} {\bibinfo {author} {\bibfnamefont {M.~S.}\ \bibnamefont
  {Church}}\ and\ \bibinfo {author} {\bibfnamefont {N.}~\bibnamefont
  {Ananth}},\ }\href {https://doi.org/10.1063/1.5117160} {\bibfield  {journal}
  {\bibinfo  {journal} {The Journal of Chemical Physics}\ }\textbf {\bibinfo
  {volume} {151}},\ \bibinfo {pages} {134109} (\bibinfo {year}
  {2019})}\BibitemShut {NoStop}%
\bibitem [{\citenamefont {Church}\ \emph {et~al.}(2018)\citenamefont {Church},
  \citenamefont {Hele}, \citenamefont {Ezra},\ and\ \citenamefont
  {Ananth}}]{Church2018}%
  \BibitemOpen
  \bibfield  {author} {\bibinfo {author} {\bibfnamefont {M.~S.}\ \bibnamefont
  {Church}}, \bibinfo {author} {\bibfnamefont {T.~J.~H.}\ \bibnamefont {Hele}},
  \bibinfo {author} {\bibfnamefont {G.~S.}\ \bibnamefont {Ezra}},\ and\
  \bibinfo {author} {\bibfnamefont {N.}~\bibnamefont {Ananth}},\ }\href
  {https://doi.org/10.1063/1.5005557} {\bibfield  {journal} {\bibinfo
  {journal} {The Journal of Chemical Physics}\ }\textbf {\bibinfo {volume}
  {148}},\ \bibinfo {pages} {102326} (\bibinfo {year} {2018})}\BibitemShut
  {NoStop}%
\bibitem [{\citenamefont {Filinov}(1986)}]{Filinov1986a}%
  \BibitemOpen
  \bibfield  {author} {\bibinfo {author} {\bibfnamefont {V.}~\bibnamefont
  {Filinov}},\ }\href {https://doi.org/10.1016/S0550-3213(86)80034-7}
  {\bibfield  {journal} {\bibinfo  {journal} {Nuclear Physics B}\ }\textbf
  {\bibinfo {volume} {271}},\ \bibinfo {pages} {717} (\bibinfo {year}
  {1986})}\BibitemShut {NoStop}%
\bibitem [{\citenamefont {Makri}\ and\ \citenamefont
  {Miller}(1987)}]{Makri1987c}%
  \BibitemOpen
  \bibfield  {author} {\bibinfo {author} {\bibfnamefont {N.}~\bibnamefont
  {Makri}}\ and\ \bibinfo {author} {\bibfnamefont {W.~H.}\ \bibnamefont
  {Miller}},\ }\href {https://doi.org/10.1016/0009-2614(87)80142-2} {\bibfield
  {journal} {\bibinfo  {journal} {Chemical Physics Letters}\ }\textbf {\bibinfo
  {volume} {139}},\ \bibinfo {pages} {10} (\bibinfo {year} {1987})}\BibitemShut
  {NoStop}%
\bibitem [{\citenamefont {Makri}\ and\ \citenamefont
  {Miller}(1988)}]{Makri1988b}%
  \BibitemOpen
  \bibfield  {author} {\bibinfo {author} {\bibfnamefont {N.}~\bibnamefont
  {Makri}}\ and\ \bibinfo {author} {\bibfnamefont {W.~H.}\ \bibnamefont
  {Miller}},\ }\href {https://doi.org/10.1063/1.455061} {\bibfield  {journal}
  {\bibinfo  {journal} {The Journal of Chemical Physics}\ }\textbf {\bibinfo
  {volume} {89}},\ \bibinfo {pages} {2170} (\bibinfo {year}
  {1988})}\BibitemShut {NoStop}%
\bibitem [{\citenamefont {Thoss}\ \emph {et~al.}(2001)\citenamefont {Thoss},
  \citenamefont {Wang},\ and\ \citenamefont {Miller}}]{Thoss2001a}%
  \BibitemOpen
  \bibfield  {author} {\bibinfo {author} {\bibfnamefont {M.}~\bibnamefont
  {Thoss}}, \bibinfo {author} {\bibfnamefont {H.}~\bibnamefont {Wang}},\ and\
  \bibinfo {author} {\bibfnamefont {W.~H.}\ \bibnamefont {Miller}},\ }\href
  {https://doi.org/10.1063/1.1359242} {\bibfield  {journal} {\bibinfo
  {journal} {The Journal of Chemical Physics}\ }\textbf {\bibinfo {volume}
  {114}},\ \bibinfo {pages} {9220} (\bibinfo {year} {2001})}\BibitemShut
  {NoStop}%
\bibitem [{cor()}]{corrcode}%
  \BibitemOpen
  \href {https://doi.org/10.5281/zenodo.3945531} {\bibinfo {title} {Sc-corr:
  open-source software for the calculation of sc-ivr and sc-ivr based
  approximations to quantum real-time correlation functions.}}\BibitemShut
  {Stop}%
\bibitem [{\citenamefont {Colbert}\ and\ \citenamefont
  {Miller}(1992)}]{Colbert1992a}%
  \BibitemOpen
  \bibfield  {author} {\bibinfo {author} {\bibfnamefont {D.~T.}\ \bibnamefont
  {Colbert}}\ and\ \bibinfo {author} {\bibfnamefont {W.~H.}\ \bibnamefont
  {Miller}},\ }\href {https://doi.org/10.1063/1.462100} {\bibfield  {journal}
  {\bibinfo  {journal} {J. Chem. Phys.}\ }\textbf {\bibinfo {volume} {96}},\
  \bibinfo {pages} {1982} (\bibinfo {year} {1992})}\BibitemShut {NoStop}%
\bibitem [{\citenamefont {Meyer}\ and\ \citenamefont
  {Miller}(1979)}]{Meyer1979}%
  \BibitemOpen
  \bibfield  {author} {\bibinfo {author} {\bibfnamefont {H.~D.}\ \bibnamefont
  {Meyer}}\ and\ \bibinfo {author} {\bibfnamefont {W.~H.}\ \bibnamefont
  {Miller}},\ }\href {https://doi.org/10.1063/1.437910} {\bibfield  {journal}
  {\bibinfo  {journal} {The Journal of Chemical Physics}\ }\textbf {\bibinfo
  {volume} {70}},\ \bibinfo {pages} {3214} (\bibinfo {year}
  {1979})}\BibitemShut {NoStop}%
\bibitem [{\citenamefont {Stock}\ and\ \citenamefont
  {Thoss}(1997)}]{Stock1997}%
  \BibitemOpen
  \bibfield  {author} {\bibinfo {author} {\bibfnamefont {G.}~\bibnamefont
  {Stock}}\ and\ \bibinfo {author} {\bibfnamefont {M.}~\bibnamefont {Thoss}},\
  }\href@noop {} {\bibfield  {journal} {\bibinfo  {journal} {Physical Review
  Letters}\ }\textbf {\bibinfo {volume} {78}},\ \bibinfo {pages} {578}
  (\bibinfo {year} {1997})}\BibitemShut {NoStop}%
\bibitem [{\citenamefont {Kelly}\ \emph {et~al.}(2012)\citenamefont {Kelly},
  \citenamefont {van Zon}, \citenamefont {Schofield},\ and\ \citenamefont
  {Kapral}}]{Kelly2012}%
  \BibitemOpen
  \bibfield  {author} {\bibinfo {author} {\bibfnamefont {A.}~\bibnamefont
  {Kelly}}, \bibinfo {author} {\bibfnamefont {R.}~\bibnamefont {van Zon}},
  \bibinfo {author} {\bibfnamefont {J.}~\bibnamefont {Schofield}},\ and\
  \bibinfo {author} {\bibfnamefont {R.}~\bibnamefont {Kapral}},\ }\href
  {https://doi.org/10.1063/1.3685420} {\bibfield  {journal} {\bibinfo
  {journal} {The Journal of Chemical Physics}\ }\textbf {\bibinfo {volume}
  {136}},\ \bibinfo {pages} {084101} (\bibinfo {year} {2012})}\BibitemShut
  {NoStop}%
\bibitem [{\citenamefont {Topaler}\ and\ \citenamefont
  {Makri}(1994)}]{Topaler1994}%
  \BibitemOpen
  \bibfield  {author} {\bibinfo {author} {\bibfnamefont {M.}~\bibnamefont
  {Topaler}}\ and\ \bibinfo {author} {\bibfnamefont {N.}~\bibnamefont
  {Makri}},\ }\href {https://doi.org/10.1063/1.468244} {\bibfield  {journal}
  {\bibinfo  {journal} {Journal of Chemical Physics}\ }\textbf {\bibinfo
  {volume} {101}},\ \bibinfo {pages} {7500} (\bibinfo {year}
  {1994})}\BibitemShut {NoStop}%
\bibitem [{\citenamefont {Lambert}\ and\ \citenamefont
  {Makri}(2012{\natexlab{a}})}]{Lambert2012a}%
  \BibitemOpen
  \bibfield  {author} {\bibinfo {author} {\bibfnamefont {R.}~\bibnamefont
  {Lambert}}\ and\ \bibinfo {author} {\bibfnamefont {N.}~\bibnamefont
  {Makri}},\ }\href {https://doi.org/10.1063/1.4767931} {\bibfield  {journal}
  {\bibinfo  {journal} {Journal of Chemical Physics}\ }\textbf {\bibinfo
  {volume} {137}},\ \bibinfo {pages} {22A552} (\bibinfo {year}
  {2012}{\natexlab{a}})}\BibitemShut {NoStop}%
\bibitem [{\citenamefont {Lambert}\ and\ \citenamefont
  {Makri}(2012{\natexlab{b}})}]{Lambert2012b}%
  \BibitemOpen
  \bibfield  {author} {\bibinfo {author} {\bibfnamefont {R.}~\bibnamefont
  {Lambert}}\ and\ \bibinfo {author} {\bibfnamefont {N.}~\bibnamefont
  {Makri}},\ }\href {https://doi.org/10.1063/1.4767980} {\bibfield  {journal}
  {\bibinfo  {journal} {Journal of Chemical Physics}\ }\textbf {\bibinfo
  {volume} {137}},\ \bibinfo {pages} {22A553} (\bibinfo {year}
  {2012}{\natexlab{b}})}\BibitemShut {NoStop}%
\bibitem [{\citenamefont {{Loho Choudhury}}\ and\ \citenamefont
  {Gro{\ss}mann}(2020)}]{LohoChoudhury2020}%
  \BibitemOpen
  \bibfield  {author} {\bibinfo {author} {\bibfnamefont {S.}~\bibnamefont
  {{Loho Choudhury}}}\ and\ \bibinfo {author} {\bibfnamefont {F.}~\bibnamefont
  {Gro{\ss}mann}},\ }\href {https://doi.org/10.3390/condmat5010003} {\bibfield
  {journal} {\bibinfo  {journal} {Condensed Matter}\ }\textbf {\bibinfo
  {volume} {5}},\ \bibinfo {pages} {3} (\bibinfo {year} {2020})}\BibitemShut
  {NoStop}%
\end{thebibliography}%

\end{document}